\RequirePackage{rotating}
\documentclass[11pt,twocolumn]{aastex62}
\usepackage{amssymb,amsmath}
\usepackage{apjfonts}

\newcommand{\msun}{M$_{\odot}$}

\newcommand{\gadget}{\texttt{GADGET}}

\newcommand{\gadthree}{\texttt{GADGET-3}}
\newcommand{\sunrise}{\texttt{SUNRISE}}

\usepackage{graphics}
\usepackage{amsmath}
\usepackage{relsize}
\usepackage{natbib}
\usepackage{longtable}
\usepackage{rotating} 

\begin{document}

\title{
Accurate Identification of Galaxy Mergers with Imaging}

\author{R. Nevin}
\affil{Department of Astrophysical and Planetary Sciences, University of Colorado, Boulder, CO 80309, USA}

\author{L. Blecha}
\affil{Department of Physics, University of Florida, Gainesville, FL 32611, USA}

\author{J. Comerford}
\affil{Department of Astrophysical and Planetary Sciences, University of Colorado, Boulder, CO 80309, USA}

\author{J. Greene}
\affil{Department of Astrophysical Sciences, Princeton University, Princeton, NJ 08544, USA}

\begin{abstract}

Merging galaxies play a key role in galaxy evolution, and progress in our understanding of galaxy evolution is slowed by the difficulty of making accurate galaxy merger identifications. We use \texttt{GADGET-3} hydrodynamical simulations of merging galaxies with the dust radiative transfer code \texttt{SUNRISE} to produce a suite of merging galaxies that span a range of initial conditions. This includes simulated mergers that are gas poor and gas rich and that have a range of mass ratios (minor and major). We adapt the simulated images to the specifications of the SDSS imaging survey and develop a merging galaxy classification scheme that is based on this imaging. We leverage the strengths of seven individual imaging predictors ($Gini$, $M_{20}$, concentration, asymmetry, clumpiness, S\'ersic index, and shape asymmetry) by combining them into one classifier that utilizes Linear Discriminant Analysis. It outperforms individual imaging predictors in accuracy, precision, and merger observability timescale ($>2$ Gyr for all merger simulations). We find that the classification depends strongly on mass ratio and depends weakly on the gas fraction of the simulated mergers; asymmetry is more important for the major mergers, while concentration is more important for the minor mergers. This is a result of the relatively disturbed morphology of major mergers and the steadier growth of stellar bulges during minor mergers. Since mass ratio has the largest effect on the classification, we create separate classification approaches for minor and major mergers that can be applied to SDSS imaging or adapted for other imaging surveys. 

\end{abstract}

\section{Introduction}

In the current $\Lambda$ cold dark matter ($\Lambda$CDM) framework for structure formation in the universe, galaxies form as gas cools at the center of dark matter halos (e.g., \citealt{White1978,White1991,Cole2008}). These galaxies then grow through gas accretion and mergers from small, irregular galaxies with high rates of star formation to large, quiescent galaxies with lower rates of star formation in the local universe (e.g., \citealt{Glazebrook1995,Lilly1995,Giavalisco1996}).

Simultaneously, supermassive black holes (SMBHs), which are found at the centers of all massive galaxies, have accumulated mass over time. Both SMBHs and galaxies grow through the accretion of gas; SMBHs that are actively accreting gas are known as active galactic nuclei (AGNs) and can be among the most luminous objects in the universe. Observational correlations suggest a co-evolution between SMBHs and their host galaxies (\citealt{Magorrian1998,Ferrarese2000,Gebhardt2000}), but it remains unclear which processes are most important for triggering AGNs and star formation.

Observational work has identified three main processes that drive evolution, but disagrees on the relative import of each process. Tidal torques from major mergers (where the mass ratio of the galaxies is less than 1:4) can drive gas accretion; some work indicates that these tidal torques from major mergers are primarily responsible for fueling both star formation (\citealt{Mihos1994,Mihos1996}) and rapid SMBH growth (\citealt{DiMatteo2005,Hopkins2005,Ellison2011,Koss2012,Treister2012,Satyapal2014}). Other work suggests that minor mergers or continuous `cold flow' gas accretion are the most important mechanism for shaping the morphologies of galaxies, driving star formation, and contributing to the mass growth of SMBHs (e.g., \citealt{Noeske2007,Daddi2007,Cisternas2011,Kocevski2012,Kaviraj2013,Villforth2014}). Yet other studies find that secular instabilities driven by disks and spiral arms in the local universe, as well as highly irregular morphologies and high gas fractions in the high redshift universe, can dominate galaxy evolution. These secular instabilities can grow pseudo-bulges locally and contribute to significant gas inflows and disk and bulge growth in high redshift galaxies (e.g., \citealt{Bournaud2016} and references therein). Many details of these processes that could drive evolution remain unclear, such as when and how these processes operate on AGNs and galaxies.

One main reason these details are unknown is that it is difficult to build a clean observational sample of galaxy mergers (major and minor). Imaging studies that rely upon one or a couple of imaging predictors can fail to accurately identify mergers, which leads to inconclusive results (e.g., \citealt{Conselice2014} and references therein). Recent work has relied increasingly upon non-parametric tools to identify merging galaxies from imaging surveys, such as the $Gini$-$M_{20}$ method or the $CAS$ (Concentration-Asymmetry-Clumpiness) method (\citealt{Lotz2004,Conselice2003}). These methods are each individually limited by different merger initial conditions, such as mass ratio and gas fraction, and by merger stage. For instance, while identifying merging galaxies using asymmetry tends to be more sensitive to early-stage mergers, $Gini-M_{20}$ tends to identify late-stage mergers. Additionally, previous simulations of galaxy mergers have demonstrated that the merger observability timescale varies strongly for different non-parametric tools (e.g., \citealt{Lotz2008,Lotz2010b,Lotz2010a}). 

We combine the sensitivities of different imaging predictors to create an imaging classification method that is better able to identify merging galaxies over a larger range of merger initial conditions and merger stages. In Nevin et al. (2019, in prep), we will incorporate kinematic predictors as well.

It is challenging to identify galaxy mergers directly from observations because each merger is observed at only a single viewing angle and moment in time, whereas the full duration of a merging event is several Gyr. Since the observational signatures of a merger depend so heavily on the merger initial conditions and stage of the merger, we create our classification scheme from hydrodynamics simulations that cover a range of merger initial conditions. In this way, we determine the fundamental capabilities of different imaging predictors. We utilize the \texttt{GADGET-3} smooth-particle hydrodynamical code coupled with the \texttt{SUNRISE} dust radiative transfer code to construct mock observations of the simulated galaxies. From these mock observations, we create the imaging classification and determine its accuracy and precision for identifying galaxy mergers of different gas fractions, mass ratios, and merger stages. We tailor our classification for SDSS imaging, although the code will be publicly available and can be easily modified for different imaging surveys.

The remainder of this paper is organized as follows. We describe the hydrodynamics and radiative transfer simulations, techniques for matching the simulated galaxies to SDSS's specifications, and merger classification scheme in Section \ref{methods}. In Section \ref{results} we describe the performance of the classification scheme and the sensitivities of the individual imaging predictors. We compare the technique to previous imaging methods for merger identification and discuss the implications for merging galaxy identification in imaging surveys in Section \ref{discuss}. We present our conclusions in Section \ref{conclude}. A cosmology with $\Omega_m = 0.3$, $\Omega_{\Lambda}=0.7$, and $h=0.7$ is assumed throughout.

\section{Methods}
\label{methods}

We create the imaging classification scheme from simulated galaxy mergers, which we introduce in Section \ref{sims} and \ref{sunrise}. In order to develop the classification for SDSS imaging data, we `SDSS-ize' the simulations to create mockup images matching SDSS specifications in Section \ref{image_prep}. Next, we determine the separation of the stellar bulges to assign galaxy merger stage in Section \ref{bulge}. Finally, we develop the imaging classification scheme using LDA (Linear Discriminant Analysis) in Section \ref{classify}.

\subsection{N-body/Hydrodynamics Merger Simulations}
\label{sims}
To develop our imaging classification scheme, we begin with a suite of simulated merging and isolated galaxies. Specifically, we use two of the high-resolution simulations from \citet{Blecha2018}, to which we have added three new simulations to cover a larger parameter space of initial conditions. We also have a set of isolated galaxies that is matched by stellar mass and gas fraction to each merger simulation.

These simulations were carried out with \texttt{GADGET-3} \citep{Springel2003,Springel2005}, a smoothed-particle hydrodynamical (SPH) and N-body code that conserves energy and entropy and includes sub-resolution models for physical processes such as radiative heating and cooling, star formation and supernova feedback, and a multi-phase interstellar medium (ISM). All simulations have a baryonic mass resolution of $2.8\times 10^4$ \msun\ and a gravitational softening length of 23 pc. SMBHs are modeled as gravitational "sink" particles that accrete via an Eddington-limited Bondi-Hoyle \citep{bonhoy44} prescription. AGN feedback is also incorporated by coupling 5\% of the accretion luminosity ($L_{\mathrm{bol}} = \epsilon_{\mathrm{rad}}\dot{M}c^2$) to the gas as thermal energy. We assume a radiative efficiency $\epsilon_{\mathrm{rad}} = 0.1$ for accretion rates $\dot M > 0.01 \dot M_{\mathrm {Edd}}$ (where $\dot M_{\mathrm {Edd}}$ is the Eddington limit); below this we assume radiatively inefficient accretion following \cite{Narayan2008}. \gadget\ has been used for many studies concerning merging galaxies (e.g., \citealt{DiMatteo2005,Snyder2013,Blecha2011,Blecha2013}).

The merger progenitor galaxies include a dark matter halo, a disk of gas and stars, a stellar bulge in some cases, and a central SMBH. The initial conditions for each simulated galaxy merger are given in Table \ref{tab:table1}, and the initial conditions for the matched isolated galaxy simulations are given in Table \ref{tab:table1iso}. In this work, we focus primarily on the effects of varying the merger mass ratio and initial gas fraction, since these two parameters have been shown to have the largest effect on the morphology and star formation rates of merging galaxies in previous work (\citealt{Cox2008,Lotz2008,Lotz2010b,Lotz2010a,Blecha2013}). We include three major merger simulations (q0.5\_fg0.3, q0.333\_fg0.3, and q0.333\_fg0.1) with mass ratios 1:2, 1:3, and 1:3, respectively. The initial progenitor gas fractions in these simulations (defined as $M_{\rm gas,disk} / (M_{\rm gas,disk} + M_{\rm *,disk})$), which are identical for both merging galaxies in a given simulation, are 0.3, 0.3, and 0.1. These major merger simulations have a bulge-to-total mass (B/T) ratio of 0. We design the two 1:3 major merger simulations to have different gas fractions but identical mass ratios to investigate the effects of varying gas fractions on the morphology of mergers. We also create two minor merger simulations (q0.2\_fg0.3\_BT0.2 and q0.1\_fg0.3\_BT0.2), both of which have a B/T ratio of 0.2. These two minor mergers have initial gas fractions of 0.3 and mass ratios of 1:5 and 1:10, respectively. We design these two minor mergers to have a gas fraction of 0.3 so that we can directly compare mass ratios of 1:2, 1:3, 1:5, and 1:10 across simulations with identical gas fractions. We further justify our choice of initial conditions in Appendix \ref{ICs}.

\begin{table*}[h!]
  \begin{center}
    \caption{Key parameters of our suite of high-resolution \gadthree\ galaxy merger simulations. }
    \label{tab:table1}
    \begin{tabular}{c|c|c|c|c}
      
     Model & $M_{\rm tot}$ &   Stellar Mass & Gas Fraction & Mass Ratio  \\
      &  [$10^{11}$ \msun] &    [$10^{10}$ \msun] &   & \\
   \hline 
      
      q0.5\_fg0.3 & 20.8  & 5.9 &0.3 &1:2 \\
      q0.333\_fg0.3 & 18.7 &  5.2 & 0.3 & 1:3   \\
      
      q0.333\_fg0.1 & 18.7 &  6.3 &0.1 & 1:3   \\
       
q0.2\_fg0.3\_BT0.2 & 16.8  & 5.0 & 0.3 & 1:5 \\
      
      q0.1\_fg0.3\_BT0.2 & 15.1  & 4.6 & 0.3 & 1:10 \\

    \end{tabular}
  \end{center}
\end{table*}

\begin{table*}[h!]
  \begin{center}
    \caption{Key parameters of the matched sample of isolated galaxies. These are matched to the mass of the primary or secondary galaxy in the merger for each simulation and the gas fraction. The gas fraction is the same for each merger progenitor in a given simulation.}
    \label{tab:table1iso}
    \begin{tabular}{c|c|c|c|c}
      
     Model & $M_{\rm tot}$ &  Stellar Mass & Gas Fraction& Matched Model(s) \\
      &  [$10^{11}$ \msun] &    [$10^{10}$ \msun] & & \\
   \hline 

      m1\_fg0.3 & 13.9 &  3.9 & 0.3&q0.5\_fg0.3, q0.333\_fg0.3\\
      m0.5\_fg0.3 & 6.9 &  2.0 & 0.3&q0.5\_fg0.3 \\
      m1\_fg0.1 & 14.0 &  4.7 & 0.1&q0.333\_fg0.1\\
      m0.333\_fg0.1 & 4.7 &  1.6 & 0.1&q0.333\_fg0.1\\
      m1\_fg0.3\_BT0.2 & 13.7 &  4.2 & 0.3&q0.2\_fg0.3\_BT0.2, q0.1\_fg0.3\_BT0.2\\

    \end{tabular}
  \end{center}
\end{table*}

\subsection{Radiative Transfer Simulations}
\label{sunrise}

In order to directly compare the simulated galaxies with observations, we use the 3D, polychromatic, Monte-Carlo dust radiative transfer code \texttt{SUNRISE} (\citealt{Jonsson2006,jonsso10}) to produce resolved UV to IR spectra and broadband images.

It has been used extensively in combination with \texttt{GADGET} galaxy merger simulations (e.g., \citealt{Lotz2008, Lotz2010a,Lotz2010b,Wuyts2010,Narayanan2010}). 

Age- and metallicity-dependent spectral energy distributions for each star particle are calculated using the STARBURST99 stellar population synthesis models (\citealt{Leitherer1999}).  Emission from HII regions (including dusty photodissociation regions) around young stars is calculated by applying MAPPINGSIII models (\citealt{Groves2008}) to newly formed star particles, based on their age, metallicity, and surrounding gas pressure. The AGN spectrum is determined using the SMBH accretion rate and the luminosity-dependent templates of \citet{Hopkins2007}. 

To calculate the dust distribution, we use the \citet{Draine2007} Milky Way dust model with R$_{\rm V} =3.1$ and assume that 40\% of gas-phase metals are in dust (\citealt{dwek98}). A 3D adaptively-refined grid is placed on the simulation domain to map the gas-phase metal distribution. Following \citet{Snyder2013} and \citet{Blecha2018}, we assume that gas in the cold phase of the \gadthree\ multi-phase ISM model has a negligible volume filling factor and therefore does not contribute to the attenuation of radiation. While this may not be an appropriate choice for extremely gas-rich, high-redshift galaxies that produce extreme IR and sub-mm luminosities \citep[e.g.,][]{hayward11,snyder13}, it is a reasonable assumption for the low-redshift analog galaxy simulations in our suite. 

\sunrise\ performs Monte Carlo radiative transfer through this grid, computing emission from stars, HII regions, and AGN, as well as energy absorption (including dust self-absorption) to obtain the emergent, attenuated resolved SEDs for seven different isotropically distributed viewing angles.

\begin{figure*}
\centering
\includegraphics[scale=0.3]{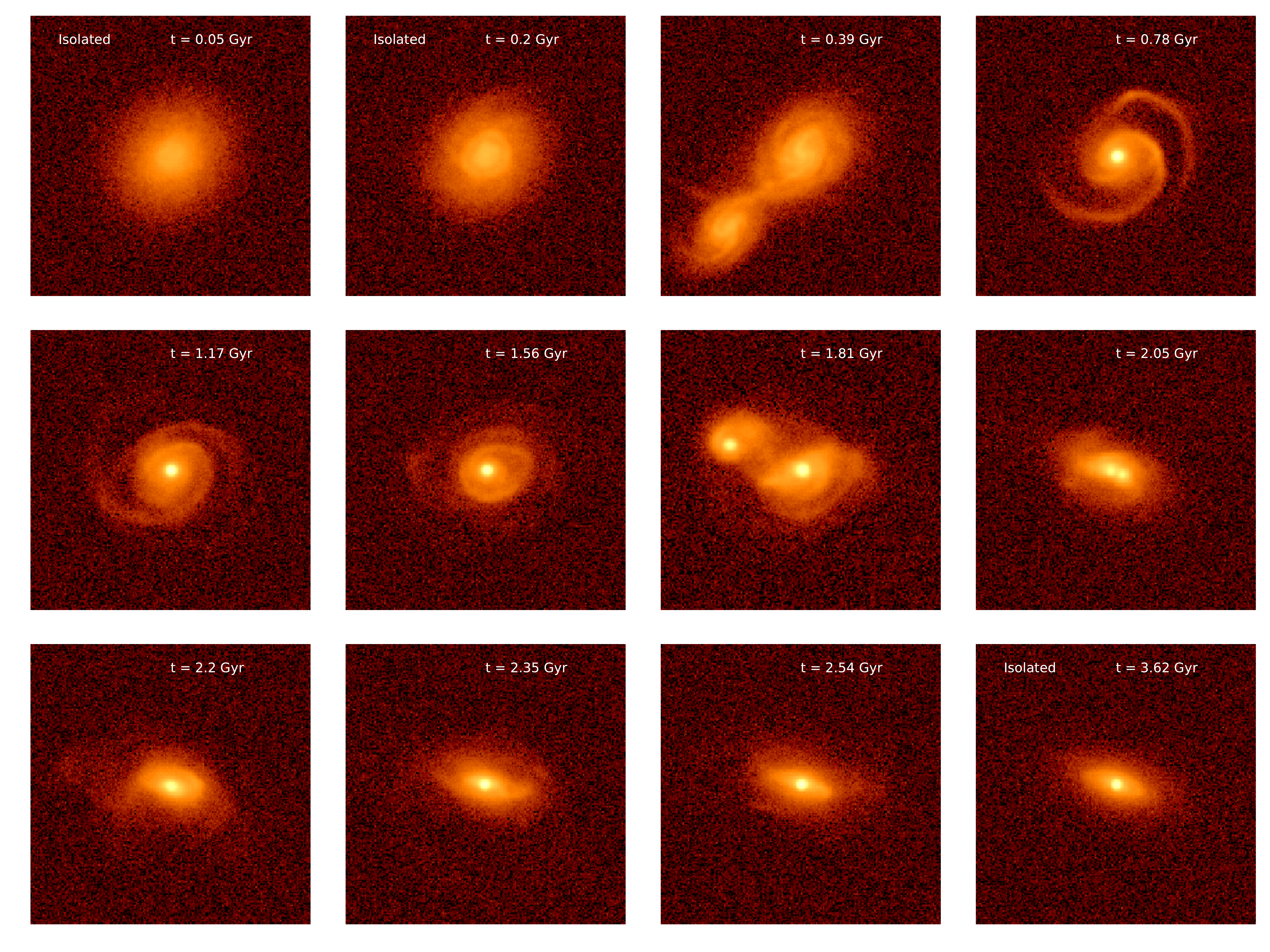}

\centering
\includegraphics[scale=0.7]{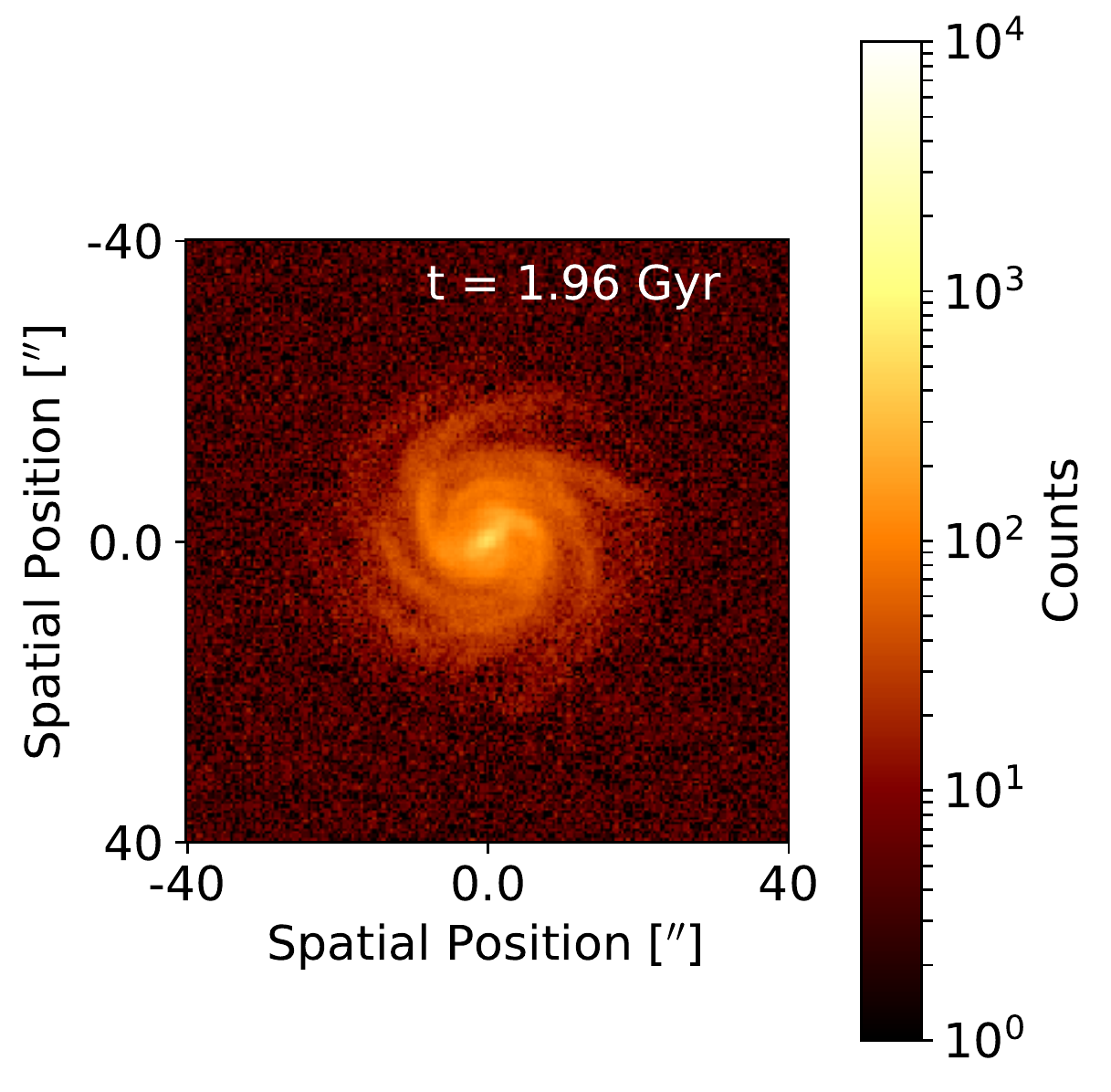}
\caption{Time series of $r-$band `SDSS-ized' images from the q0.5\_fg0.3 merger simulation (mass ratio 1:2, gas fraction 0.3). To SDSS-ize simulated images, we convert to counts, convolve to the seeing limit of the survey, rebin to the SDSS imaging pixelscale, and add background noise typical of SDSS imaging. All images are centered on the brightest \texttt{Source Extractor} selected source and cut to the $80\farcs0$ ($\sim$50 kpc) SDSS imaging camera field of view. The merger images at t = 0.05, 0.2, and 3.62 Gyr are included as isolated galaxies in the analysis. The merger images at t = 0.39, 0.78, 1.17, and 1.56 Gyr are early-stage mergers, the images at t = 1.81 and 2.05 Gyr are late-stage mergers, and the images at t = 2.2, 2.35, and 2.54 Gyr are post-coalescence mergers. The bottom middle image is an isolated galaxy snapshot that is matched to the q0.5\_fg0.3 simulation for mass and gas fraction. }
\label{fig:simulation}
\end{figure*}

For each merger simulation, we perform \sunrise\ calculations on snapshots at $\sim50-100$ Myr intervals during the merger. The spatial resolution of all images and resolved spectra is 167 pc, which exceeds the resolution of the SDSS survey (see Section \ref{image_prep}). We divide each merger simulation into early-stage, late-stage, and post-coalescence snapshots based on the projected separations of the stellar bulges in the images. We describe this process in more depth in Section \ref{bulge}. Briefly, early-stage mergers are defined as the snapshots with average stellar bulge separations $\Delta \mathrm{x}\geq$ 10 kpc. Late-stage mergers are defined to have stellar bulges with separations of $1\ \mathrm{kpc}< \Delta \mathrm{x} < 10\ \mathrm{kpc}$. Post-coalescence snapshots are those in which two stellar bulges are no longer resolvable in SDSS ($\Delta \mathrm{x} \leq 1 $ kpc) since the spatial resolution of SDSS is 1-2 kpc. For each merger simulation, we run \sunrise \ at $\sim100$ Myr intervals in the early stage of the merging galaxies, at $\sim50$ Myr intervals in the late stage, and at $\sim100$ Myr intervals for the post-coalescence stage. This creates a roughly equal number of $\sim5-10$ \sunrise \ snapshots per merger stage. In Figure \ref{fig:simulation}, we show $r$-band images for early-stage, late-stage, and post-coalescence snapshots from the 1:2 major merger gas-rich simulation q0.5\_fg0.3.

We also simulate isolated galaxies with matched stellar mass and gas fraction for each merger simulation (Table \ref{tab:table1iso}). Additionally, because the progenitor galaxies are still isolated and undisturbed in the very early stages of the merger simulations, we include snapshots prior to first pericentric passage in our sample of isolated galaxy snapshots as well. We confirm, using the supplemental outputs of \texttt{SUNRISE}, that the star formation rate and AGN luminosity have yet to be affected by the merger in these snapshots. Additionally, the imaging predictors are not yet significantly different than the matched sample of isolated galaxies.

We also include merger snapshots at times > 0.5 Gyr after final coalescence as isolated galaxies. Our motivation for this is twofold. First, after > 0.5 Gyr following final coalescence, the simulated galaxies begin to lose tidal features but remain centrally concentrated when compared to the isolated matched sample of galaxies. If we include these post-coalescence galaxies in the analysis as mergers, the technique becomes overly sensitive to the central concentration of galaxies and is most efficient at identifying early-type galaxies. Second, since we wish to develop a tool that best identifies galaxy mergers in the early, late, and beginning of the post-coalescence stage, we terminate the merger period at 0.5 Gyr after final coalescence for all simulations. We find that this choice of cutoff time allows the sensitivity of our merger detection technique to decay smoothly during the post-coalescence stage. We include an isolated galaxy snapshot in Figure \ref{fig:simulation} as well as several isolated snapshots prior to first pericentric passage and 0.5 Gyr following final coalescence.

Broadband images for each snapshot are produced for seven isotropically-distributed viewpoints. We focus on the SDSS $r-$band filter, since the $r-$band is a good tracer of stellar populations in low redshift galaxies. Since we next plan to incorporate kinematic predictors into the analysis (Nevin et al. (2019, in prep)), we will apply the classification technique to the MaNGA (Mapping Nearby Galaxies at Apache Point) survey, which is an integral field spectrograph (IFS) survey of a subsample of $\sim$10,000 SDSS galaxies. We therefore place the simulated galaxies at the average redshift of the MaNGA survey ($\langle z \rangle \sim 0.03$) and extract the $r-$band images, which we process further to match the specifications of the SDSS survey in Section \ref{image_prep}.

To understand the range of redshift and surface brightness for which our merger classification can return consistent classifications, we experiment with adjusting the surface brightness and redshift of the simulated images. The consistency of the classification is closely tied to the behavior of the imaging predictors, which are sensitive to both resolution and the average S/N per pixel (<S/N>). For instance, \citet{Lotz2004} find that $Gini$, $M_{20}$, $C$, $A$, and $S$ are reliable to 10\% for <S/N> $\ge$ 2 and systematically decrease with <S/N> below this level. We implement a <S/N> cutoff of 2.5 (which is calculated for all pixels within the segmentation mask) because the measurements of the imaging predictors (especially $A$ and $S$) from \texttt{statmorph} are unreliable below this threshold (Vicente Rodriguez-Gomez, private communication). For instance, $A$ systematically decreases to negative values below this threshold. We also use this <S/N> cutoff value to assess the magnitude limit of the method (described below). 

We find that the surface brightness of the simulated galaxies changes over the course of each simulation. This happens as the galaxies brighten and dim with star formation and AGN activity as the merger proceeds. This corresponds to a range in $r-$band apparent magnitude from $14-16$ (at $z = 0.03$). We convert from surface brightness to $r-$band magnitude using the conversion in Section \ref{image_prep} to convert to units of nanomaggies, which we then convert to apparent magnitude using the Petrosian radius as the aperture. We experiment with dimming the images to determine how the $r-$band Petrosian magnitude of a mock image relates to the S/N per pixel. The classification becomes significantly different (since the mock images begin to drop below a <S/N> value of 2.5) when the $r-$band Petrosian magnitudes are $\ge$17. In other words, the classification only works for $r-$band magnitudes $\ge$17 and it should not be used for fainter galaxies.  For context, typical SDSS galaxies from this paper (in Section \ref{galzoo}) have <S/N> values between 5-10, which corresponds roughly to $r-$band magnitudes of $\sim$16. Since SDSS imaging has flux limit of 17.77 in the $r-$band, the LDA classification technique applies to the majority of galaxies in the SDSS photometric catalog (\citealt{Strauss2002}).

Likewise, we move the simulated galaxies to higher redshifts while maintaining the same surface brightness and find that the predictor coefficients in the classification change significantly at $z \sim 0.5$. The average redshift of the SDSS photometric survey is $z \sim 0.4$, so the LDA technique should still function well for the majority of SDSS galaxies (\citealt{Sheldon2012}).

\begin{figure*}
    \centering
    \includegraphics[scale=0.85,trim=2.5cm 0cm 0cm 0cm]{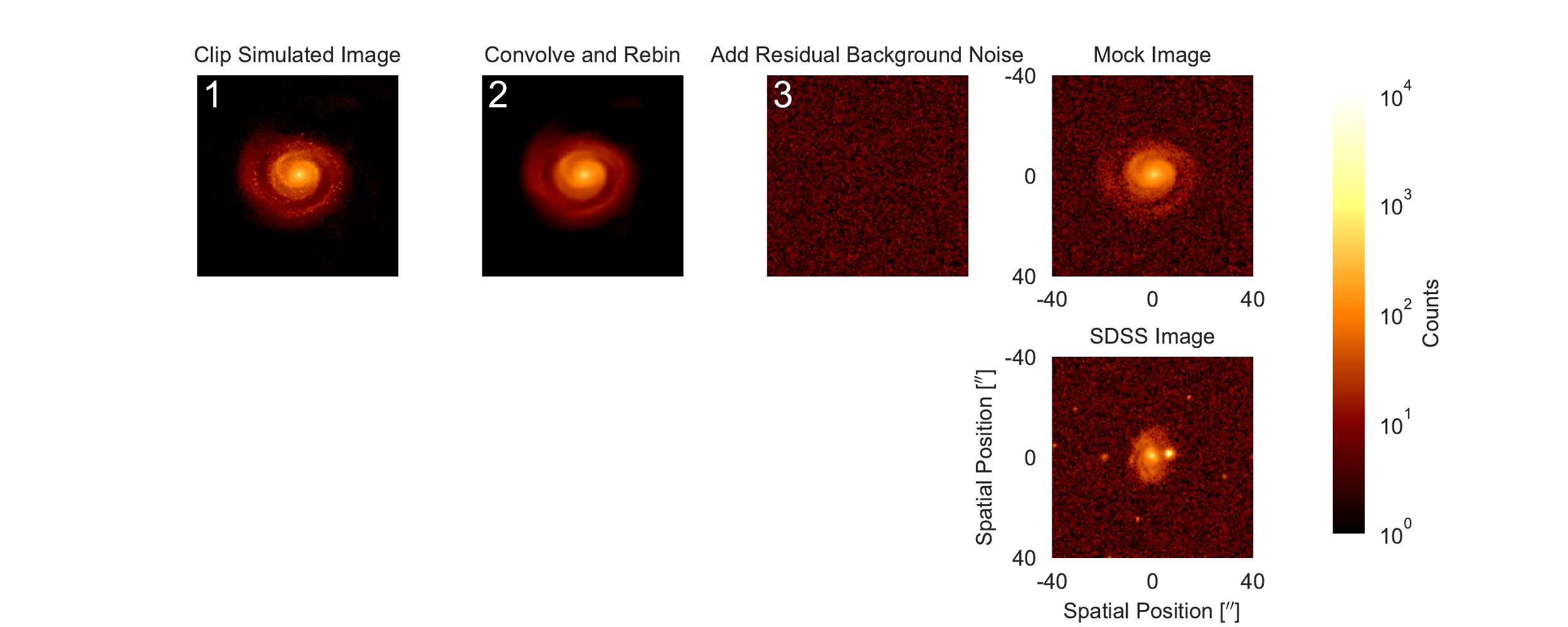}
    \caption{Steps of the process to create mock images from the simulated images (1-3). We first clip the simulated image to the $80\farcs0$ field of view (1). We then convolve the image with the $1\farcs43$ FWHM PSF and rebin to the $0\farcs396$ pixelscale of SDSS imaging (2). Finally, we add residual background noise that is characteristic of SDSS imaging (3) to create a mock image (upper right panel). We compare to a SDSS image (lower right panel) that has been centered on a galaxy and cut to the same $80\farcs0$ field of view as the mock image. }
    \label{fig:my_label}
\end{figure*}

\subsection{SDSS-izing images from the simulations}
\label{image_prep}

In order to construct a classification scheme that can be applied directly to SDSS galaxies, we first `SDSS-ize' or degrade the simulated images to match the specifications of the SDSS survey. In this section, we describe the relevant SDSS imaging properties and data products. Then, we provide a detailed description of the process of SDSS-izing the simulated images. Finally, we detail how we determine the stage of the merger snapshots.

The process of SDSS-izing the simulated images to create mock images that match the specifications of the SDSS imaging involves the following steps (Figure \ref{fig:my_label}):
\begin{enumerate}
    \item Clip the images.
    \item Convolve and rebin to the spatial resolution and pixelscale of SDSS imaging.
    \item Introduce residual background noise.
    \item Create an error image.
\end{enumerate}

To complete these steps, we utilize the imaging properties (i.e., noise, instrumental gain, sky levels, etc) of SDSS imaging, which are described in \citet{Albareti2017} and \citet{Blanton2011}. The SDSS imaging procedure involves producing large field images that are composed of six long rectangular images of the sky called `camcols'. The camcols are then further split into individual filters ($u$, $g$, $r$, $i$, and $z$) and six smaller `frame' images. Frame images are the basic data product of SDSS; these images are background subtracted and include an extension with background sky levels, instrumental gain, dark variance, and a calibration factor to convert between flux and photoelectrons. The frame images can be further cut to postage stamp images (in our case, with a field of view of $80\farcs0 \times 80\farcs0$). The most recent SDSS data release (DR13) uses a specific NASA-Sloan Atlas (NSA) reprocessing of the original SDSS DR7 imaging data, which includes a new background subtraction that improves the photometry of large galaxies (\citealt{Blanton2011}). We use DR13 imaging properties to compare to SDSS-ized images below. The median seeing, which is the effective width (FWHM) of the PSF, for SDSS imaging is $1\farcs43$ and the pixel scale is $0\farcs396$ pix$^{-1}$ (\citealt{Ivezic2004,Blanton2011}).

We start with the imaging output of \texttt{SUNRISE} for the five broadband SDSS filters ($u$, $g$, $r$, $i$, $z$). Here, we focus on the SDSS $r-$band images since they best capture light from stellar bulges for nearby galaxies. To best mimic the placement of the imaging camera from SDSS, we use aperture photometry to identify the brightest pixels over which to center the camera. We identify the brightest source using \texttt{Source Extractor}, which is a useful tool to extract sources through aperture photometry on astronomical images (\citealt{Bertin1996}). \texttt{Source Extractor} separates an object from the background noise, applies a convolution filter to separate low surface brightness sources from spurious detections, and deblends sources. We use a detection threshold of 1.5$\sigma$ above the local sky background and a minimum group number of two pixels to trigger a detection. We use a normal convolution kernel with size $3 \times 3$ pixels, a FWHM of two pixels, and a deblending threshold of 32 (the recommended value for \texttt{Source Extractor}). The output from \texttt{Source Extractor} includes x and y positions of sources and aperture photometry, which includes Petrosian radii and corresponding fluxes. We determine the brightest source from these fluxes and then clip the image in a $80\farcs0$ square around this source. We select an $80\farcs0$ square cutout because it allows us to accurately determine the image background for the extraction of the imaging predictors. Some of our simulations snapshots have smaller fields of view (down to $50\farcs0$) since the simulated galaxies are in the edge of the simulation field of view. We include these smaller snapshots in interest of maximizing the temporal resolution of our method. We find that very few mock images have a smaller field of view than $80\farcs0$ and that this does not affect the imaging predictors for these snapshots.

After clipping the mock images, we convolve them with a PSF with FWHM $1\farcs43$, which is the median PSF for the $r-$band (\citealt{Ivezic2004}). Then, we rebin the images to the pixelscale of SDSS ($0\farcs396$ pixel$^{-1}$).

We then convert to flux units typical of SDSS, introduce residual background noise, and produce an error image, as outlined below. The units of the simulated image are surface brightness (W/m/m$^2$/sr). We convert to flux density in nanomaggies:
$$\mathrm{nanomaggy} = \mathrm{Janskys} \times 3.631 \times 10^{6}$$
where we first convert to Janskys using the pixelscale and angular diameter distance of a simulated galaxy at the average redshift of the MaNGA survey, $z\sim0.03$. Again, we use the average redshift of the MaNGA survey since we plan to further develop the kinematic technique for MaNGA IFS in Nevin et al. (2019, in prep). 

Then, we extract a nanomaggy to data number (dn) conversion rate from each frame image ($c$). This conversion rate is used to produce a mock image in units of counts (from here on dn is synonymous with counts). We find an average $c$ value of 0.005 with a standard deviation of 0.0002. The conversion rate varies little across the frames and camcols.

In order to introduce background noise to the mock images, we characterize the residual background of the SDSS frame images using bilinear interpolation. We also determine other imaging properties such as background sky levels (prior to background subtraction), instrumental gain, and dark variance from the frame images. Since the gain and background sky levels vary in complicated ways across the frames and camcols (Michael Blanton, private communication), we characterize these values based upon several locations from the larger frame images.

For instance, we use 50 postage stamps (from the frame images) that are selected to belong to all six camcols and locations on the frame images. We extract a region from the background and characterize its mean and standard deviation. The postage stamp images have already been background-subtracted, so this region is characteristic of the residual background of SDSS images following the sky subtraction step. We find that the typical residual background has a mean of 0.33 dn (counts) with a standard deviation of 5.63 dn. After conducting an improved background subtraction for the SDSS-III DR8 imaging data (which is the same imaging reduction used for DR13), \citet{Blanton2011} find a residual standard deviation of 0.02 nanomaggies in the $r-$band photometry. This is $\sim4$ dn, so our standard deviation of 5.63 dn is a good approximation of the residual noise.

We reintroduce this background into our images by adding a standard normal with a mean of 0.33 and a standard deviation of 5.63 dn to each pixel. This mock image is used in the calculation of the imaging predictors in Section \ref{imgpred}. We use both the conversion factor, $c$, and the residual background value, $\mathrm{bg_{resid}}$, to produce an image that is representative of a SDSS image (in counts):
$$\mathrm{dn} = \mathrm{nanomaggies}/c + \mathrm{bg_{resid}}$$

We use the images in units of dn for display purposes and for the extraction of the imaging predictors.

Finally, we create an error image. To calculate the photometric uncertainty, we use the average gain and dark variance from the $r$-band frame images (4.7 photoelectrons per dn and 1.2 dn$^2$, respectively) in combination with the simulated galaxy image to produce an error image in dn: 

$$\sigma_{\mathrm{dn}} = \sqrt{\mathrm{(dn+bg_{sky})} /\mathrm{gain} + \mathrm{dark var}}\ $$
where we also include the background counts prior to background subtraction ($\mathrm{bg_{sky}}$). The photometric uncertainty is dominated by the galaxy flux except for low surface brightness features such as tidal tails, where the background sky dominates. To determine the background sky level, we extract a region from each sky image and measure the average value. We find that this value varies between frame images and that the mean background value is 121.2 dn with a standard deviation of 37.4.

Figures \ref{fig:my_label} and \ref{fig:simulation} show examples of simulated images after the image has been spatially convolved, rebinned, and the residual background has been introduced to match the specifications of the SDSS survey. 

\subsection{Measuring Stellar Bulge Separations}
\label{bulge}

We use \texttt{Source Extractor} and \texttt{GALFIT} (\citealt{Bertin1996,Peng2002}) to identify, pinpoint, and measure the separation of stellar bulges from the SDSS-ized $r-$band images. Using \texttt{Source Extractor}, we first determine if there are one or two stellar bulges within the field of view and pinpoint their locations. We eliminate spurious detections from \texttt{Source Extractor} using the above prescription for a detection threshold (1.5$\sigma$ above sky) combined with a normal convolution kernel (a $3\times3$ pixel mask with a FWHM of 2 pixels). We avoid the detection of star forming regions by requiring that the flux within the measured Petrosian radius of the secondary source be greater than 10\% of the primary source. 

Under these prescriptions \texttt{Source Extractor} performs well, detecting the primary and secondary stellar bulges for four of the merger simulations without spurious detections or detections of star forming regions. To ensure that \texttt{Source Extractor} is not detecting star forming regions, we require that the location of the regions identified by \texttt{Source Extractor} correspond to the locations of the SMBHs tracked by \gadget.

\texttt{Source Extractor} fails to accurately identify the secondary source for the q0.1\_fg0.3\_BT0.2 simulation. Since we require that the flux of the secondary source detected by \texttt{Source Extractor} be greater than 10\% of the flux of the primary source, the 1:10 minor merger often falls below this level. We do not lower the 10\% detection cutoff since we wish to avoid star forming region detection, so we use the locations of the SMBHs for the q0.1\_fg0.3\_BT0.2 simulation to identify the secondary sources in order to determine merger stage.

Then, we use \texttt{GALFIT}, which is a two-dimensional fitting algorithm that extracts structural components from images of galaxies. It can fit one or more two-dimensional models such as exponential disks, S\'{e}rsic profiles, Gaussian profiles, or Moffat functions to the light profile of a galaxy. We use \texttt{GALFIT} to fit a S\'{e}rsic profiles to each source identified by \texttt{Source Extractor} and extract the projected separations of the stellar bulges (if there are two). With the \texttt{GALFIT} output in hand, we average the projected separation of the stellar bulges for all viewpoints of a given snapshot of a merger and use this average to determine the merger stage. Again, if the average separation is $\Delta x \geq$ 10 kpc the merger is early-stage, if the separation is $1\ \mathrm{kpc}< \Delta x < 10\ \mathrm{kpc}$ the merger is late-stage, and separations $\Delta x \leq 1$ kpc are post-coalescence.

\subsection{Creating the Classification Scheme}
\label{classify}

Using the simulated galaxies, we know a priori whether a galaxy is a merging or nonmerging galaxy. In this section, we discuss the preparation of the imaging parameters that we use as an input to a supervised Linear Decomposition Analysis (LDA). We refer to these imaging parameters as `predictors' from here on because they help predict whether a galaxy is undergoing a merger. We also describe the LDA technique, which allows us to determine which imaging predictors are critical for best separating the classes of merging and nonmerging galaxies for each simulation.

\subsubsection{Imaging Predictors}
\label{imgpred}
In this section, we first describe the imaging predictors and then the methods used to extract them from the SDSS-ized galaxy images. We discuss their weaknesses and strengths; no one imaging predictor is the best determination of a merging galaxy. Instead they are sensitive to different orientations, merger stages, and mass ratios. The statistical power of the LDA methodology allows us to select the most successful predictors for various types of merging systems. We discuss these results in Section \ref{results}.

There are two main approaches to identifying a galaxy merger from imaging: parametric and nonparametric modeling of the surface brightness of the galaxy image. The parametric approach requires modeling the surface brightness of the galaxy using integrated light profiles such as bulges, disks, or S\'{e}rsic profiles. Since parametric modeling tends to assume a symmetric profile for the surface brightness of a galaxy, it fails for irregular galaxies as well as those with structures such as compact nuclei, spiral arms, or bars (\citealt{Lotz2004}). More recent work on merger identification has focused on nonparametric modeling of the surface brightness of galaxies. Nonparametric tools can be applied to irregular galaxies as well as the more standard early or late Hubble-type galaxies. We employ two widely used nonparametric approaches as imaging predictors: the $CAS$ (concentration ($C$), asymmetry ($A$), and clumpiness/smoothness ($S$)) morphological classification technique and the $Gini$ - $M_{20}$ method. We also use a binary variation of $A$, shape asymmetry ($A_S$) from \citet{Pawlik2016}. Finally, we incorporate one parametric approach, the S\'{e}rsic index ($n$). Overall, we utilize seven different imaging predictors, defined below: $Gini$, $M_{20}$, $C$, $A$, $S$, $n$, and $A_S$.

Concentration is defined by \citet{Lotz2004} as the ratio of light within circular radii containing 80\% and 20\% of the total flux of the galaxy: 
$$C = 5\  \mathrm{log}\Big(\frac{r_{80}}{r_{20}}\Big)$$
where $r_{80}$ is the circular radius that contains 80\% of the total flux, and $r_{20}$ is the circular radius that contains 20\% of the total flux. We use the approach from \citet{Conselice2003} that defines the total flux as that within 1.5 Petrosian radii ($r_p$) of the galaxy's center. We measure the Petrosian radius using \texttt{Source Extractor}.

A galaxy with a higher value for $C$ has more light contained within the central regions of the galaxy and is therefore more likely to be an early-type galaxy. 

The imaging rotational symmetry predictor, $A$, is from \citet{Conselice2000}:

$$A = \sum_{ij} \frac{|I(i,j)-I_{180}(i,j)|}{|I (i,j) |} - \sum_{ij} \frac{|B(i,j)-B_{180}(i,j)|}{|I (i,j)|}$$
where asymmetry is summed over all pixels ($i$,$j$), $I(i,j)$ is the image, $I_{180}(i,j)$ is the image rotated by 180$^{\circ}$ about the center, $B(i,j)$ is the background image (the background image is described in Section \ref{image_prep} and includes only the residual background typical of SDSS imaging following background subtraction), and $B_{180}$ is the background image rotated by 180$^{\circ}$ about the same center. We define the center of the galaxy as the location that minimizes the value of asymmetry as in \citet{Lotz2008}. Again, the galaxy image and background image are both masked to 1.5 r$_{p}$.

A galaxy with a higher value of $A$ has disturbed structure and/or bright tidal tails and is therefore more likely to be a galaxy undergoing a merger. $A$ is particularly good at identifying early-stage merging galaxies (following first pericentric passage) when the structure of a merging galaxy is most disturbed and tidal tails are most prominent.

The shape asymmetry, $A_S$ is measured using the same procedure as the asymmetry, but with a binary detection mask. The technique is described in detail in \citet{Pawlik2016} and \citet{Rodriguez-Gomez2018}. Since it is measured using a binary mask, $A_S$ is more sensitive to low surface brightness tidal features than $A$.

Clumpiness or smoothness ($S$) is defined by \citet{Conselice2003} and \citet{Lotz2004} to be the fraction of light within clumpy distributions in a galaxy:

$$S =  \frac{\sum_{ij} |I(i,j)-I_{S}(i,j)|}{|I (i,j) |} - B_S $$
where $I(i,j)$ is the image and $I_S(i,j)$ is the smoothed image which is smoothed using a boxcar of width 0.25 r$_{p}$. $B_S$ is the average smoothness of the background calculated in a $10 \times 10$ pixel box using the same 0.25 r$_{p}$ boxcar. $S$ is summed over all pixels ($i$, $j$) within 1.5 r$_{p}$ of the galaxy's center. However, the pixels within 0.25 r$_{p}$ of the galaxy center are excluded for the calculation of $S$ because the central regions of galaxies are highly concentrated and this elevates the value of $S$ (see \citealt{Conselice2003}).

Since $S$ measures the fraction of light from a galaxy that can be found in clumpy distributions, it identifies merging galaxies that have recently undergone star formation (e.g., \citealt{Conselice2003}). For instance, galaxies with a low value of $S$ tend to be elliptical galaxies and galaxies with a high value of $S$ are either undergoing mergers (with star formation) or undergoing bursty star formation without experiencing a merger event.

The $CAS$ morphological classification system was put forth as a method for cleanly separating galaxies based on their morphologies using their location in $CAS$ space. However, it is limited in several ways. First, concentration assumes circular symmetry and therefore fails for some irregular galaxies (\citealt{Lotz2004}). For instance, \citet{Conselice2003} find that the average value of $C$ for ULIRGs (ultraluminous infrared galaxies; L$_{\mathrm{IR}} > 10^{12} \ \mathrm{L}_{\odot}$) is not significantly different from that of Hubble sequence galaxies. This is problematic for merger identification since a significant fraction of ULIRGs (at least in the local universe) are gas rich major mergers (e.g., \citealt{Veilleux2002,Draper2012}). Second, not all mergers are asymmetric, and not all asymmetric galaxies are mergers (\citealt{Thompson2015}). Third, clumpiness is very dependent on the choice of boxcar width (smoothing length) used to smooth the image (\citealt{Andrae2011}), which has not been studied in detail. In this work, we find that clumpiness is most sensitive to viewing angle and therefore a poor merger predictor, so while we include it in the analysis, we focus more on concentration and asymmetry. This decision is supported by previous findings that focus on $C$ and $A$ alone from the $CAS$ morphology (e.g., \citealt{Lotz2008}).

The $Gini$ coefficient is used to describe the relative concentration of light in a galaxy and is insensitive to whether the light lies at the center of the galaxy. $Gini$ is sensitive to major and minor mergers and is most sensitive for face-on systems (\citealt{Thompson2015}). $Gini$ is defined by \citet{Abraham2003} and \citet{Lotz2004} as:

$$Gini = \frac{1}{|\bar{f}|n(n-1)}\sum_i^n(2i-n-1)|f_i|$$
where $\bar{f}$ is the average flux value, $n$ is the number of total pixels in the image, and $f_i$ is the flux value for each pixel where the $n$ pixels are ordered by brightness in the summation. 

$Gini$ is high for galaxies with very bright single or multiple nuclei and low for galaxies with more distributed light, such as late-type disk galaxies. Therefore, a higher value of $Gini$ will select for merging galaxies during late stage mergers (with multiple bright nuclei) as well as post-coalescence merging galaxies.

The $M_{20}$ coefficient is often combined with $Gini$ to identify merging galaxies. It measures the relative concentration of the light in a galaxy and also does not assume a central concentration. The second-order moment of the light in a galaxy ($M_{\mathrm{tot}}$) is the sum of the flux in each pixel, $f_i$, multiplied by the distance squared to the center of the galaxy:
$$M_{\mathrm{tot}} = \sum_i^n M_i = \sum_i^n f_i[(x_i-x_c)^2+(y_i-y_c)^2]$$
 where $M_i$ is the flux in a single pixel multiplied by the distance squared to the center of the galaxy. The center ($x_c$, $y_c$) is chosen to minimize the value of $M_{\mathrm{tot}}$. $M_{\mathrm{tot}}$ is a tracer for the spatial distribution of any bright areas in the galaxy. $M_{tot}$ is then used to compute $M_{20}$, which is defined by \citet{Lotz2004} to be the normalized second order moment of the brightest 20\% of the galaxy's flux:

$$M_{20} = \mathrm{log}_{10}\Big(\frac{\sum_i M_i}{M_{\mathrm{tot}}}\Big), \  \mathrm{while} \sum_i f_i < 0.2f_{tot}$$
where $f_{tot}$ is the total flux of all of the pixels that are identified by the segmentation map (defined below), and $f_i$ are the fluxes rank ordered from brightest to faintest. The division by $M_{\mathrm{tot}}$ removes all dependence on the total galaxy flux. 

$M_{20}$ is similar to $C$ but the center of the galaxy is a free parameter, allowing it to be more sensitive to spatial variations of light. Also, $M_{20}$ is always a negative value due to the logarithm. Clear mergers with multiple bright nuclei have higher values of $M_{20}$ ($M_{20} > -1$) and early-type galaxies have lower values ($M_{20} \leq -2$; \citealt{Lotz2008}). Therefore, higher values of $M_{20}$ select for merging galaxies.

Since $Gini$ and $M_{20}$ are sensitive to the ratio of low surface brightness pixels to high surface brightness pixels, we use a segmentation map to measure both of these predictors, as in \citet{Lotz2008}. The segmentation map assigns pixels to the galaxy that are above the threshold value given by the surface brightness at the Petrosian radius. We use a segmentation map instead of making S/N cuts, because galaxies with the same morphologies but different intrinsic luminosities will have different $Gini$ values if the cut is made based on S/N.

In addition to the $CAS$ and $Gini-M_{20}$ nonparametric predictors, we measure the shape asymmetry ($A_S$) for each galaxy. Shape asymmetry is similar to the imaging asymmetry we also describe above; it is calculated using the same method, but with a binary detection mask instead of the image itself. This weights all parts of the galaxy equally regardless of relative brightness, making it a useful probe of morphological asymmetry (as opposed to the asymmetry of the light distribution). It has proven useful for detecting faint asymmetric tidal features that are suggestive of a merger (\citealt{Pawlik2016}). 

Our final imaging predictor is the S\'{e}rsic index, which is used to define the exponential surface brightness profile of a galaxy:
$$I(R) = I_e\  \mathrm{exp} \Big( -b_n \Big[ \Big( \frac{R}{R_e} \Big) ^{1/n} - 1 \Big]\Big)$$
where $I(R)$ is the intensity within a given circular radius, $I_e$ is the intensity at the effective radius ($R_e$), which is the radius that contains half of the total light, and $b_n$ is a constant that depends on the S\'{e}rsic index, $n$ (\citealt{Sersic1963}). 

A S\'ersic index of $n=1$ denotes an exponential disk, indicative of a spiral galaxy, while $n=4$ denotes a de Vaucouleurs profile, indicative of an elliptical galaxy. In general, higher $n$ indicates light that is more centrally concentrated. A division between morphologies has been standardized as $n \lesssim 2.5$ for spirals and $n \gtrsim 2.5$ for ellipticals (\citealt{vanderWel2008}). \citet{Fisher2008} predict that values of $n > 2$ (steeper surface brightness profiles) are produced by major mergers.

To extract the values of $Gini$, $M_{20}$, $C$, $A$, $S$, and $A_S$ for each galaxy, we utilize the galaxy morphology tool \texttt{statmorph} (\citealt{Rodriguez-Gomez2018}). Within this tool, we invoke the segmentation map defined from the surface brightness at 1.5$r_p$, which is measured using \texttt{Source Extractor}. We measure the value of $n$ for each galaxy with \texttt{GALFIT}.

\subsubsection{Identifying Mergers with Imaging Predictors}
\label{imag_LDA}
We seek a classifier that can separate merging and nonmerging galaxies of various merger mass ratios, gas fractions, viewing angles, and merger stages. We also need to incorporate multiple different imaging predictors. LDA is uniquely suited for these purposes. LDA is able to maximize the separation between multiple classes (in our case, we only need to separate two classes, `merging' vs `nonmerging' galaxies). In this work, we use LDA as a classifier. Here, we train LDA on our SDSS-ized simulation data to determine the most important imaging predictors for each simulation. Then, we combine all simulated galaxies to prepare an LDA classifier. In a subsequent paper, we will apply the LDA classifier to the SDSS galaxies.

Past work on simulated galaxies has shown that the effectiveness of the imaging predictors depends strongly on merger stage, the initial mass ratio, and the gas fraction of the merging galaxies (\citealt{Lotz2008,Lotz2010b,Lotz2010a}). We therefore run LDA for each simulation individually so that we can compare the LDA outputs from different merger initial conditions. In this way, we are able to compare the sensitivity of different imaging predictors for minor and major mergers with low and high gas fractions at three different merger stages (early, late, post-coalescence). For each iterative LDA run, we use simulated nonmerging galaxies that are matched for gas fraction and stellar mass to the merging galaxies, since \citet{Lotz2010a} find that gas fraction can alter the performance of $CAS$ and $Gini-M_{20}$. We therefore approach the LDA classification with a set of galaxies for which we know a priori if a galaxy belongs to the nonmerging (0) or merging (1) class. We include enough nonmerging galaxies to roughly balance the number of merging galaxies. Our motivation is to achieve an accurate LDA classification by ensuring that the isolated galaxies cover a realistic range of imaging predictor space and roughly balance the number of galaxies in the merging class. We later account for the lack of merging galaxies in nature with a prior (described below). We use disk-dominated simulated galaxies to create the LDA, so it is important to note that this classification technique is most applicable to galaxies with similar properties.

The purpose of LDA is to use Bayesian likelihood to calculate a posterior probability that a galaxy belongs to a given class (for a review of LDA, see \citealt{James2013}): 
\begin{equation}
p(\pi_0|x) = \frac{e^{\hat{\delta}_0(x)}}{e^{\hat{\delta}_0(x)}+e^{\hat{\delta}_1(x)}}
\label{1}
\end{equation}
where $\pi_0$ is the prior probability of the nonmerging class (described below), $\hat{\delta}_0$ is the score of the nonmerging class, and $\hat{\delta}_1$ is the score of the merging class. The score is the relative probability that the galaxy belongs to a class, so a galaxy will be classified into the class that has the maximum score. This classifier is nonbinary; instead of classifying galaxies as simply nonmerging or merging, we will assign a probability that a galaxy belongs to each class.

When there is only one input predictor, the discriminant score for the nonmerging class is defined as:
$$\hat{\delta}_0(x) = x \cdot \frac{\hat{\mu}_0}{\hat{\sigma}^2} - \frac{\hat{\mu}_0^2}{2\hat{\sigma}^2}+\mathrm{log}(\hat{\pi}_0)$$

where $\hat{\delta}_0(x)$ is the discriminant score for class 0 (the nonmerging class) for a set of galaxies, $x$ is the list of the one measured predictor value for all simulated galaxies, $\hat{\mu}_0$ is the mean vector for the predictor for the nonmerging class, $\hat{\sigma}^2$ is the variance of the predictor for the nonmerging class, log$(\hat{\pi}_0)$ is the prior probability of belonging to the nonmerging class, and $\hat{\delta}_1(x)$ is defined the same way but for mergers.

For the priors of the two classes for the major mergers, we use $\hat{\pi}_{0} = f_{nonmerg} = 0.9$ and $\hat{\pi}_{1} = f_{merg} = 0.1$ based on the fraction of nonmerging and merging galaxies from observations and simulations (e.g., \citealt{,Rodriguez-Gomez2015,Lotz2011,Conselice2009,Lopez-Sanjuan2009,Shi2009}). We use $\hat{\pi}_{0} = f_{nonmerg} = 0.7$ and $\hat{\pi}_{1} = f_{merg} = 0.3$ for the minor mergers since minor mergers are 3-5 times more frequent in the local universe (e.g., \citealt{Bertone2009,Lotz2011}). We find that the LDA analysis is relatively insensitive to the chosen priors within a range of values ($0.1 < f_{nonmerg} < 0.9$). For a full discussion of priors see Appendix \ref{Apriors}.

For multiple predictor variables (seven in our case), the LDA score can be generalized:
$$\hat{\delta}_0(x) = x^T\Sigma^{-1}\hat{\mu}_0 - \frac{1}{2}\hat{\mu}_0^T \Sigma^{-1} \hat{\mu}_0 + \mathrm{log}(\hat{\pi}_0)$$

where $x$, $\Sigma$, and $\hat{\mu}_0$ are now vectors for the values of the predictors, covariance matrix, and mean value of each predictor, respectively. LDA assumes that the data are normally distributed, that the input predictors are independent, and that each class has identical covariance matrices. The homogeneity of covariance matrices assumption leads to the simplification: $\Sigma_0 = \Sigma_1 = \Sigma$. We examine these statistical assumptions in more detail in Appendix \ref{mva}.

We classify a galaxy as `nonmerging' if $\hat{\delta_0} > \hat{\delta_1}$ and `merging' if $\hat{\delta_1} > \hat{\delta_0}$. Since we are working in a multi-dimensional space, this is equivalent to searching for the dividing hyperplane that satisfies:

\begin{align*}x^T\Sigma_0^{-1}\hat{\mu}_0 - \frac{1}{2}\hat{\mu}_0^T \Sigma_0^{-1} \hat{\mu}_0 + \mathrm{log}(\hat{\pi}_0) = x^T\Sigma_1^{-1}\hat{\mu}_1 - \frac{1}{2}\hat{\mu}_1^T \Sigma_1^{-1} \hat{\mu}_1 \\ + \mathrm{log}(\hat{\pi}_1)
\end{align*}

The terms with the covariance matrices can be expanded fully out to yield a quadratic classifier, as is done in Quadratic Discriminant Analysis (QDA). We assume the equality of covariance matrices, which means the covariances between predictors are roughly equivalent for the nonmerging and merging class ($\Sigma_0 = \Sigma_1$). This assumption yields a linear classifier (LDA):

$$\Sigma^{-1}(\mu_0-\mu_1) + \frac{1}{2} \mu_0^T \Sigma^{-1} \mu_0 +\frac{1}{2} \mu_1^T \Sigma^{-1} \mu_1
 + \mathrm{log}(\frac{\hat{\pi}_0}{\hat{\pi}_1}) = 0$$
 
We solve for the hyperplane that satisfies the above equation, LD1:
 $$\mathrm{LD1} = \hat{\vec{w}}^T \vec{x} + \hat{w_0} = 0$$
 
where the slope, $\hat{\vec{w}}$, is the weight vector:
 $$\hat{\vec{w}} = \Sigma^{-1}(\mu_0-\mu_1)$$
 
 and the intercept is given by $\hat{w_0}$:
 $$\hat{w_0} = \frac{1}{2} \mu_0^T \Sigma^{-1} \mu_0 +\frac{1}{2} \mu_1^T \Sigma^{-1} \mu_1
 + \mathrm{log}(\frac{\hat{\pi}_0}{\hat{\pi}_1})$$
 
LD1 is also known as the first discriminant axis. Since we have only two classes (merging and nonmerging) to separate in this analysis, the second, third, and so on discriminant axes are unnecessary. Instead, we are able to focus only on one hyperplane to separate the populations.

 We run the LDA on the imaging input predictors, which are the $Gini$, $M_{20}$, $C$, $A$, $S$, $n$, and $A_S$. We specifically utilize the python package \texttt{sklearn} for this analysis. By focusing on the imaging predictors, our goal is to produce a result that is useful for observational samples of galaxies with imaging only. Since the imaging predictors are cross-correlated, meaning that combinations of two of the predictors have a linear relationship with one another (Appendix \ref{mva}), we also include `interaction' terms which are multiples of all combinations of the imaging predictors (e.g., $Gini*M_{20}$, $Gini*C$, $Gini*A$, etc). We refer to these as `interaction' terms, but they can be better thought of as multiplicative terms that allow us to explore the synergistic effects of combining predictors. These interaction terms allow us to remove cross-correlation effects from the original `primary' ($Gini$, $M_{20}$, $C$, $A$, $S$, $n$, and $A_S$) imaging predictors. We can then directly explore how these primary imaging predictors affect the classifier in Section \ref{vec}.
 
Including the interaction terms, we have 34 input terms for each run of LDA. Therefore, we first use forward stepwise selection with $k-$fold cross-validation to select the best input variables for each simulation. In brief, forward stepwise selection proceeds by introducing one predictor at a time; we choose the number of predictors that minimizes the number of misclassifications determined with cross-validation. We specifically implement $k-$fold cross-validation, which is a method to divide the full sample of merging and nonmerging galaxies (for each run) into $k$ equally sized subsamples, where $k = 10$. We then train the LDA on nine of the subsamples, and test on the tenth sample. We repeat this procedure ten times, and the mean number of misclassifications all ten test samples allows us to decide which set of input predictors to select. We proceed, adding one predictor at a time, until the minima of the misclassifications is determined. We describe this process in more detail in Appendix \ref{Akfold}. 

The input predictors that are selected by the forward stepwise selection are given in Tables \ref{tab:cv} and \ref{tab:cvinteraction}, along with their coefficient values and standard errors from the LDA run. The standard errors are obtained using $k-$fold cross-validation (Appendix \ref{Akfold}). If a predictor is selected by the forward stepwise selection but the 3$\sigma$ standard error indicates that it is consistent with zero, we eliminate it from the selected predictors. We refer to the remaining imaging predictors and imaging predictor interaction terms as `required' predictors henceforth because they are necessary to separate the merging galaxies from the nonmerging galaxies along LD1 for each simulation. LD1 is a linear combination of the selected input imaging predictors and interaction terms, with weights $\hat{\vec{w}}$ and intercept term $\hat{w_0}$. Each element of $\hat{\vec{w}}$ corresponds to an imaging predictor or interaction term, and their relative absolute values represent their degree of importance to the classification. We report and interpret these coefficients, their relative signs, and their order of importance in Section \ref{results}.

After running LDA on each simulation individually, we assess their differences in Section \ref{discuss_parameters} and \ref{gas}. Since the major and minor merger LDA runs are significantly different, we caution against combining all runs into one overall classifier. We do attempt, however, to combine all simulations into one classifier and find that it does not adequately separate merging from nonmerging galaxies for all merger simulations. Instead, we create two classifiers, one from the combined major merger simulations and one from the combined minor merger simulations, that will be used to classify the SDSS galaxies. They could also function as a diagnostic tool to determine the mass ratio of the merging galaxies.

In subsequent work, we will calculate the value of LD1, or the score of a given galaxy, using the linear combination of the terms from $\hat{\vec{w}}$ and the input predictors and $\hat{w_0}$ given in Section \ref{results}. For example, LD1 for the combined overall run for major mergers is:

\begin{equation}
\begin{aligned}
\mathrm{LD1}_{\mathrm{major}} = &\ \ \ \ 0.69\ Gini +3.84\  C + 5.78\  A + 13.14\  A_S \\
&-3.68\  Gini*A_S -6.5\ C*A_S -6.12\ A*A_S\\
&-0.81
\label{eq2}
\end{aligned}
\end{equation}

where all predictor inputs must be standardized before using this equation. To standardize, we subtract the mean and divide by the standard deviation of the set of all predictor values.

Likewise, LD1 for the minor merger combined simulation is:

\begin{equation}
\begin{aligned}
\mathrm{LD1}_{\mathrm{minor}} = &\ \ \ \ 8.64\ Gini +14.22\  C + 5.21\ A + 2.53\ A_S \\
&-20.33\ Gini*C -4.32\ A*A_S  \\
& -0.87
\label{eq3}
\end{aligned}
\end{equation}

The decision boundary for LD1 for the major merger combined run is 1.16 and 0.42 for the combined minor merger run; all galaxies with values of LD1 greater than this value will be classified as merging. This decision boundary is the halfway mark between the mean of the merging and nonmerging galaxy distributions. From here on, we use `LD1' to describe the linear combination of predictor coefficients for each run of LDA.

LD1 is a hyperplane, so it is unable to capture complicated non-linearities in the imaging predictors. For instance, there is some migration for different merger stages that occurs for predictors such as $M_{20}$, where merging galaxies occupy different regions of predictor space for different phases of the merger. Since the LDA captures the bulk behavior of each imaging predictor, it searches for the overall trend for all stages within each merger simulation and is unable to describe these non-linearities.

 \begin{figure*}
\centering
\hspace{-1cm}

\includegraphics[scale=0.19,trim=1cm 0cm 0cm 0cm]{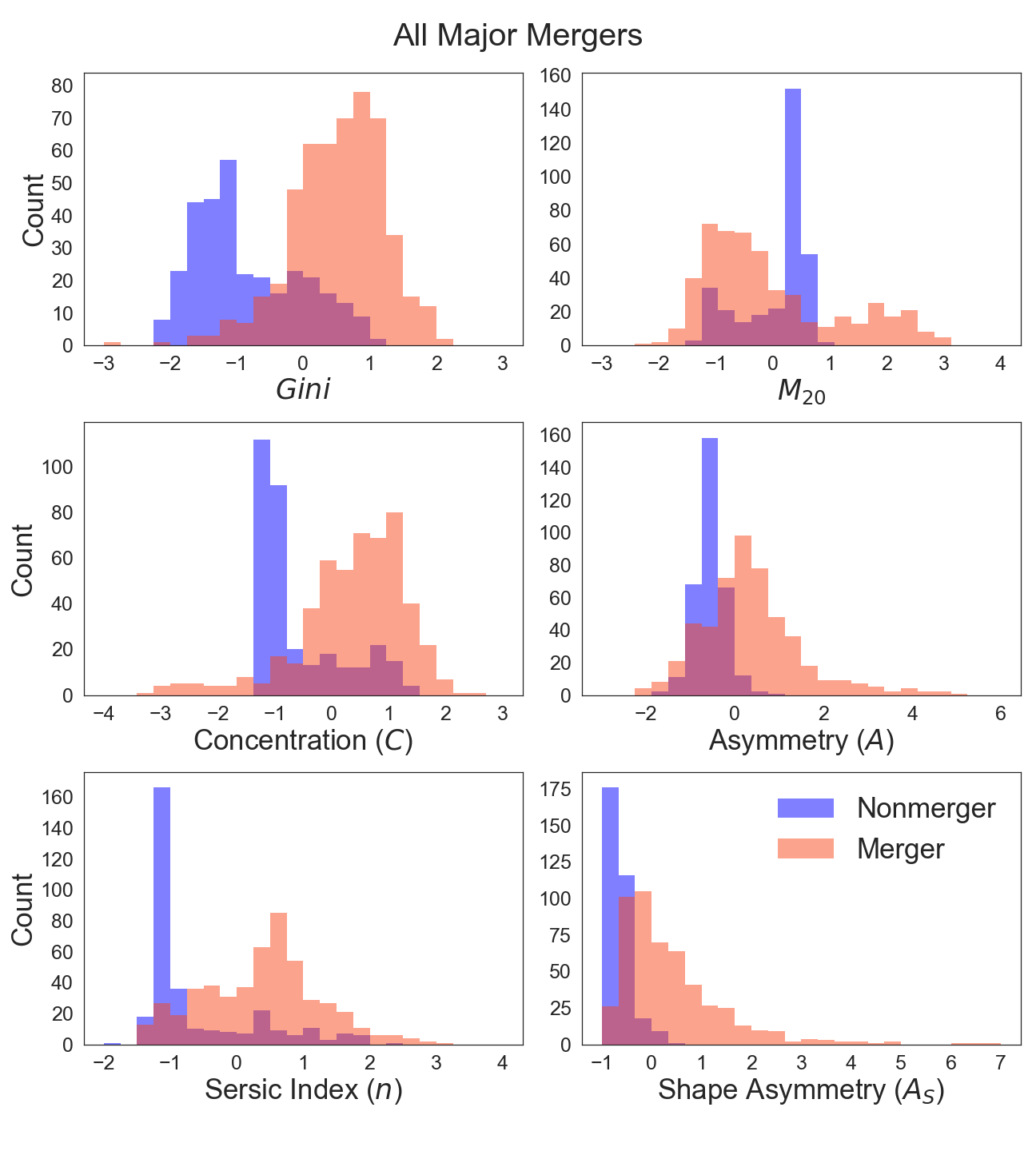}\hspace{0.5cm}
\includegraphics[scale=0.19,trim=1cm 0cm 2cm 0cm]{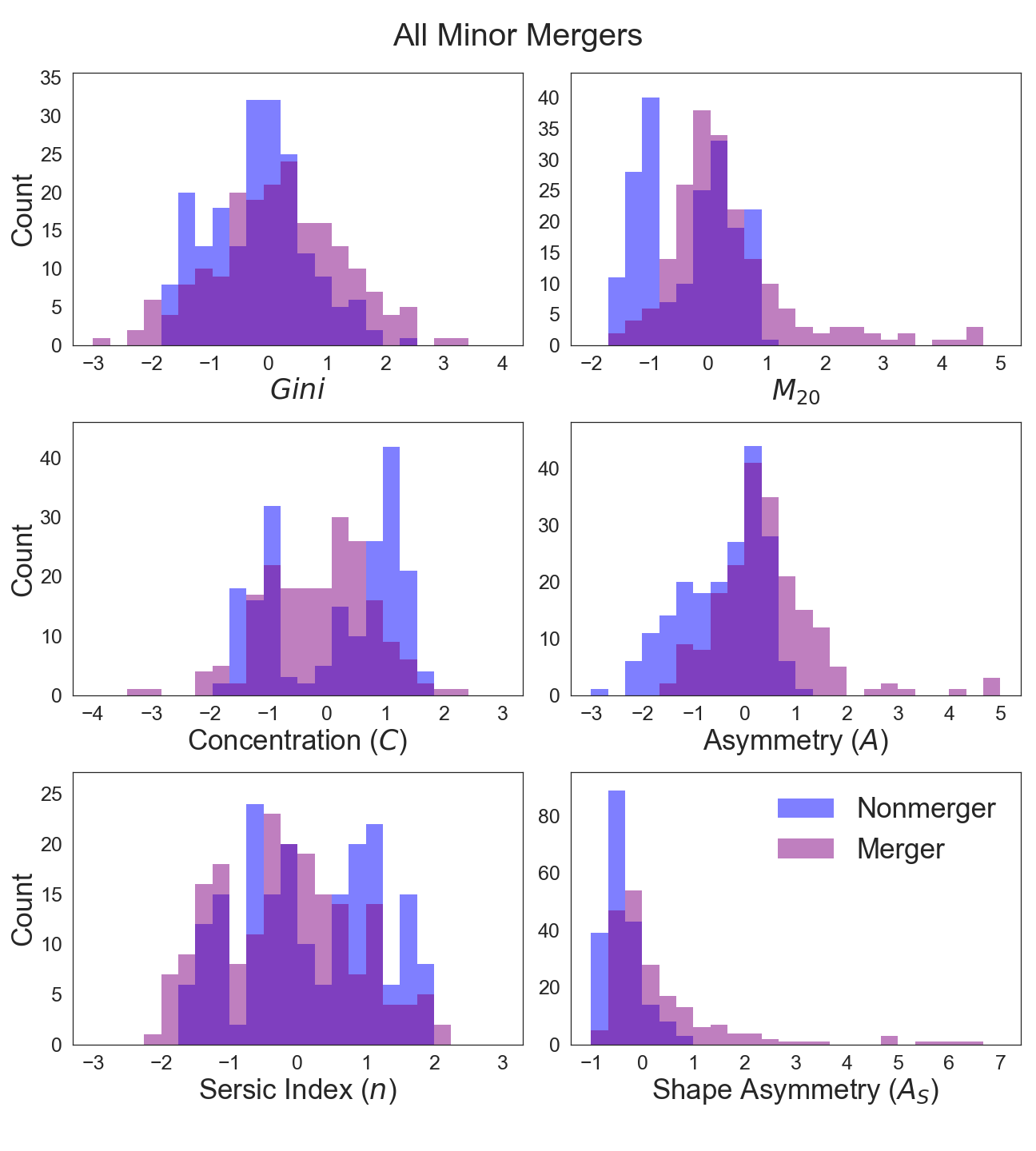}
\caption{Individual histograms for each imaging predictor for the combined major merger simulations (left) and combined minor merger simulations (right). We show for all the simulations combined that we are unable to cleanly separate the nonmerging (blue) and merging (pink and purple for major and minor mergers, respectively) galaxies using any individual imaging predictor. The y-axis is the `Count' or number of merger simulation snapshots in each bin. The x-axis of each subplot is the standardized predictor value for the seven imaging predictors, where the mean of the combined merging and nonmerging populations for each predictor is 0 and the standard deviation is 1. Standardizing the input predictor values acts to stabilize the LDA but has no effect on the relative value of each predictor; for example, a greater value of $Gini$ corresponds to an increased likelihood that a given galaxy is a merger. This statement is valid both for the measured $Gini$ value and the standardized $Gini$ value given here. }
\label{histograms}
\end{figure*}

\begin{figure*}
\includegraphics[scale=0.33]{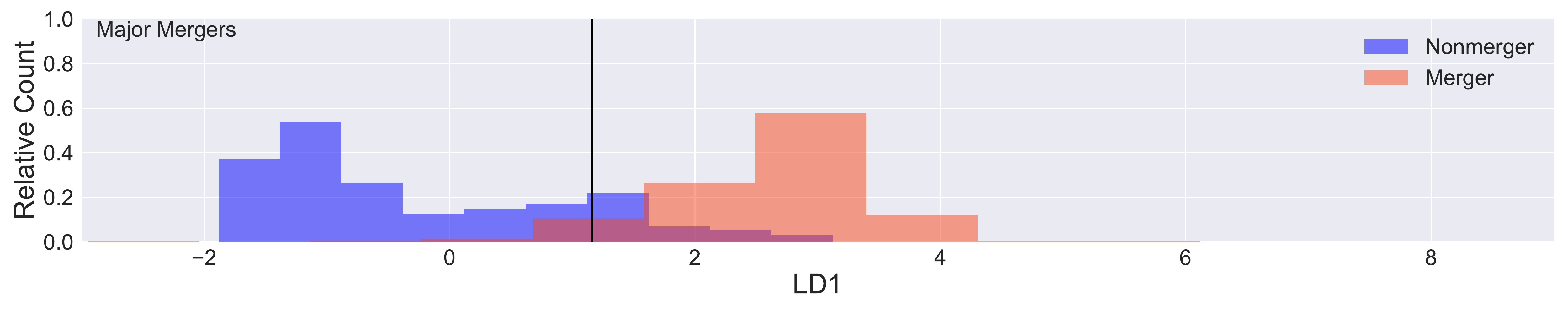}
\includegraphics[scale=0.33]{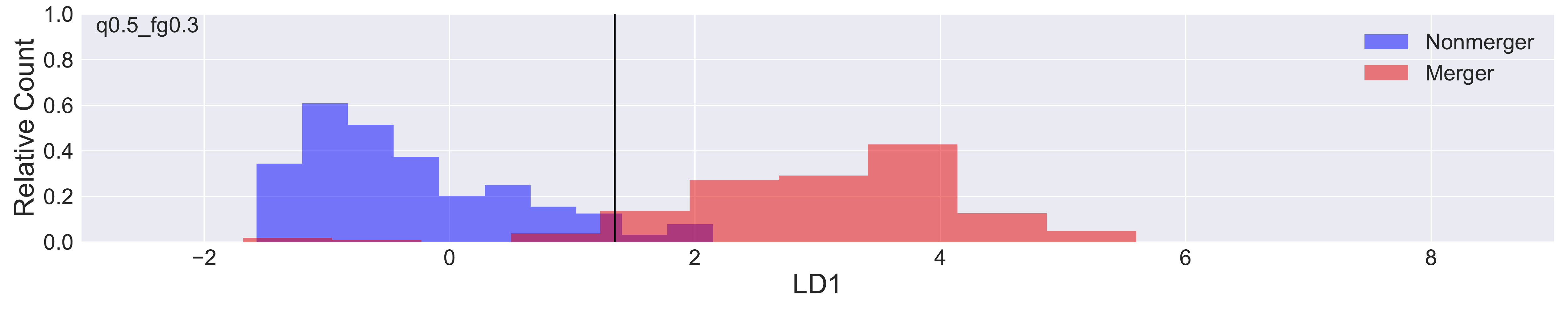}

\includegraphics[scale=0.33]{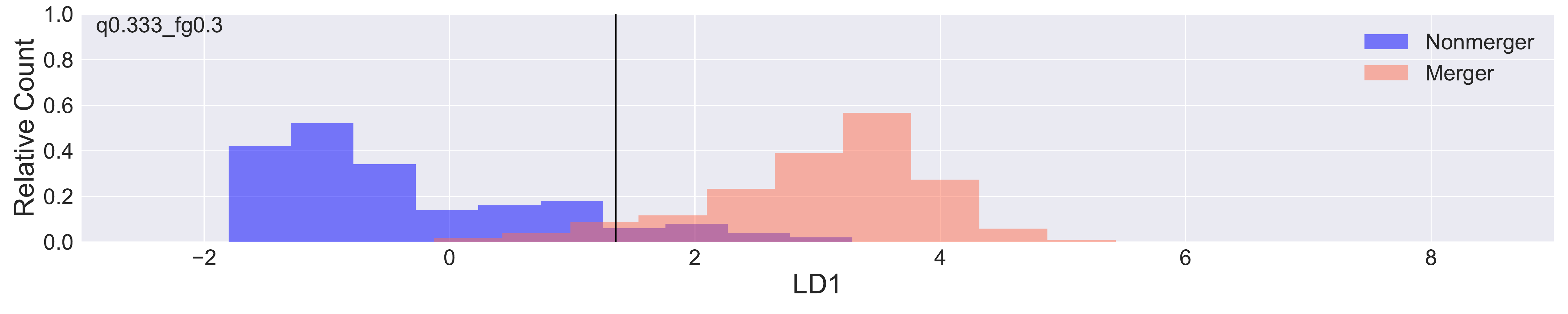}
\includegraphics[scale=0.33]{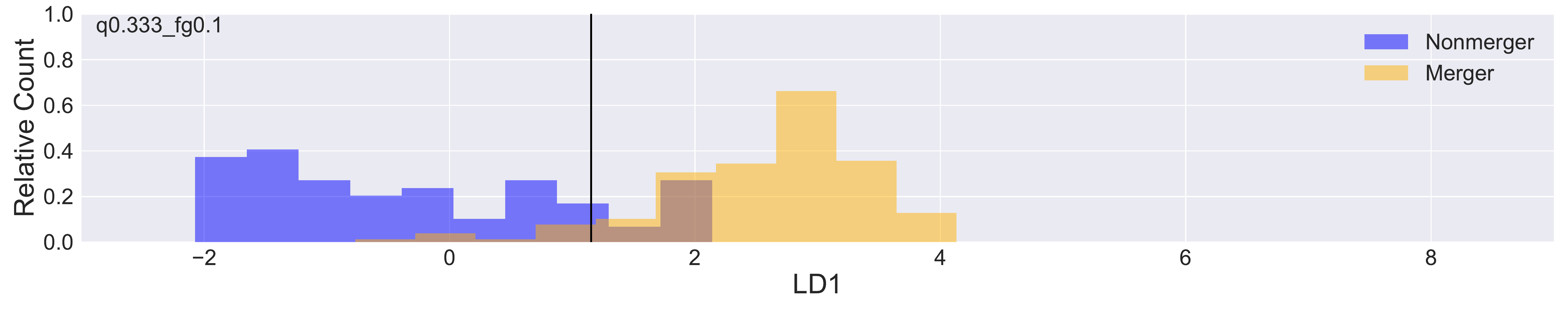}

\caption{Histograms of the distribution of score values for galaxies in the nonmerging and merging class for all of the major merger simulations and the combined major merger run (top). The vertical black line marks the decision boundary - the halfway mark between the mean of the nonmerging and merging distributions. A galaxy with a score above (to the right) of this value has a higher probability of being a merging galaxy and a galaxy with a score to the left of this value has a higher probability of being a nonmerging galaxy. The decision boundaries are in similar locations for all major merger runs of LDA because the separation of the merging and nonmerging populations is so similar. The histograms have different colors to distinguish different LDA runs and the blue bars are the matched nonmerger sample for each run.}
\label{after_histograms}
\end{figure*}

\begin{figure*}

\includegraphics[scale=0.33]{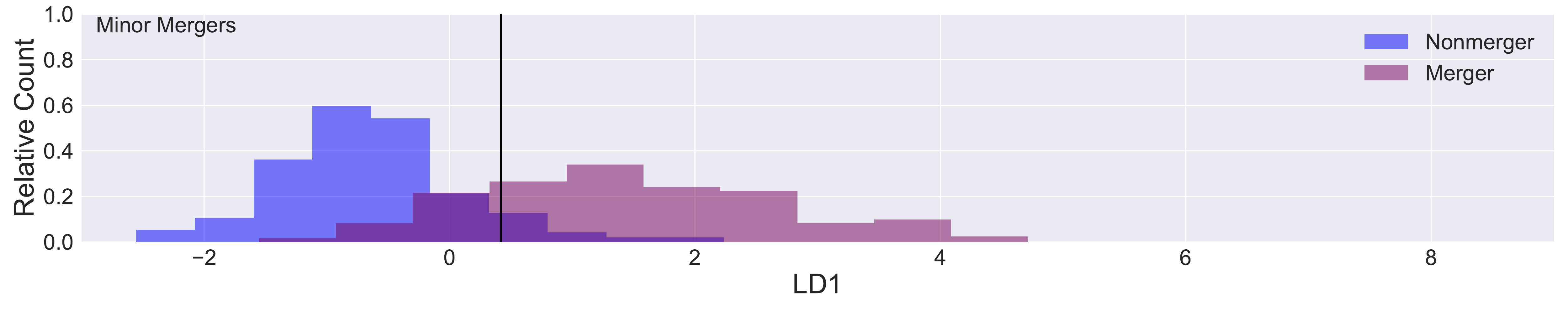}
\includegraphics[scale=0.33]{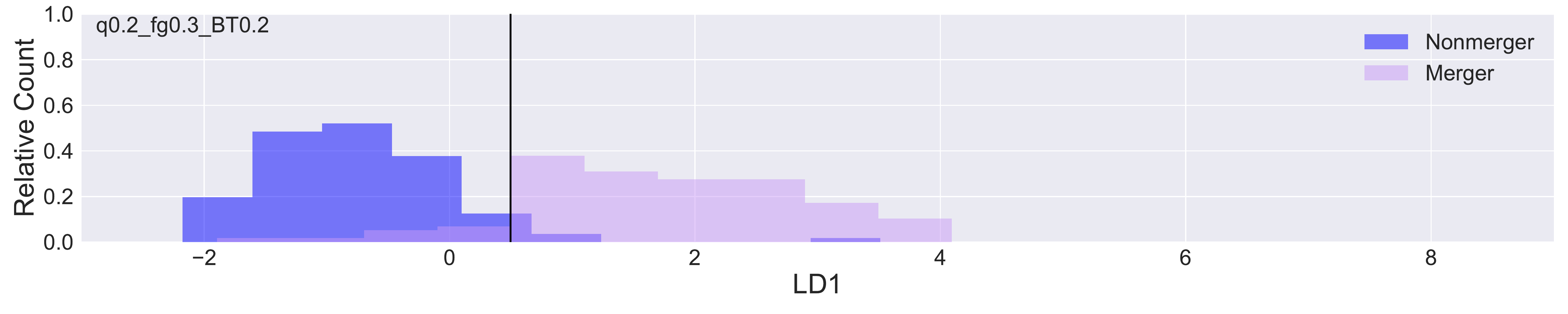}

\includegraphics[scale=0.33]{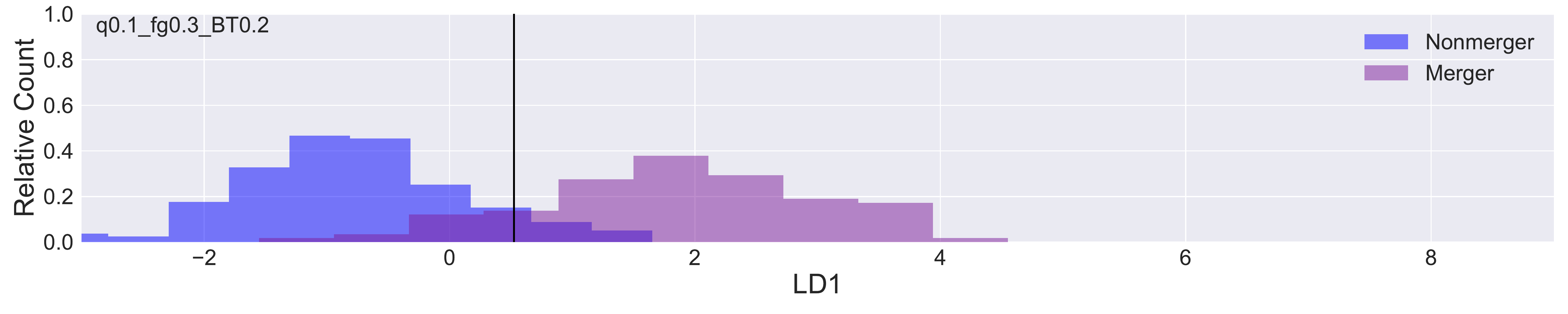}

\caption{Same as Figure \ref{after_histograms} but for all of the minor merger runs of LDA and the combined minor merger run (top). The decision boundaries for these simulations are lower than for the major merger runs because the separation between classes is not as extreme for the minor mergers.}
\label{after_histograms_2}
\end{figure*}
 
LDA adequately separates nonmerging from merging galaxies. For instance, Figure \ref{histograms} presents the histograms for the imaging predictors for all simulations, and it is clear that the imaging predictors are each individually unable to separate the populations of merging and nonmerging galaxies. After performing LDA, we find that we are able to more cleanly separate the two classes using the first discriminant axis LD1 (Figures \ref{after_histograms} and \ref{after_histograms_2}).

While it is possible to classify a galaxy as merging or nonmerging given a decision boundary and a value of LD1, we use the posterior probability that a galaxy belongs to a given class from Equation \ref{1}. Since we standardize the input predictors to train the LDA, classifying galaxies after the determination of LD1 is complicated. Instead of simply plugging in measured values of predictors into LD1, it is necessary to apply the same standardization used in this work prior to classification. 

We discuss the statistical assumptions made by LDA in Appendix \ref{mva}. We discuss the coefficients of LD1 in Section \ref{vec} and the implications for each run and the combined run. Finally, we demonstrate in Appendix \ref{accuracy} that LDA classification is able to accurately separate the classes of merging and nonmerging galaxies.

\section{Results}
\label{results}

After running LDA for each galaxy merger, we compare the results. We describe our methodology to compare the LDA classifications from different simulations in Section \ref{vec}. Finally, we compute the observability timescales for $Gini-M_{20}$, $A$, $A_S$, and the LDA technique in Section \ref{timescale}. We describe the LDA classification in more statistical detail in Appendices \ref{Apriors}, \ref{mva}, \ref{Akfold}, and \ref{accuracy}, where we include an investigation of the merging galaxy priors used, a multivariate analysis of the assumptions of LDA, a description of the k-fold error estimation, and an examination of the accuracy and precision of the tool, respectively.

\subsection{Analyzing the LD1 Coefficients}
\label{vec}

Since we run LDA on each merger simulation individually, and LD1 is a vector, we produce different values for each coefficient of LD1. An advantage of LDA is that we are able to directly interpret the relative weights of each individual predictor (Tables \ref{tab:cv} and \ref{tab:cvinteraction}) for each simulation. We focus on the primary predictors, which are in Table \ref{tab:cv}, since they are a more straightforward way to interpret the influence of the imaging predictors than the interaction terms in Table \ref{tab:cvinteraction}. We compare the values of these primary coefficients of LD1 for each simulation. The coefficients have positive or negative values; since a larger value of LD1 indicates that a galaxy is a merger, a positive coefficient indicates that increasing the corresponding predictor increases the likelihood that the galaxy is a merger. Our goal is to determine if the classification is significantly different for different simulations and if it differs for different merger initial conditions.

\begin{sidewaystable*}

    \vspace{5cm}
  \begin{center}
    \caption{LD1 predictor coefficients with 1$\sigma$ confidence intervals. Bolded values are significantly greater than zero (to 3$\sigma$). We include only the predictors that are selected by the forward stepwise selection; for example, in q0.5\_fg0.3, $M_{20}$, $C$, $A$, and $A_S$ are excluded by this selection. $\hat{\vec{w}}$ and $\hat{w_0}$ are the weight vector (composed of the predictor coefficients) and the intercept, respectively. Together, they describe the LD1 hyperplane that best separates the populations of merging and nonmerging galaxies for each simulation. The coefficients have positive or negative values;  a positive coefficient indicates that increasing the corresponding predictor increases the likelihood that the galaxy is a merger. }
    \label{tab:cv}
    \begin{tabular}{c|ccccccc|c}
      
      & \multicolumn{6}{c}{\Large{$\hat{\vec{w}}$}} & & \Large{$\hat{w_0}$} \\
      \hline
      Simulation & $Gini$ & $M_{20}$ & $C$ & $A$ & $S$ & $n$ & $A_{S}$ &  \\

\hline

All Major & \textbf{0.69 $\pm$ 0.21} & -- & \textbf{3.84 $\pm$ 0.23} & \textbf{5.78 $\pm$ 0.21} & -- & -- & \textbf{13.14 $\pm$ 0.61} & \textbf{-0.81 $\pm$ 0.05}\\

All Minor & \textbf{8.64 $\pm$ 1.14} & -- & \textbf{14.22 $\pm$ 1.66} & \textbf{5.21 $\pm$ 0.26} & -- & -- & \textbf{2.53 $\pm$ 0.2} & \textbf{-0.87 $\pm$ 0.04}\\

q0.5\_fg0.3 & -- & 0.75 $\pm$ 0.29 & \textbf{-0.82 $\pm$ 0.16} & \textbf{9.93 $\pm$ 0.39} & -- & -- & \textbf{5.89 $\pm$ 0.19} & \textbf{-2.76 $\pm$ 0.05}\\

q0.333\_fg0.3 & \textbf{4.18 $\pm$ 0.22} & -- & -- & \textbf{6.15 $\pm$ 0.68} & -- & -- & \textbf{22.17 $\pm$ 1.2} & \textbf{-0.44 $\pm$ 0.14}\\

q0.333\_fg0.1 & -- & -- & \textbf{5.38 $\pm$ 0.19} & \textbf{5.66 $\pm$ 0.28} & -- & -- & \textbf{11.41 $\pm$ 0.39} & \textbf{-0.56 $\pm$ 0.1}\\

q0.2\_fg0.3\_BT0.2 & \textbf{19.34 $\pm$ 2.89} &-4.08 $\pm$ 3.29 & \textbf{24.69 $\pm$ 1.87} & \textbf{5.88 $\pm$ 0.43} & -- & -- & \textbf{3.97 $\pm$ 0.31} & \textbf{-0.87 $\pm$ 0.07}\\

q0.1\_fg0.3\_BT0.2 & \textbf{11.39 $\pm$ 1.17} & -- & \textbf{33.27 $\pm$ 2.0} & \textbf{29.74 $\pm$ 2.33} & -- & -- & -5.05 $\pm$ 1.95 & \textbf{-1.75 $\pm$ 0.11}\\

    \end{tabular}
  \end{center}
     
\end{sidewaystable*}

\begin{sidewaystable*}

\vspace{9cm}
  \begin{center}
     \caption{Same as Table \ref{tab:cv} but for the interaction terms of all runs. }
    \label{tab:cvinteraction}
    \begin{tabular}{c|ccccccccc}
      
      & \multicolumn{6}{c}{\Large{$\hat{\vec{w}}$}} & & \\
      
      \hline
      Simulation & $Gini*M_{20}$ & $Gini*C$ & $Gini*A$ & $Gini*S$ & $Gini*n$ & $Gini*A_S$ & $M_{20}*C$ &  $M_{20}*A$ & $M_{20}*S$\\
\hline

All Major  & -- & -- & -- & -- & -- & \textbf{-3.68 $\pm$ 0.93} & -- & -- & --\\

All Minor   & -- & \textbf{-20.33 $\pm$ 2.53} & -- & -- & -- & -- & -- & -- & --\\

q0.5\_fg0.3 & -- & -- & -- & -- & -- & -- & \textbf{-2.01 $\pm$ 0.43} & -- & --\\

q0.333\_fg0.3  & -- & -- & -- & -- & -- & \textbf{-19.09 $\pm$ 1.14} & -- & -- & --\\

q0.333\_fg0.1 & -- & -- & -- & -- & -- & -- & -- & -- & --\\

q0.2\_fg0.3\_BT0.2 & 2.98 $\pm$ 3.57 & \textbf{-38.04 $\pm$ 2.88} & -- & -- & -- & -- & -- & -- & --\\

q0.1\_fg0.3\_BT0.2  & -- & \textbf{-39.29 $\pm$ 2.88} & \textbf{-27.95 $\pm$ 2.48} & -- & -- & \textbf{28.81 $\pm$ 2.04} & -- & -- & --\\

\hline
\hline
\hline
       & $M_{20}*n$ & $M_{20}*A_S$ & $C*A$ & $C*S$ & $C*n$ & $C*A_S$ & $A*S$ &  $A*n$ & $A*A_S$\\
\hline

All Major & -- & -- & -- & -- & -- & \textbf{-6.5 $\pm$ 0.5} & -- & -- & \textbf{-6.12 $\pm$ 0.27}\\

All Minor & -- & -- & -- & -- & -- & -- & -- & -- & \textbf{-4.32 $\pm$ 0.39}\\

q0.5\_fg0.3 & -- & -- & -- & -- & -- & -- & -- & -- & \textbf{-9.52 $\pm$ 0.44}\\

q0.333\_fg0.3  & -- & -- & -- & -- & -- & -- & -- & -- & \textbf{-6.03 $\pm$ 0.91}\\

q0.333\_fg0.1  & -- & -- & -- & -- & -- & \textbf{-8.57 $\pm$ 0.34} & -- & -- & \textbf{-5.92 $\pm$ 0.3}\\

q0.2\_fg0.3\_BT0.2  & -- & -- & -- & -- & -- & -- & -- & -- & \textbf{-5.21 $\pm$ 0.47}\\

q0.1\_fg0.3\_BT0.2 & -- & -- & \textbf{7.16 $\pm$ 0.73} & -- & -- & \textbf{-20.28 $\pm$ 0.86} & -- & -- & \textbf{-6.88 $\pm$ 0.34}\\

\hline
\hline
\hline

 & $S*n$ & $S*A_S$ & $n*A_S$ & & & & \\
\hline

All Major & -- & -- & --\\

All Minor & -- & -- & --\\

q0.5\_fg0.3& -- & -- & --\\

q0.333\_fg0.3  & -- & -- & --\\

q0.333\_fg0.1 & -- & -- & --\\

q0.2\_fg0.3\_BT0.2 & -- & -- & --\\

q0.1\_fg0.3\_BT0.2 & -- & -- & --\\

    \end{tabular}
  \end{center}
    
\end{sidewaystable*}

We use stratified $k$-fold cross-validation (Appendix \ref{Akfold}) to determine the standard error on the coefficients of LD1 that are selected by forward stepwise selection. Briefly, we randomly split the sample into ten parts, where nine parts are the training sample and the tenth part is the test sample. Stratified $k-$fold cross-validation ensures that the percentage of merging and nonmerging galaxies in the test set matches that of the full sample. We perform this operation ten times and then calculate the mean value and standard deviation (standard error) for the LD1 coefficients and intercept ($\hat{\vec{w}}$ and $\hat{w}_0$) from the ten iterations of training and test sets. 

\begin{figure*}
\hspace{-1cm}
\includegraphics[scale=0.3]{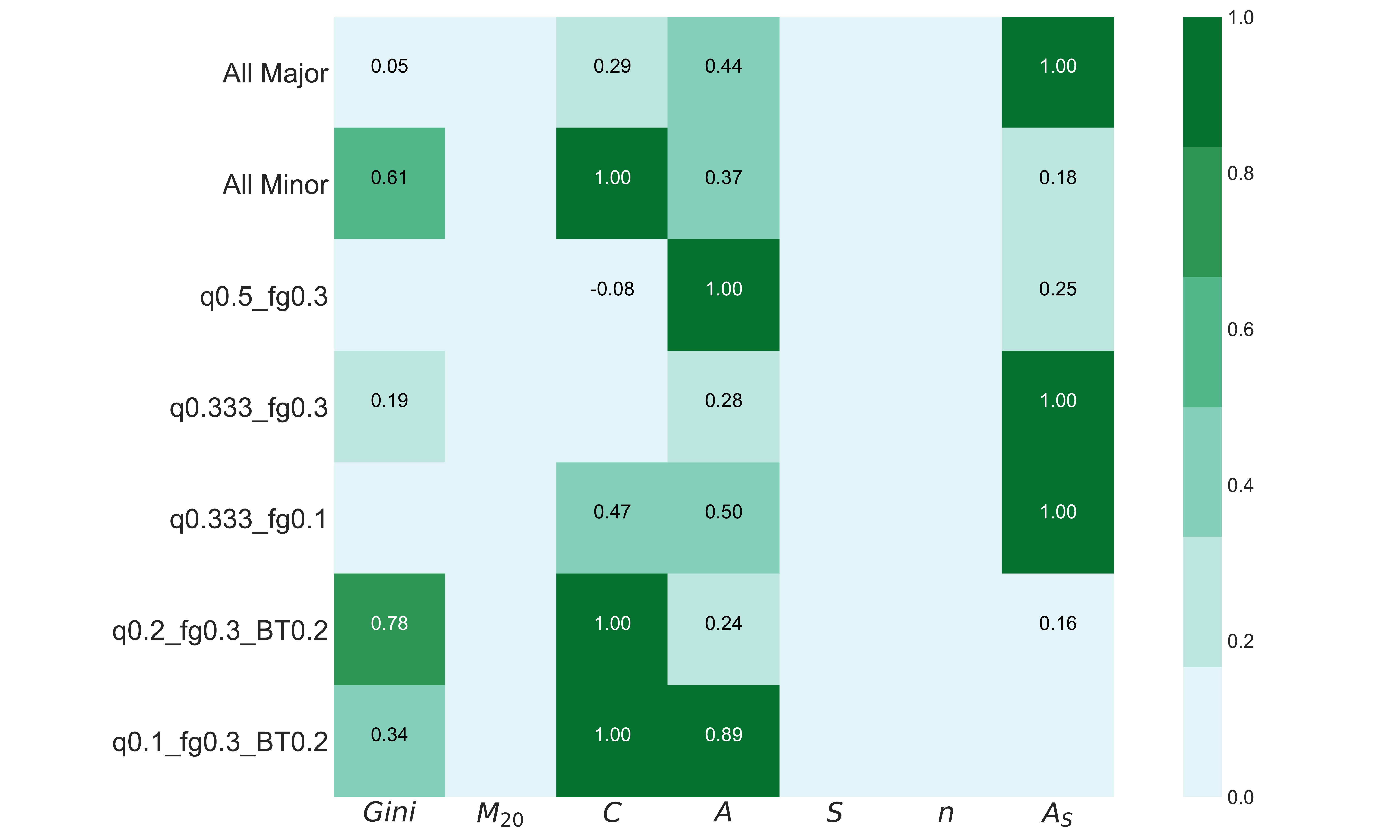}
\caption{Visualization of the LD1 coefficient values of each predictor, weighted relative to the maximum primary predictor for each run. This includes each LDA run for each individual simulation and all of the major merger simulations combined (first row) and all of the minor merger simulations combined (second row). We find that the relative importance of predictors changes between all simulations.}
\label{color_predictors}
\end{figure*}

For both Table \ref{tab:cv} and Table \ref{tab:cvinteraction}, include the predictors that are selected by the forward stepwise selection. Additionally, we bold the input predictors that are significant (to 3$\sigma$ above zero) according to their errors provided by $k-$fold cross-validation. We use both of these predictor selection techniques to determine which predictors are selected and significant (we exclude all other predictors from our analysis and discussion). We show a visualization in Figure \ref{color_predictors} of the order of importance of the individual primary imaging predictors for each simulation. Overall, we can only discard the clumpiness ($S$) primary predictor from our analysis; it is always either excluded by the forward stepwise selection of predictors or $<3\sigma$ above zero.

There are significant differences between the rankings of imaging predictors for each simulation. For instance, we find that the major merger simulations (q0.5\_fg0.3, q0.333\_fg0.3, and q0.333\_fg0.1) have different rankings of predictor importance; $A_S$ and $A$ are more important for the major mergers. For minor mergers, $A_S$ is unimportant, while $C$ and $Gini$ become very important.

We interpret the sign of each coefficient individually for each simulation in Section \ref{discuss}, comparing to previous work. We further interpret the relative importance of the coefficients for different merger initial conditions and discuss that the value of the predictors evolve as the merger progresses in Section \ref{discuss}.

\begin{figure*}
\vspace{-2cm}\includegraphics[scale=0.8]{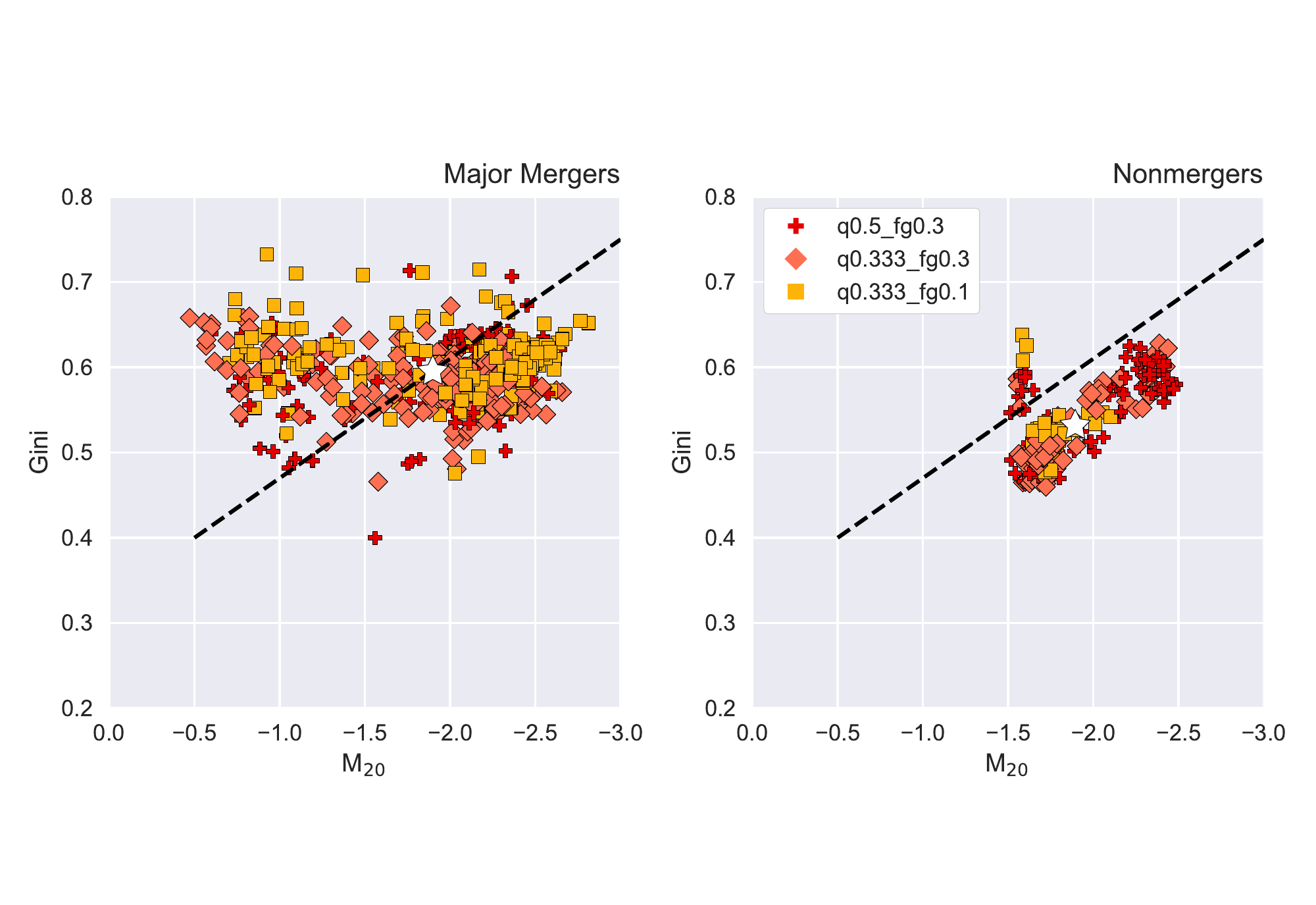}\vspace{-1.5cm}
\vspace{-6cm}\includegraphics[scale=0.8]{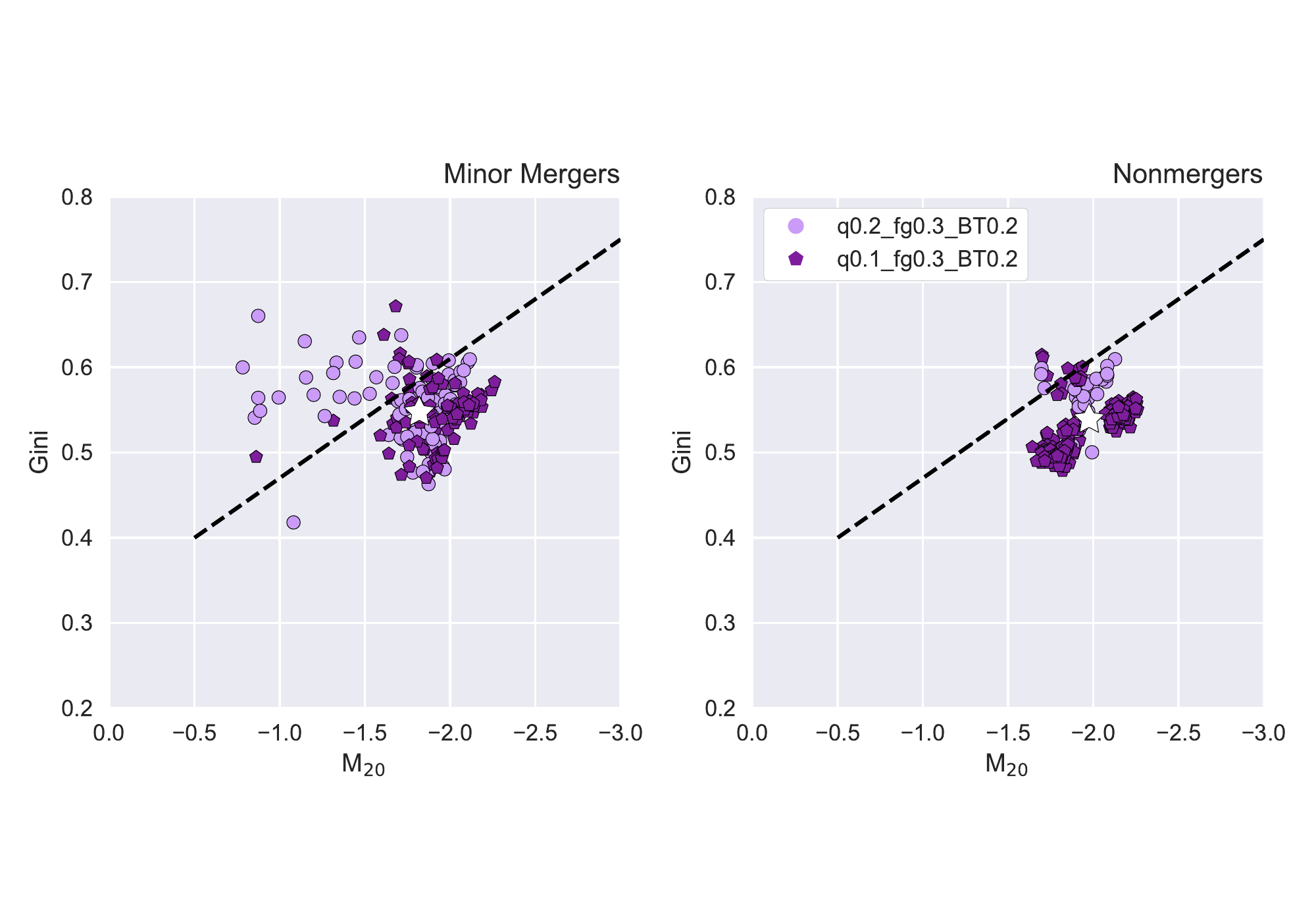}
\vspace{5cm}
\caption{Viewpoint-averaged values at each snapshot of a merger for the three major merger simulations (top) and two minor merger simulations (bottom) in the $Gini-M_{20}$ two-dimensional predictor space. Red is q0.5\_fg0.3, pink is q0.333\_fg0.3, yellow is q0.333\_fg0.1, light purple is q0.2\_fg0.3\_BT0.2, and dark purple is q0.1\_fg0.3\_BT0.2. The stars mark the mean of the data presented on each plot. We show the merging galaxies in the left plots and the matched nonmerging isolated galaxies on the right. We include the cuts in predictor space from Section \ref{timescale} used to identify merging galaxies in other work. The cut in $Gini-M_{20}$ is $Gini > 0.14 \ M_{20} + 0.33$; all galaxies above this cut are identified as mergers in other work (e.g., \citealt{Lotz2008}).}
\label{gini_m20}
\end{figure*}

\begin{figure*}
\vspace{-2cm}\includegraphics[scale=0.8]{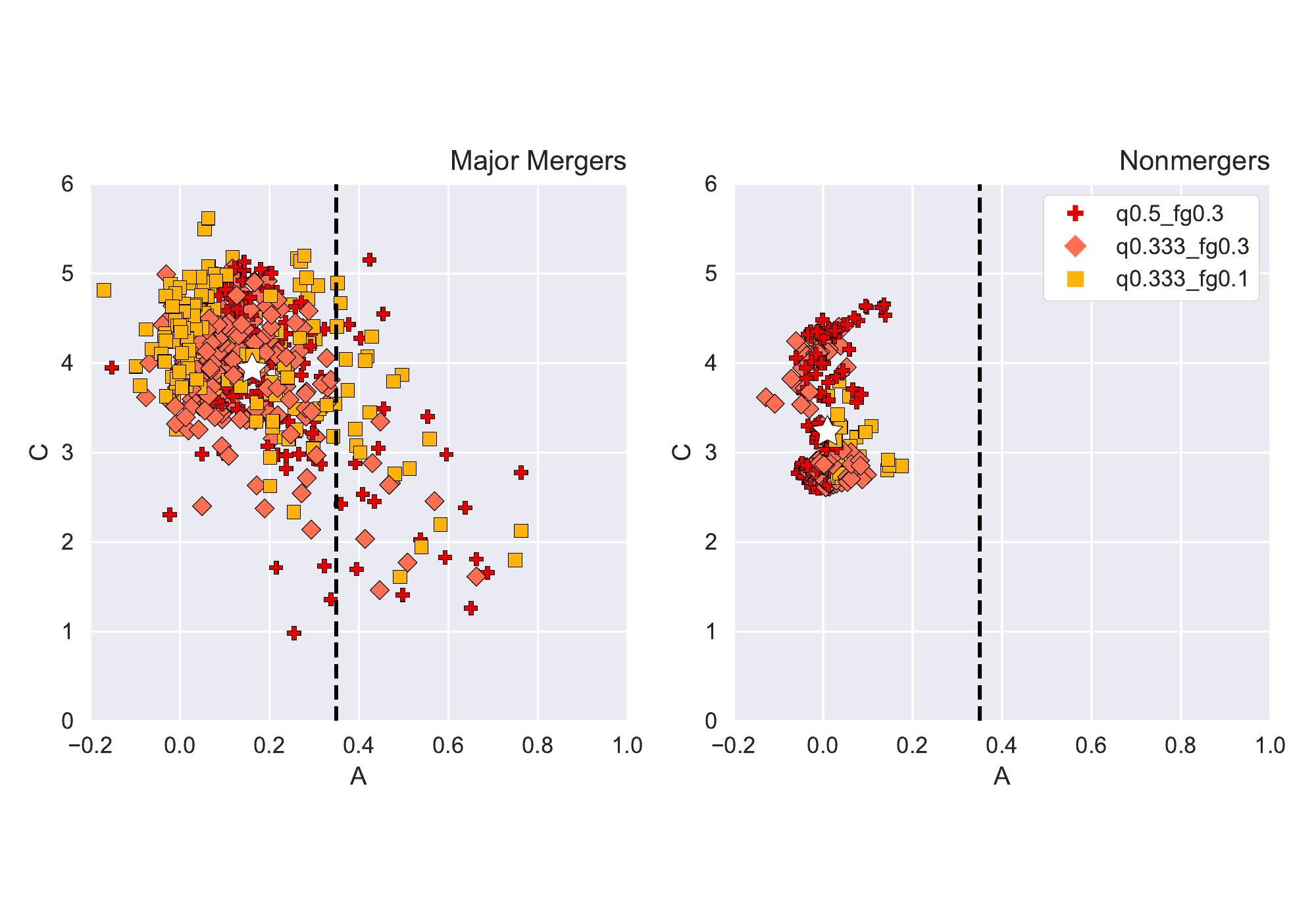}\vspace{-1.5cm}
\vspace{-6cm}\includegraphics[scale=0.8]{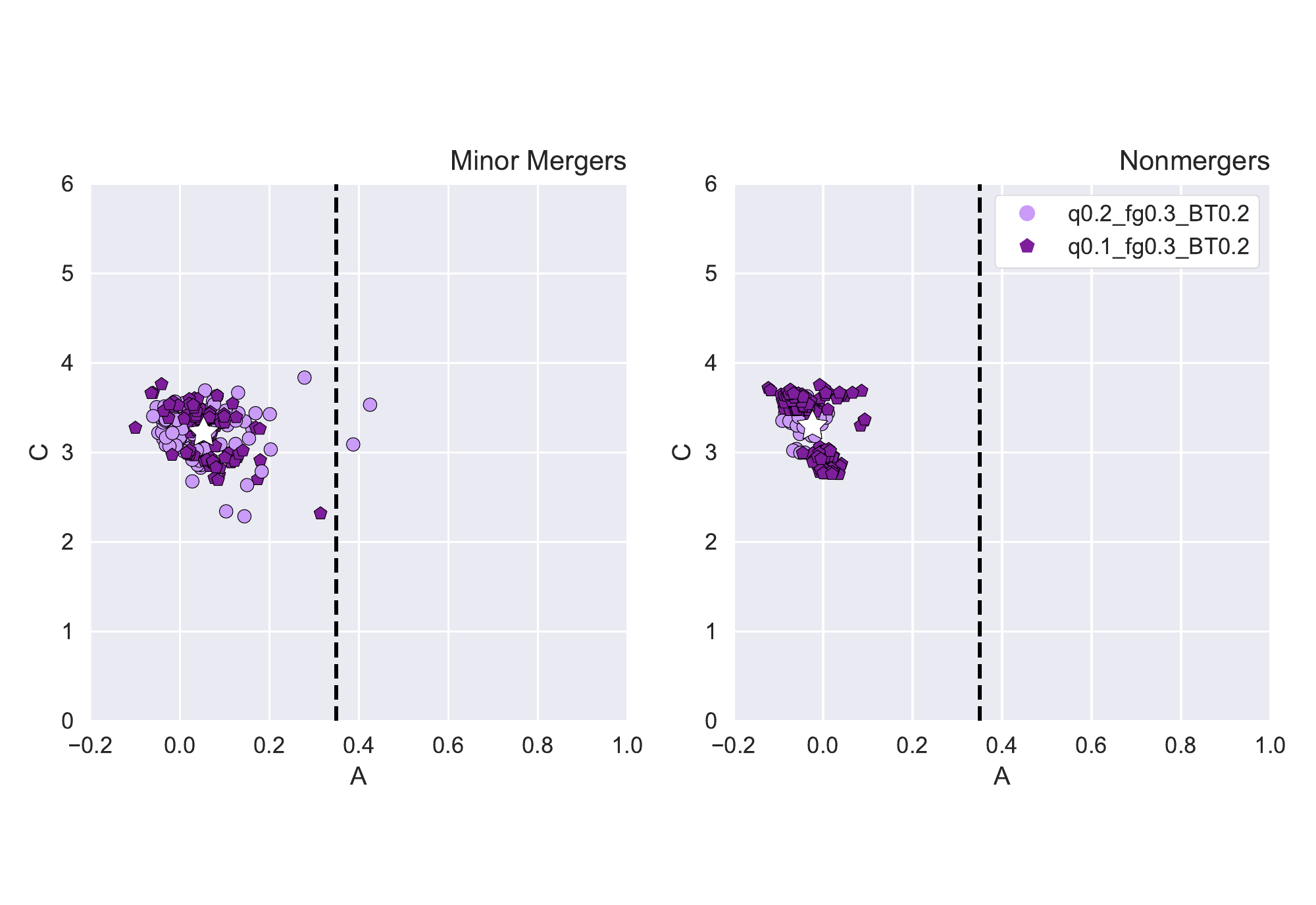}
\vspace{5cm}
\caption{Same as Figure \ref{gini_m20} but for the $CA$ predictor space. The cut in $CA$ space is $A > 0.35$; all galaxies to the right of this line are identified as mergers in other work (e.g., \citealt{Conselice2003}).}
\label{C_A}
\end{figure*}

\begin{figure*}
\vspace{-2cm}\includegraphics[scale=0.8]{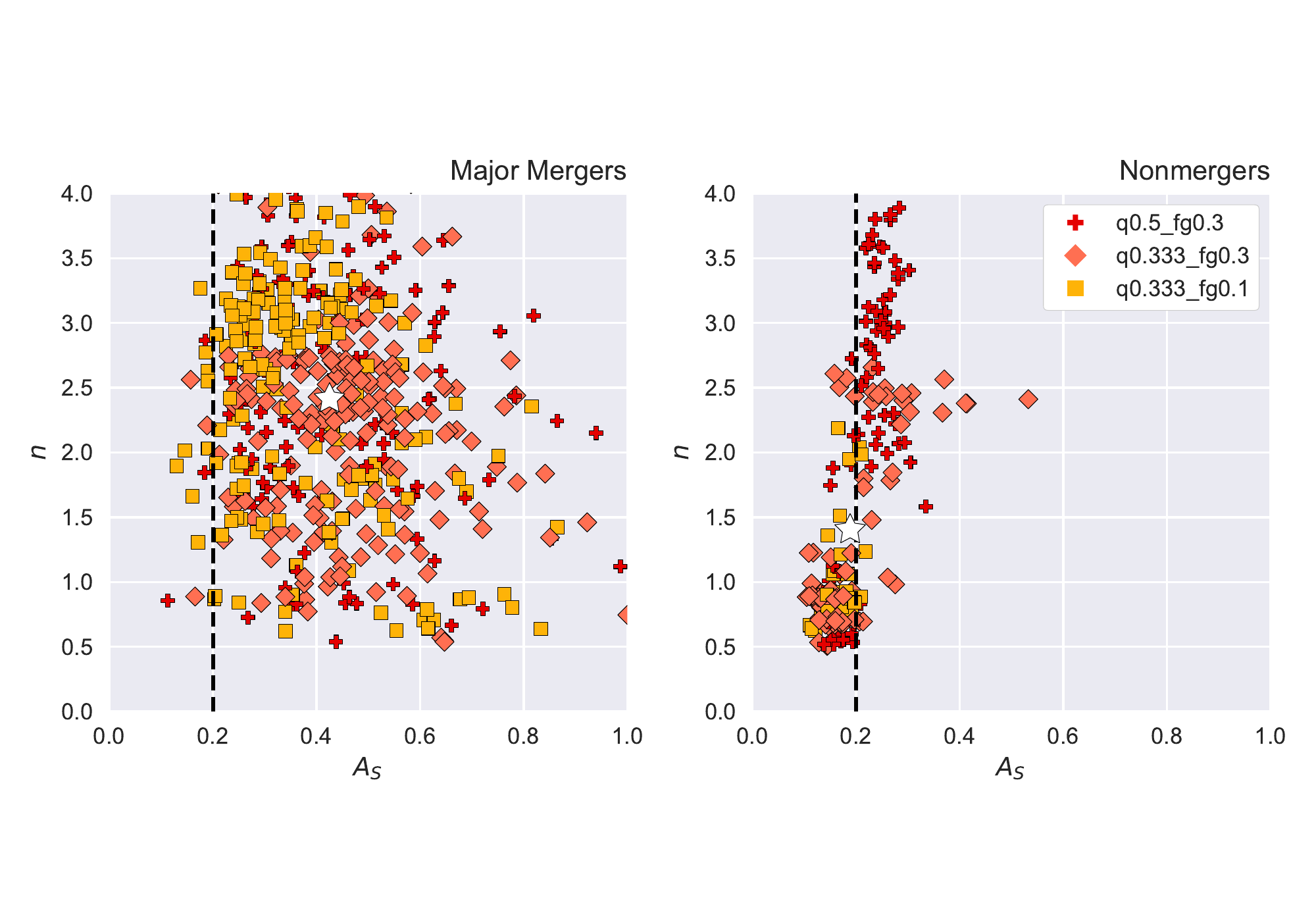}\vspace{-1.5cm}
\vspace{-6cm}\includegraphics[scale=0.8]{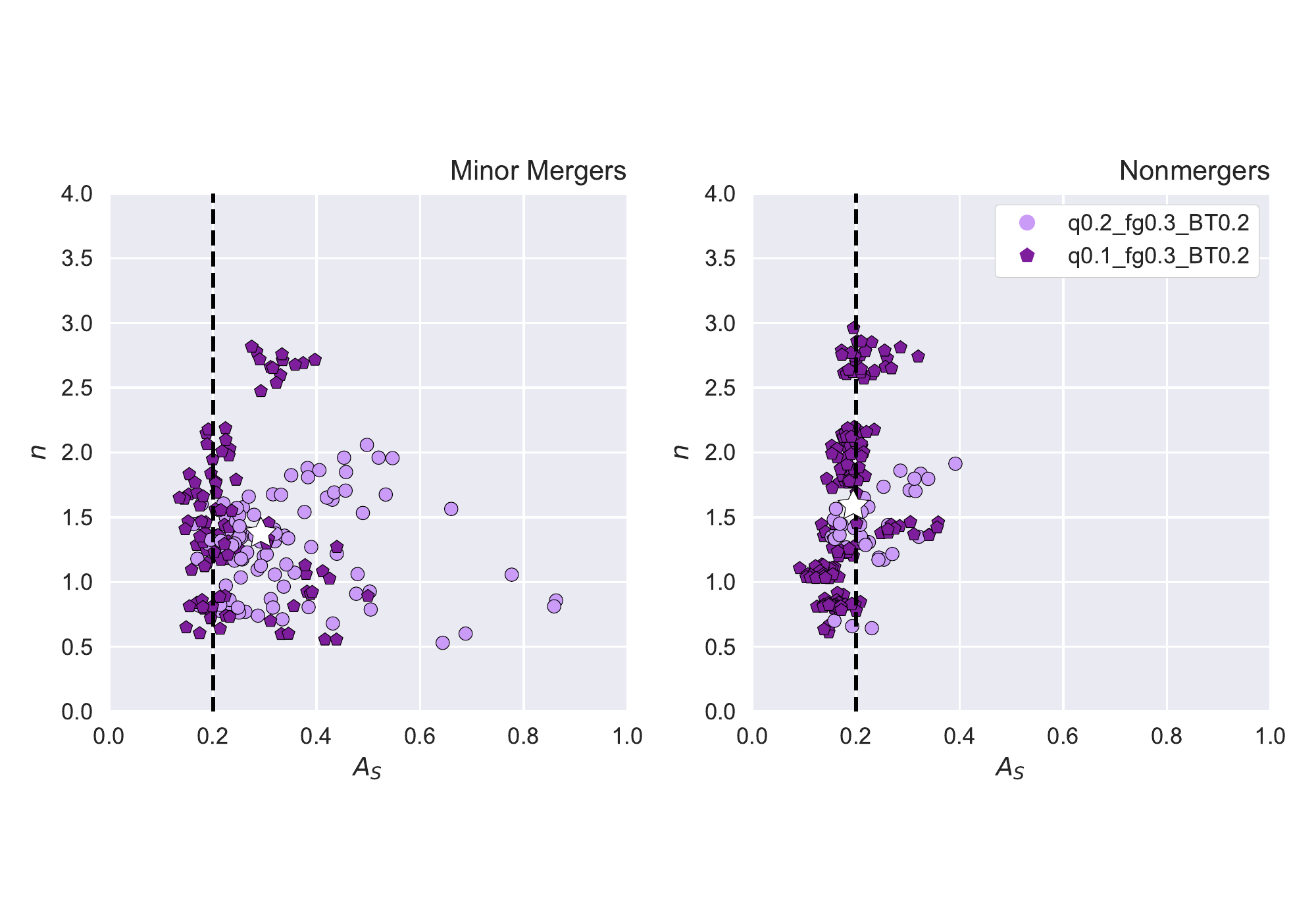}
\vspace{5cm}
\caption{Same as Figure \ref{gini_m20} but for the $n-A_S$ diagrams. The dashed line for the $n-A_S$ diagram is for $A_S > 0.2$; all galaxies to the right of this cut are identified as mergers in \citet{Pawlik2016}.}
\label{n_A_S}
\end{figure*}

\subsection{Observability Timescales}
\label{timescale}
To compare our new LDA technique to previous work that identifies merging galaxies, we calculate the observability timescales of $Gini-M_{20}$, $A$, $A_S$, and the LDA technique for the simulated galaxies. We focus on these particular predictors because past work has defined cuts for $Gini-M_{20}$, $A$, and $A_S$, and classified galaxies lying above these thresholds as merging. Likewise, the observability timescale of $Gini-M_{20}$, $A$, and $A_S$ are measured from these cuts in predictor space, where a simulated galaxy is `identifiable' as a merger for the duration of the time it spends above these thresholds. For $Gini-M_{20}$, \citet{Lotz2008} use:

$$Gini > -0.14\ M_{20} + 0.33$$ 
where everything above the line is defined as a merger.
The asymmetry cut is defined by \citet{Conselice2003}:
$$A > 0.35$$
where galaxies with $A$ values above 0.35 are mergers.
The shape asymmetry cut is from \citet{Pawlik2016}:
$$A_S > 0.2$$
where galaxies with $A_S$ values above this cut are mergers.

We show these cuts in predictor space in Figures \ref{gini_m20}, \ref{C_A}, and \ref{n_A_S}, respectively, for the combined major and minor merger simulations. We plot $C$ with $A$ in Figure \ref{C_A} to include the evolution of $C$ although there are is no formal cut in predictor space for this predictor. For the same reason, we plot $n$ against $A_S$ in Figure \ref{n_A_S}. In these three predictor space plots, we are able to show all of the predictors (we only exclude $S$ because it is unimportant to the analysis).

For each snapshot in each merger simulation, we determine the viewpoint-averaged value for $Gini-M_{20}$, $A$, and $A_S$. If a given snapshot exceeds the cut threshold for a merging galaxy, we designate that snapshot as `identifiable'. By combining all identified snapshots, we determine the observability timescale, which we list in Table \ref{tab:timescale}. If zero snapshots were successfully identified, the observability timescale is less than the time resolution (i.e., < 0.1 Gyr). The timescale of observability from the LDA technique is shown in Figure \ref{mountain}; we label a snapshot of a merger as identifiable if the viewpoint-averaged mean of LD1 is above the decision boundary (shown with a horizontal black line).

\begin{table*}
  \begin{center}
    \caption{Observability timescales in Gyr for the four different merger identification techniques compared in Section \ref{timescale} as well as the total time of the merger.}
    \label{tab:timescale}
    \begin{tabular}{c|c|c|c|c|c}

      Simulation & Total Merger Time &LDA & $Gini-M_{20}$ & $A$ & $A_S$  \\
      \hline
     
      q0.5\_fg0.3 & 2.20 &1.96 & 0.59  & < 0.1 & 2.20 \\
      q0.333\_fg0.3 & 2.64 &2.45 & 0.34 & < 0.1 & 2.64\\
      q0.333\_fg0.1 & 2.83 &2.05& 0.78  & < 0.1 &2.34 \\
      
      q0.2\_fg0.3\_BT0.2 &3.52 & 3.52 & 0.19 & < 0.1 & 3.52 \\
 	q0.1\_fg0.3\_BT0.2 & 9.17 & 8.78 & 0.73 & < 0.1 & 7.79 \\

    \end{tabular}
  \end{center}

\end{table*}

For all simulations, we find that the timescale of observability for the LDA technique is longer than the individual $Gini-M_{20}$, $A$, and $A_S$ timescales of observability. The overall trend is $\sim0.2-0.8$ Gyr observability timescales in $Gini-M_{20}$, very short timescales of observability for $A$ (< 0.1 Gyr), and longer observability timescales in $A_S$ that are > 1 Gyr. The observability window for LDA comprises $\sim80-90$\% of the total length of the merger event, which translates to $2.0-2.5$ Gyr timescales for the major mergers and $3.5-8.8$ Gyr timescales for the minor mergers. 

Overall, the LDA observability timescale dominates because it relies upon multiple different imaging predictors that are sensitive to the merging galaxies at different stages of the merger. However, for the major mergers, the $A_S$ timescale is comparable to the LDA observability timescale. We discuss these trends, how observability timescales scale with the merger initial conditions, and how these timescales compare to previous work in Section \ref{discuss_time}.

\section{Discussion}
\label{discuss}

We explore the behavior of the individual predictors in the LDA technique. Since we remove correlations between predictors with the interaction terms, we are able to discuss the positive or negative signs of the primary predictors (we refer to $Gini$, $M_{20}$, $C$, $A$, $S$, $n$, and $A_S$ as the `primary predictors') in Section \ref{discuss_sign}. We also compare these results to past work with these imaging predictors and discuss how their values change for merging and nonmerging galaxies. Then, we discuss the strengths of the LDA technique. First, we focus on the increased observability timescale of the LDA technique in Section \ref{discuss_time} and how it is sensitive to different stages of the merger. We also discuss how different imaging predictors change in sensitivity throughout the timeline of a merger. Second, we focus on how the classification changes for different mass ratios and gas fractions in Section \ref{discuss_parameters} and Section \ref{gas}, respectively. Finally, we assess the overall accuracy and precision of the LDA technique in Section \ref{discuss_accuracy} and test it on a subsample of SDSS galaxies in Section \ref{SDSS}.  

\subsection{The signs of the LD1 coefficients are consistent with previous work}
\label{discuss_sign}

\begin{figure*}
\centering
\includegraphics[scale=0.3]{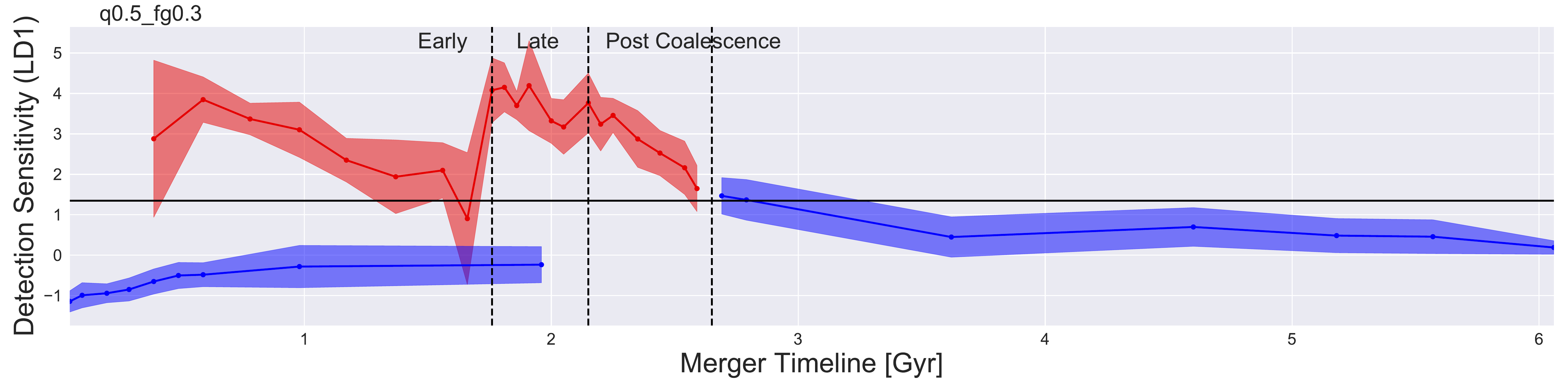}
\includegraphics[scale=0.3]{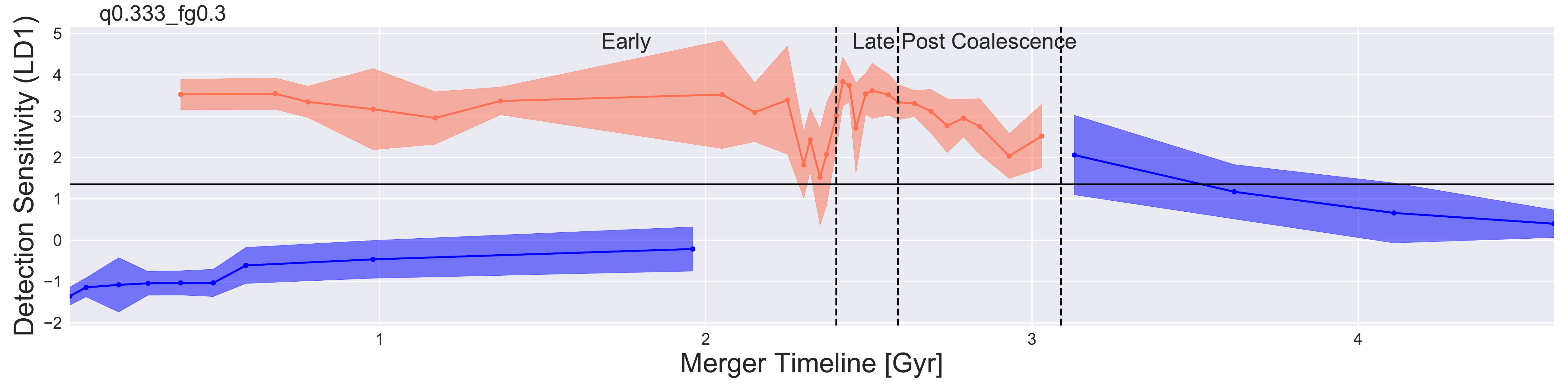}
\includegraphics[scale=0.3]{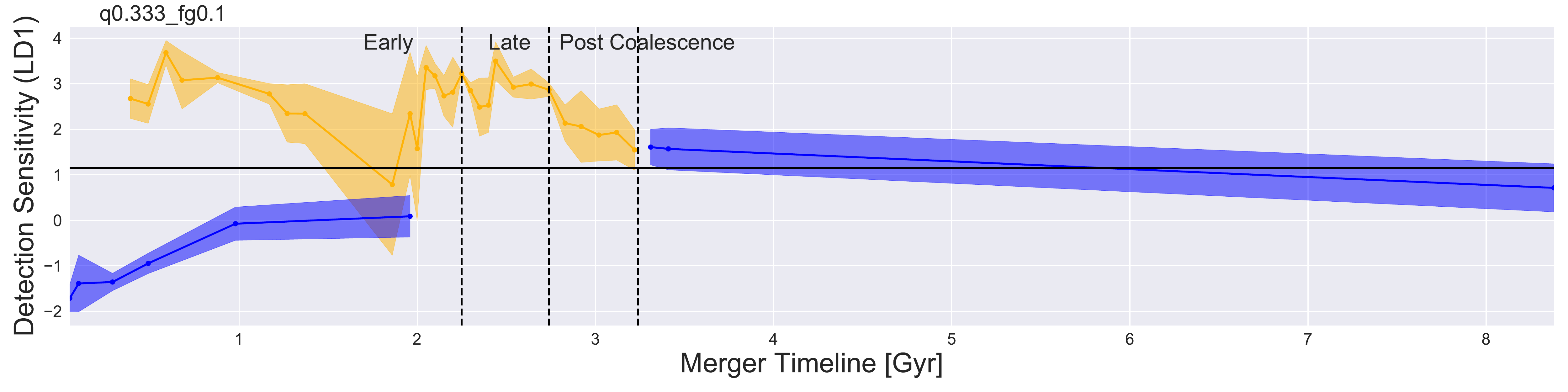}

\includegraphics[scale=0.3]
{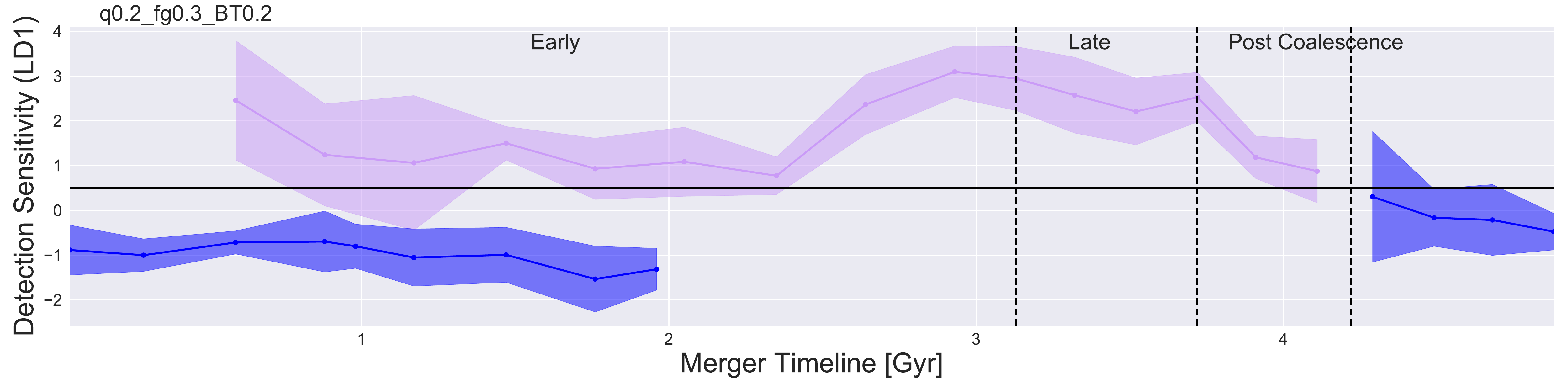}
\includegraphics[scale=0.3]
{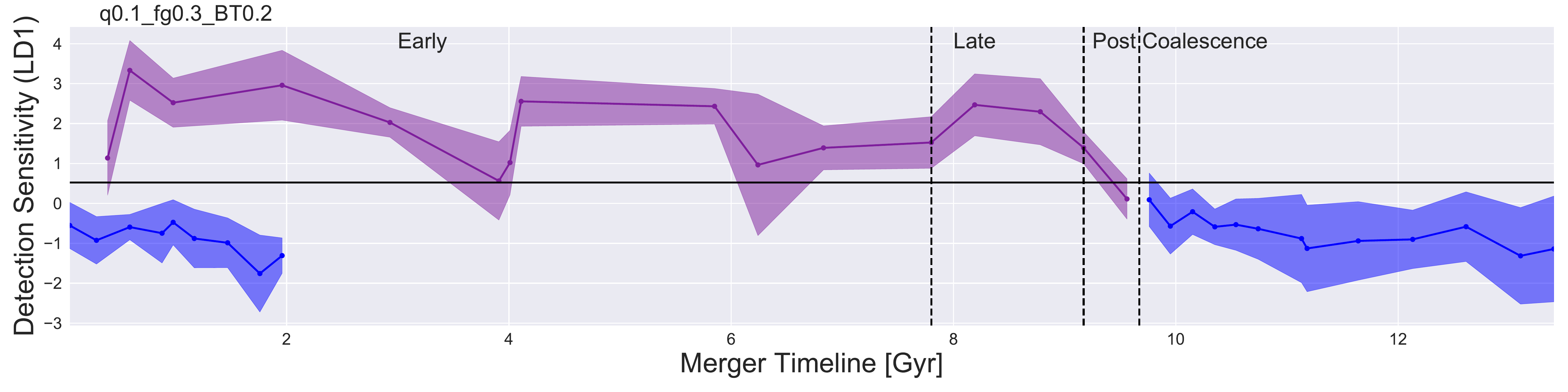}

\caption{Average value of LD1 (mean for all viewpoints) of each snapshot of a merger for all galaxy merger simulations with confidence intervals 1$\sigma$ above and below the mean. The horizontal black line is the decision boundary; galaxies above this line are classified as mergers and galaxies below this line are nonmergers. LD1 and the decision boundary are different for each merger. The vertical dashed lines mark the beginning of the late stage, the beginning of the post coalescence stage, and the end of the merger (0.5 Gyr following final coalescence). The blue line and confidence intervals are for the matched samples of nonmerging galaxies for each simulation. LD1 provides sensitive identification of merger morphology at many stages throughout the merger.}
\label{mountain}
\end{figure*}

One of the strengths of LDA is that we can independently interpret the behavior of each predictor. We compare the primary coefficients of LD1 to previous work by \citet{Conselice2003},  \citet{Lotz2008}, and \citet{Lotz2010b,Lotz2010a} in terms of the signs (positive or negative) of the predictor coefficients.

In Figures \ref{after_histograms} and \ref{after_histograms_2}, a higher value of LD1 indicates that a galaxy is more likely to be identified as a merger. Since LD1 is linear, we can interpret the individual signs of the coefficients in a similar way. If a coefficient is positive, this indicates that a higher value of the coefficient will increase the probability that a galaxy is classified as a merger and vice versa. We provide Figures \ref{gini_m20}, \ref{C_A}, and \ref{n_A_S} to visually compare the location in predictor space of the population of merging galaxies relative to the population of nonmerging galaxies. Figures \ref{ani_fg3_m12} and \ref{ani_fg3_m15} examine the time evolution of the values of individual predictors for the q0.5\_fg0.3 and q0.2\_fg0.3\_BT0.2 runs, respectively. We select these two runs since they are representative of the predictor evolution for a typical major and minor merger simulation.

Since this discussion relies upon the time evolution of predictors, we quickly recap the definitions of merger stage. A merger begins at first pericentric passage and ends 0.5 Gyr following the final coalescence of the nuclei. An early-stage merger is one where the separation of the stellar bulges is $\Delta x \geq$ 10 kpc, a late-stage merger is $1\ \mathrm{kpc}< \Delta x < 10\ \mathrm{kpc}$, and a post-coalescence merger is $\Delta x \leq 1$ kpc.

Overall, we conclude that the positive/negative signs of the individual predictor coefficients are as expected from past studies of merger identification. We discuss the predictor coefficients in more detail and how they change for different mass ratios and gas fractions in Sections \ref{discuss_parameters} and \ref{gas}.

\subsubsection{$Gini$}
\begin{sloppypar}
The $Gini$ coefficients are significant and positive for the combined major and minor merger simulations, as well as q0.333\_fg0.3, q0.2\_fg0.3\_BT0.2, and q0.1\_fg0.3\_BT0.2, which is unsurprising because a higher $Gini$ index has been shown to identify merging galaxies with one or more bright nuclei (e.g., \citealt{Conselice2014} and references therein).

\end{sloppypar}

\subsubsection{$M_{20}$}
The $M_{20}$ coefficient is insignificant for all runs. Interestingly, the value of $M_{20}$ for the mergers evolves with time; this behavior can be examined in Figure \ref{ani_fg3_m12}, which shows the evolution of all the imaging predictors with time for the q0.5\_fg0.3 simulation. This time evolution is especially apparent for the major merger simulations. Early stage mergers evolve to the left towards the merger region of the $Gini-M_{20}$ diagram as their concentration decreases early in the merger (recall, $M_{20}$ is similar to $C$ but does not depend on the location of the center of the galaxy). This leftward migration towards the merger domain would correspond to a negative value for the $M_{20}$ coefficient.

Then, in the post-coalescence stages, the merging galaxies evolve away from the merger region on the $Gini-M_{20}$ diagram, to the right. \citet{Lotz2008} also find this trend in which galaxies evolve away from the merger region of the $Gini-M_{20}$ diagram for the later stages of a merger. This rightward migration makes sense because post-coalescence galaxies begin to lose visually disturbed features such as tidal tails and appear more concentrated in their light distributions. This evolution of $M_{20}$ in both directions for major mergers leads to a washing out of any dominant trend of $M_{20}$ for the major merger simulations.

\subsubsection{Concentration}

The central concentration of light, $C$, is important for all LDA runs except q0.333\_fg0.3, where it is insignificant. The value of the $C$ coefficient is positive for all runs except q0.5\_fg0.3, where it is negative. A positive $C$ coefficient indicates that mergers tend to have a higher value of $C$. We first discuss the overall behavior of $C$ and then focus on the nuances of $C$, such as the decrease of $C$ during the early stages of major mergers. 

Since $C$ is positive for the majority of merger simulations, we can conclude that, in general, merging galaxies have more centrally concentrated light than isolated galaxies.  \citet{Lotz2008} find that concentration is not a strong predictor of a merger but that it is higher for the later stages of a merger. This is expected given that mergers tend to build elliptical galaxies, which has been shown in detail for major mergers (e.g., \citealt{Bendo2000,Bournaud2005}). It has additionally been shown that minor mergers can contribute to stellar bulge growth and drive a less dramatic transformation of galaxy morphology (e.g., \citealt{Walker1996,Cox2008}). We discuss $C$ in more detail for different mass ratios in Section \ref{discuss_parameters}. 

We observe a gradual increase of $C$ with the progression of the merger from the beginning of the early stage to the end of the post-coalescence stage. We examine Figure \ref{ani_fg3_m12} for the time evolution of the $C$ predictor for the q0.5\_fg0.3 run. The value of $C$ for q0.5\_fg0.3  demonstrates an increase with a slight decrease during the early and late stages of the merger. It remains heightened for the nonmerging snapshots following final coalescence. This overall increase is typical behavior for the rest of the merger simulations and happens for the minor merger simulations, without the dip during the end of the early and beginning of the late stages (Figure \ref{ani_fg3_m15}). The increase of $C$ throughout the lifetime of each individual merger simulation leads to positive coefficients of the $C$ predictor in the LDA technique. However, the dip in $C$ values for q0.5\_fg0.3 is pronounced during the early stages and results in a negative coefficient of $C$ in the LDA.

\subsubsection{Asymmetry and shape asymmetry}
The LD1 coefficients for the asymmetry ($A$) and shape asymmetry ($A_S$) predictors both have positive values for all simulations ($A_S$ is insignificant only for q0.1\_fg0.3\_BT0.2). This indicates that the more asymmetric a galaxy, the more likely we are to identify it as a merger. Asymmetry shows this same relationship in \citet{Lotz2008} and \citet{Conselice2003}, where the value increases for mergers.

\subsubsection{Clumpiness}

Clumpiness ($S$) is insignificant for LD1 for all simulations. This result is anticipated given that \citet{Lotz2008} find clumpiness to be a less powerful predictor, but disagrees with \citet{Conselice2003}, who find that clumpiness is higher for merging galaxies. However, the sample of merging galaxies from \citet{Conselice2003} is built from local luminous and ultraluminous infrared galaxies (LIRGs and ULIRGs), both of which are inherently very high in clumpiness. Thus, it is expected that we do not see the same importance of $S$ for the merging galaxies in this work.

\subsubsection{S\'ersic index}
\begin{sloppypar}
The S\'ersic index, $n$, is also unimportant for all simulations. If $n$ is higher for merging galaxies, this indicates that merging galaxies have steeper light profiles. The evolution of $n$ is closely tied to that of $C$, which is unsurprising given that these predictors are correlated (Appendix \ref{mva}). $n$ evolves towards higher values for later stages in the merger, where only a single nucleus is present. The key difference between $C$ and $n$ is that $n$ has a smaller separation in value between merging and nonmerging galaxies for most simulations, so it is an unimportant coefficient for the classification.
\end{sloppypar}

\subsection{LDA lengthens the timescale of observability of merging galaxies}
\label{discuss_time}

The various LDA predictors evolve with time over the course of a galaxy merger. By incorporating seven different imaging predictors, we are able to capture a longer timeline for merging galaxies with the LDA technique than with individual predictors. In this section we discuss the time evolution of the imaging predictors and how this limits their observability timescales. We also compare the estimates of observability time of different imaging predictors to past work.

\begin{figure*}
\centering
\includegraphics[scale=0.19, trim=10cm 0cm 0cm 0cm]{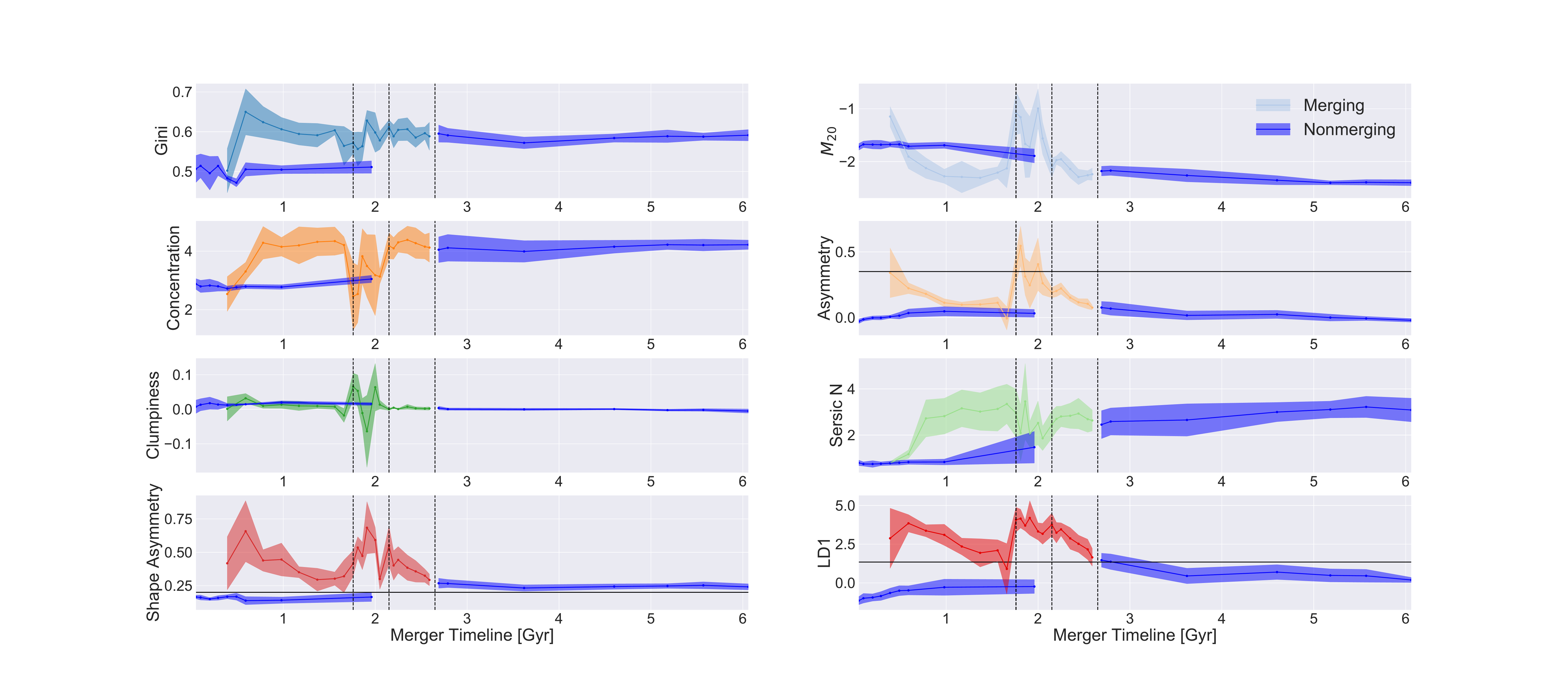}
\caption{Time evolution of the imaging predictors for the q0.5\_fg0.3 simulation. The LD1 sensitivity is shown in the bottom right panel for comparison; the dashed vertical lines mark the beginning of the late stage, beginning of the post-coalescence stage, and end of the merger. We plot the nonmerging galaxies in dark blue for comparison purposes. We also plot horizontal lines for the $A_S$ and $A$ cutoffs in the literature (0.2 and 0.35, respectively) and for the decision boundary for LD1. The dark blue lines are for the matched sample of isolated galaxies. The most powerful predictors for the q0.5\_fg0.3 simulation are $A_S$, $A$, and $C$. }
\label{ani_fg3_m12}
\end{figure*}

\begin{figure*}
\centering
\includegraphics[scale=0.19, trim=10cm 0cm 0cm 0cm]{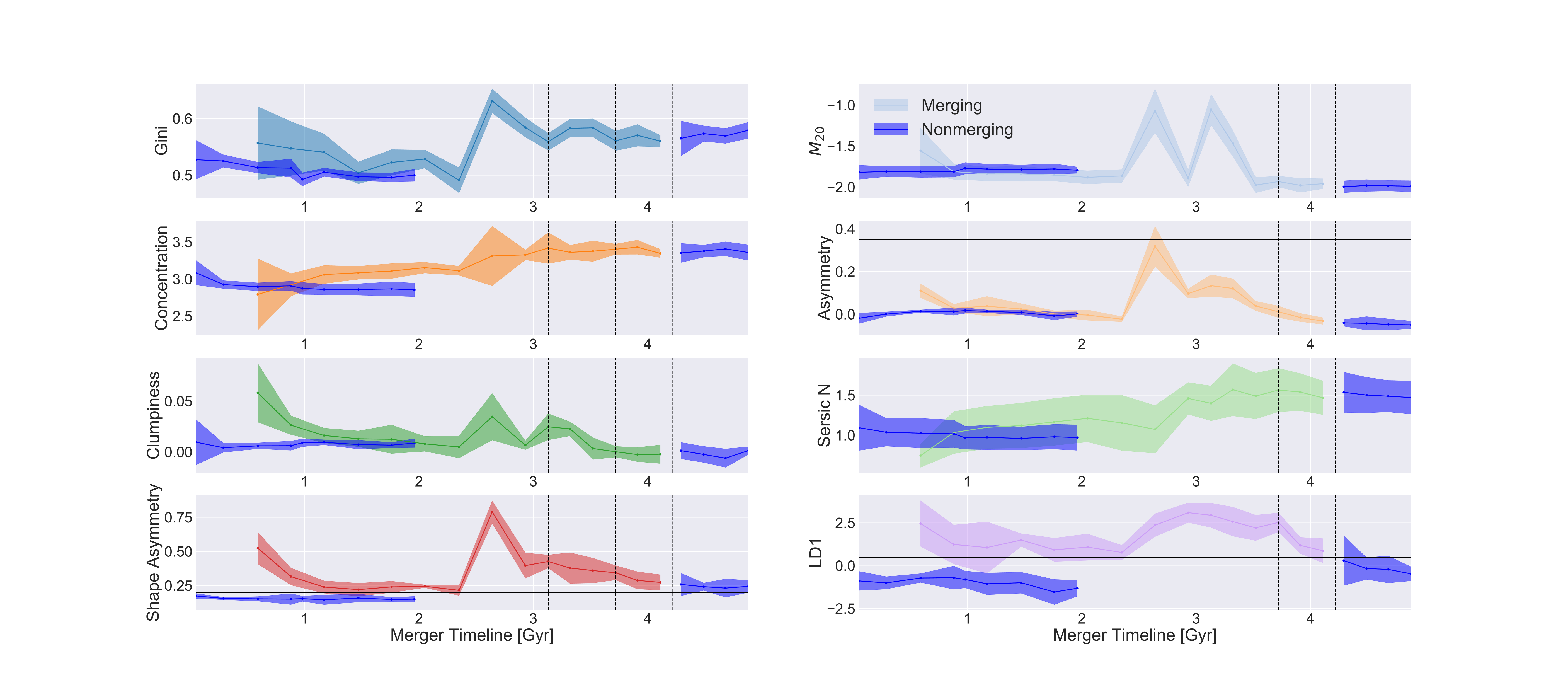}
\caption{Same as Figure \ref{ani_fg3_m12} but for the q0.2\_fg0.3\_BT0.2 simulation. The most powerful predictors for the q0.2\_fg0.3\_BT0.2 simulation are $C$, $Gini$, $A$, and $A_S$.}
\label{ani_fg3_m15}
\end{figure*}

We show the time evolution of the individual predictors (and LD1) in Figures \ref{ani_fg3_m12} and \ref{ani_fg3_m15} for the q0.5\_fg0.3 and q0.2\_fg0.3\_BT0.2 simulations, respectively. We include the cutoff values of $A$ and $A_S$; if a galaxy exceeds these values it is `identifiable' as a merger as in Section \ref{timescale}. We show one major and one minor merger simulation to demonstrate the main differences between the time evolution of the predictors for different mass ratios.

Using the $Gini-M_{20}$ cut in predictor space from Section \ref{timescale}, most of the simulated merging galaxies would be identified as merging by the cut in $Gini-M_{20}$ during the early and late stages of merging, but for a shorter total time than with the LDA technique. The $Gini-M_{20}$ 0.59 Gyr timeframe (indicated by the spike in $M_{20}$ values) for the q0.5\_fg0.3 major merger is shown in Figure \ref{ani_fg3_m12}. The q0.333\_fg0.3 and q0.333\_fg0.1 simulations are also  identified by this cut during the early and late stages of merging for a similar timeframe. However, as the mass ratio begins to increase for minor mergers, the observability timescale of the merger from the $Gini-M_{20}$ technique decreases. For instance, the q0.2\_fg0.3\_BT0.2 merger is identified by this cut during the early and late stages of merging, but only for a 0.19 Gyr timeframe (also indicated by a spike in $M_{20}$ values in Figure \ref{ani_fg3_m15}). These results are consistent with previous work; \citet{Lotz2008} find that $Gini-M_{20}$ is most sensitive to mergers during the first pass (early stage) and the final coalescence of the nuclei (late stage), and \citet{Lotz2010b} show that $Gini-M_{20}$ is sensitive to merger mass ratios less than 1:9. Also, our $0.2-0.8$ Gyr $Gini-M_{20}$ timescale of observability for the simulations with a mass ratio > 1:9 is consistent with the $0.2-0.6$ Gyr timescale of observability from \citet{Lotz2008}.

The $A$ cutoff identifies some of the early and late stages of the major mergers, but has an even shorter timescale of observability than $Gini-M_{20}$. This behavior is apparent in Figure \ref{ani_fg3_m12} when the $A$ value exceeds the 0.35 cutoff value during the beginning of the early stage and in a spike during the late stage of the merger. This is consistent with \citet{Lotz2008}, where the first passage and final coalescence (during the late stage) show the largest asymmetries. While the $A$ value exceeds 0.35 for more snapshots in the major merger simulations, we find a $< 0.1$ Gyr timescale for both major and minor mergers. This $< 0.1$ Gyr timescale for minor mergers can be seen in Figure \ref{ani_fg3_m15}, where the $A$ value only approaches the cutoff value for one snapshot. \citet{Lotz2010b} find an $A$ timescale of $0.2-0.4$ Gyr for major mergers and then less than 0.06 Gyr for minor mergers. While we have a shorter timescale of continuously heightened $A$ values for the major merger simulations, we find that the major mergers result in more snapshots where the value of $A$ exceeds 0.35, which is consistent with the longer observability timescale of $A$ for major mergers from \citet{Lotz2010b}.

$A_S$ has a longer timescale of observability than $A$ and $Gini-M_{20}$ for both major and minor mergers. The merging galaxies evolve to have large values of $A_S$ at various times throughout the early, late, and post-coalescence stages of the merger. $A_S$ identifies the major mergers at nearly all points throughout the simulation, expanding the sensitivity of the LDA technique in time. It only fails to identify the major mergers at some post-coalescence stages. $A_S$ is notably much better at identifying the minor mergers as mergers than both $A$ and $Gini-M_{20}$ and it is most sensitive to the early and late stages of these mergers.   Overall, $A_S$ shows less dependence on time in the merger and is a more consistent identifier of merging galaxies during the early, late, and early post-coalescence stages. This makes sense because $A_S$ is sensitive to faint tidal features; it should therefore be more successful than $A$ at identifying disturbed structures at all times. 

Finally, we focus on the time evolution of $C$, which is not assigned a cutoff value in the literature but which has significant importance within the LDA technique. We find that all mergers show elevated values of $C$, especially for the post-coalescence stages, meaning that $C$ is critical within the LDA technique for capturing the post-coalescence snapshots in time. $Gini$ exhibits a similar behavior to $C$ for the minor mergers, becoming most enhanced during the late and post-coalescence stages. 

\citet{Snyder2018} apply a random forest classifier to the Illustris galaxies and find that features that rely on concentration are more important for selecting recent mergers while features that rely on asymmetries are more important for selecting galaxies that are about to merge. While the Illustris simulation is a cosmological merger tree simulation, it is informative that the results are consistent with the time sensitivities of various imaging predictors in this work.

Unlike the individual imaging predictors, we find that the sensitivity of the LDA depends only minimally on merger stage. It is slightly less sensitive for the very early stages and very late post coalescence stages of the merger; this is expected since the galaxies often appear visually to be isolated galaxies prior to first pericentric passage and after coalescence. As discussed in Section \ref{sims}, we use the very early and very late stages of the merger (prior to first passage and > 0.5 Gyr following final coalescence) as isolated galaxies in this analysis, so these galaxies are very similar in imaging to galaxies at an adjacent point in time. This explains why the 1$\sigma$ confidence intervals overlap with the decision boundary for many of these very early-stage and very late-stage snapshots in Figure \ref{mountain}.

The individual imaging classification techniques are sensitive to different stages of a merger. For instance, $A$ and $Gini-M_{20}$ identify early and late-stage mergers, $A_S$ identifies early-stage, late-stage, and some post-coalescence mergers, and $C$ is most sensitive to post-coalescence mergers. LDA is able to combine these imaging techniques into one more complete classifier that maintains sensitivity throughout the lifetime of a merger.

\subsection{The coefficients of LD1 change with mass ratio}
\label{discuss_parameters}

When we examine the relative importance of various predictors for merger simulations with varying mass ratios, we determine that $A_S$ and $A$ are relatively more important for the major mergers and that $C$ and $Gini$ are relatively more important for the minor mergers. $A$ is important for all merger simulations.

First, we address the major mergers, where $A_S$ and $A$ are both important coefficients and indicators of disturbed visual morphology. The $A$ coefficient has a normalized value of $0.28-1.0$ for the three major mergers and the combined major merger runs, indicating that it is one of the most important primary predictors. It is less important for the minor mergers and the combined minor merger simulation, but its relative importance is still high $\sim0.24-0.89$ (Figure \ref{color_predictors}). This result agrees with \citet{Lotz2010b}, who finds that $A$ is a good probe of major mergers with mass ratios between 1:1 and 1:4. This is because the major mergers have more disturbed morphologies, especially during the early stages of the merger. However, the $A$ predictor remains important for the minor mergers, where the visual morphology is less disturbed.

$A_S$ is more sensitive (than $A$) to faint tidal tails in galaxies. The $A_S$ coefficient ranges in normalized values from $0.25-1.0$ for the major mergers and $0.16-0.18$ for the minor mergers. Since both $A$ and $A_S$ track visual morphology, it is significant that while $A$ is important for all runs, $A_S$ is less important for the minor mergers. This suggests that the more disturbed visual morphology of major mergers is best identified with both measures of asymmetry. On the other hand, minor mergers rely more on measures of concentration like $C$ and $Gini$, so while $A$ is still an important predictor for them, it is less dominant.

Next we address $C$ and $Gini$, where the importance to the minor mergers can be attributed to two main factors: scatter and necessity. The major mergers show more scatter in $C$ values while the minor mergers show a general trend of $C$ enhancement as the mergers progress. In Figure \ref{C_A}, the major mergers range from values of $1-5.5$ while the minor mergers only span $2-4$ in $C$. Upon examination of the predictor values with time (Figures \ref{ani_fg3_m12} and \ref{ani_fg3_m15}), we verify that $C$ increases steadily with time for the minor mergers and reaches a peak value of $\sim3-3.5$. For major mergers, $C$ shows a general increase with time but also decreases during the most visually disturbed epochs of the merger (early and late stage). However, the major mergers do ultimately build more concentrated remnants than the minor mergers, with $C$ values peaking at $4-4.5$.

$Gini$ is heightened for both major and minor mergers in Figures \ref{ani_fg3_m12} and \ref{ani_fg3_m15}. The LDA for the major mergers was able to rely upon stronger predictors such as $A$ and $A_S$ to fully separate the populations of merging and nonmerging galaxies. However, the minor mergers are less distinguishable from nonmerging galaxies using these predictors. Therefore, $Gini$ becomes more important for the minor mergers. This result is consistent with \citet{Lotz2010b}, who find that $Gini-M_{20}$ remains effective for identifying minor mergers down to mass ratios of 1:9. 

Figure \ref{color_predictors} also shows an important difference between the 1:2 major merger and the 1:3 major mergers. Between the two mass ratios, measurements of concentration ($C$ and $Gini$) become slightly more important in the 1:3 major mergers, initiating the trend towards minor mergers.

Our findings regarding $C$ make sense given the current understanding of galaxy morphological evolution. To first order, equal mass ratio major mergers build large elliptical galaxies (e.g., \citealt{Bendo2000,Bournaud2005}) while intermediate mass ratio mergers (down to 1:10) are predicted to build galaxies with spiral-like morphologies and elliptical-like kinematics (e.g., \citealt{Jog2002,Bournaud2004}). Very high mass ratio mergers (minor mergers, $\geq$ 1:10) build disturbed spiral-like galaxies (e.g., \citealt{Naab2014} and references therein). However, some work has found that multiple minor mergers can still build elliptical surface brightness profiles in the remnant galaxy (\citealt{Bournaud2007,Jesseit2009,Bois2011}) and that even one merger of a very minor mass ratio can build stellar bulges in the remnant (\citealt{Cox2008}). 

Statistically, minor mergers are much more common than major mergers, accounting for three times as many mergers (e.g., \citealt{Bertone2009,Lotz2011}). The ubiquitous nature of minor mergers increases their relative importance for galaxy evolution. For example, \citet{Ownsworth2014} find that the majority of stellar mass is added to galaxies by star formation (24\% of stellar mass) and minor mergers (34\% of stellar mass), whereas major mergers only account for 17\% of the total galaxy stellar mass at z = 0.3. Other observations of local elliptical galaxies and the dearth of major mergers indicates that minor mergers may be more important than previously thought for building local elliptical galaxies with more concentrated light profiles (\citealt{Trujillo2009,Taylor2010}).

The above picture of galaxy morphological evolution is consistent with the differences between $C$ for the major and minor merger simulations. $C$ is more important for the minor mergers because it exhibits less scatter than for the major mergers, where the dip in $C$ weakens the overall strength of the coefficient in the LDA technique. The major merger remnants show a greater overall enhancement of $C$ by the end of the merger, building galaxies with $C$ $\sim4-5$ that are more consistent with the classical picture of large elliptical galaxies. Visually, the galaxy remnants for major and minor mergers still have disk profiles with a more concentrated center, indicative of an enhancement of the stellar bulge. This means that by the end of the merger, minor and major mergers have both enhanced the concentration of the light profile of the galaxy and therefore contributed to the morphological evolution of galaxies.

\subsection{The coefficients of LD1 do not change (significantly) with gas fraction}
\label{gas}

Next, we examine differences between the gas rich and gas poor simulations. We specifically compare q0.333\_fg0.3 to q0.333\_fg0.1, which are both 1:3 mass ratio major mergers matched for all properties except gas fraction. The important predictors for the gas poor merger are 
$A_S$, $A$, and $C$, while $A_S$, $A$, and $Gini$ are important for the gas rich merger. Overall, these coefficients are very similar between the gas rich and gas poor simulations. Both rely upon measurements of asymmetry ($A$ and $A_S$). Both simulations also rely upon measurements of concentration ($Gini$ and $C$). While $C$ is more important for the gas poor merger, $Gini$ is more important for the gas rich merger.

\citet{Lotz2010a} establish that $Gini-M_{20}$ is weakly dependent on gas fraction whereas $A$ is relatively enhanced for gas rich mergers. Since $A$ and $A_S$ are relatively important for both gas rich and gas poor simulations, we conclude that the dominant initial condition must be mass ratio for $A$ to remain important for the gas poor major merger. 

While there is little difference in $A_S$ in terms of the coefficients for the gas rich and gas poor simulations, there is a small difference in the observability timescales for $A_S$ and $Gini-M_{20}$. The observability times for $A_S$ is shorter for the gas poor major merger in Table \ref{tab:timescale}. The timescale of observability for $A_S$ for q0.333\_fg0.1 is 2.34 Gyr while it is 2.64 Gyr for q0.333\_fg0.3. Since the overall merger timescales are so similar for these two simulations, the $A_S$ observability timescale is significantly longer for the gas rich major merger. The $Gini-M_{20}$ timescale is also significantly different for the gas rich and gas poor simulations; it is longer for the gas poor major merger (0.78 Gyr) and shorter for the gas rich major merger (0.34 Gyr). This result is consistent with \citet{Lotz2010a}, where the result is that the $Gini-M_{20}$ observability timescale decreases slightly with increasing gas fraction while the $A$ observability timescale increases with gas fraction. The $Gini-M_{20}$ timescale may be decreasing slightly with gas fraction here for similar reasons as stated in \citet{Lotz2010a}, that the increased dust obscuration at the central nuclei lowers the $Gini$ values for gas rich simulations. The gas poor simulation has the longest $Gini-M_{20}$ observability timescale of all the simulations.

\citet{Kormendy2009} suggest that both dry and wet mergers (gas poor and gas rich, respectively) build up the bulge mass of a galaxy, contributing to its elliptical morphology (or an increased value of $C$). Overall, dry mergers are more important for building elliptical galaxies but wet mergers can also increase the central concentration of a galaxy since they drive starbursts that contribute to bulge growth. The $C$ value for the gas rich major merger increases with time to a peak value of $\sim4.3$ while the gas poor major merger increase to a peak value of $\sim4.7$. The gas poor major merger seems to have a slightly higher concentration, possibly reflecting the tendency for dry mergers to build galaxies with elliptical morphologies. However, these values are not significantly different when we take into consideration the viewpoint-averaged standard deviation for these snapshots. Therefore, the difference in gas fraction is not producing a significant difference in the concentration of the remnant. 

Overall, while the timescale of observability for $A_S$ is longer for the gas rich major merger and the timescale of observability of $Gini-M_{20}$ is longer for the gas poor major merger (which is consistent with \citet{Lotz2010a}), the differences in LDA coefficients are most pronounced for mergers of different mass ratios (Section \ref{discuss_parameters}). This is why we choose to separate the combined runs by mass ratio as opposed to gas fraction.

\subsection{The LDA technique is accurate and precise at identifying merging galaxies}
\label{discuss_accuracy}
The accuracy and precision of the LDA technique are very high (the accuracy is 85\% and 81\% and precision is 97\% and 94\% for the major merger and minor merger combined simulations, respectively; Table \ref{tab:pa} in Appendix \ref{accuracy}). We use accuracy to determine the relative number of true detections (TP, true positives, and TN, true negatives) to all detections (includes FP, false positives, and FN, false negatives); the accuracy is defined as (TP + TN)/(TP + TN + FP + FN). If the accuracy is high, this means that our method does a good job of minimizing the number of false positives and false negatives. We use precision to determine the relative number of true positives (TP) to all positive detections, including false positives (FP); the precision is defined as (TP/(TP+FP)). If the precision is high, then there are a low percentage of false positives, which is desirable because false positives are nonmerging galaxies that are incorrectly included as mergers. We want to avoid contamination in the sample of merging galaxies when we apply the technique to real imaging. 

In this section we compare the accuracy and precision of the LDA technique to that of past work, which utilizes pair studies or cuts in $Gini-M_{20}$ space, $A$, or $A_S$. We also compare to other work that has utilized machine learning tools to identify merging galaxies.

First we compare the LDA accuracy and precision to that of close pair studies with SDSS. \citet{Darg2010} compare Galaxy Zoo classifications of major mergers to all SDSS galaxy pairs with a projected separation of < 30 kpc and a line of sight velocity offset < 500 km s$^{-1}$. After visually examining all 2308 close pair objects, they find that 28\% of objects are chance superpositions and/or have no signs of interaction. While this is an imperfect comparison to our work since close pair studies only capture a brief snapshot of a merger, the overall result is that 28\% of the close pairs are false positives. \citet{Darg2010} also estimate than only 20\% of advanced mergers identified in Galaxy Zoo are pairs in SDSS, which is a small fraction of true positives. In comparison to the LDA technique, pair studies have low accuracy since many mergers are missed by the technique and low precision since there is also a significant fraction of false positives.

Next, we directly compare the accuracy and precision of the LDA technique to that of the cuts in predictor space introduced in Section \ref{timescale}. We measure the accuracy and precision of the $Gini-M_{20}$, $A$, and $A_S$ cuts for the simulations and find that precision remains high. This means that these methods do not incorrectly identify nonmergers as mergers. In fact, they have the opposite problem and fail to identify merging galaxies as such, leading to low accuracies. We find that $Gini-M_{20}$ has accuracies from 60\% to 70\%, $A$ has accuracies from  40\% to 60\%, and $A_S$ has accuracies from 70\% to 90\%. The low accuracy of the $A$ predictor agrees with \citet{Conselice2003} who find that the fraction of mergers (defined to be a sample of ULIRGs) that are correctly identified by $A$ is $\sim50$\%. 

We also find that there is a difference in accuracy for different mass ratios. For example, the minor mergers fall at the bottom of the accuracy ranges given above. This is worrisome because the cuts in predictor space are preferentially selecting major mergers, which are much less numerous than minor mergers. In contrast, the LDA accuracy changes by less than 10\% between all simulations, ranging from $\sim85 - 90$\% accuracy for all simulations. Using $Gini-M_{20}$, $A$, or $A_S$ in isolation is not sensitive enough to correctly and consistently identify mergers of all mass ratios at all merger stages.

Our LDA technique is more accurate and precise than individual imaging predictor classifiers and is comparable to other techniques that combine many different imaging predictors. For instance, \citet{Snyder2018} and \citet{Goulding2018} use random forest classifiers with a collection of similar parametric and nonparametric imaging predictors to classify merging and nonmerging galaxies and find similar accuracies and precisions.

\citet{Snyder2018} use the Illustris cosmological simulation to produce synthetic deep Hubble Space Telescope images of merging galaxies at 12 timesteps over a range of redshifts ($0.5 < z < 5$). While \citet{Snyder2018} work with dissimilar galaxies to our low redshift SDSS galaxies, we are able to roughly compare the two methods because both rely upon similar imaging predictors. For instance, \citet{Snyder2018} use a binary classification that relies upon $Gini$, $M_{20}$, $A$, and $C$ as inputs (among other imaging predictors). They find a similar accuracy and precision of their classifier when they test it using the simulated Illustris galaxies. The result is a classifier that relies on different imaging predictors for different merger stages, similar to the LDA technique in our work. The random forest has a completeness of $\sim$70\% and a purity of 10\% at $z = 0.5$ and 60\% at $z = 3$. Completeness is defined as TP/(TP+FN), which is defined as recall in our work, and purity is TP/(TP+FP), which is defined as precision here.

Since the isolated galaxy sample in \citet{Snyder2018} is different than the sample of isolated disks used in this work, we are unable to directly compare the false positive and false negative rates. We instead discuss one relative strength of the LDA technique. Some of the false negatives in \citet{Snyder2018} result from the failure of the method to detect some minor mergers and some of the false positives result from a narrow temporal definition of the duration of the merger, which is restricted to 500 Myr. One relative strength of the LDA technique is that it is built from high temporal resolution simulations and is therefore able to use a more complete definition of merging galaxies. It also extends the definition of merging galaxies beyond the 500 Myr timeframe used in \citet{Snyder2018}. This is a general strength of high temporal resolution isolated simulations over large cosmological simulations.

\citet{Goulding2018} use a random forest to create a non-binary classifier that separates their sample of Hyper Suprime-Cam (HSC) galaxies into subsamples of major mergers, minor mergers and irregulars, and non-interacting galaxies. The input imaging predictors include $Gini$, $C$, $A$, $S$, and $n$ for the galaxy images as well as the residual images after subtracting a \texttt{GALFIT} surface brightness model. They visually classify galaxies in the HSC sample to test the performance of the classifier and find that the major mergers suffer from mild contamination ($\sim$10\%) with an overall completeness for the merger samples of 75\%. The LDA technique has a comparable result with 4\% contamination and 79\% completeness for major mergers. The minor mergers are more difficult to distinguish from isolated and major mergers in \citet{Goulding2018} and therefore have increased contamination and decreased completeness. The LDA technique on the other hand only suffers from 6\% contamination and 66\% completeness for minor mergers. It should be noted that \citet{Goulding2018} create and test the random forest method on real galaxy images, so this is an imperfect comparison, simply meant to roughly compare the accuracy and precision of different imaging merger identification methods. Additionally, since the LDA technique is developed from disk-dominated galaxies, its accuracy and precision apply best to galaxy samples that match the properties of the simulated galaxies used to construct the technique.

\subsection{Testing the Technique on SDSS Galaxies}
\label{SDSS}

\begin{figure*}
\centering
    \includegraphics[scale=0.5,trim={3.5cm 0 3.5cm 0},clip]{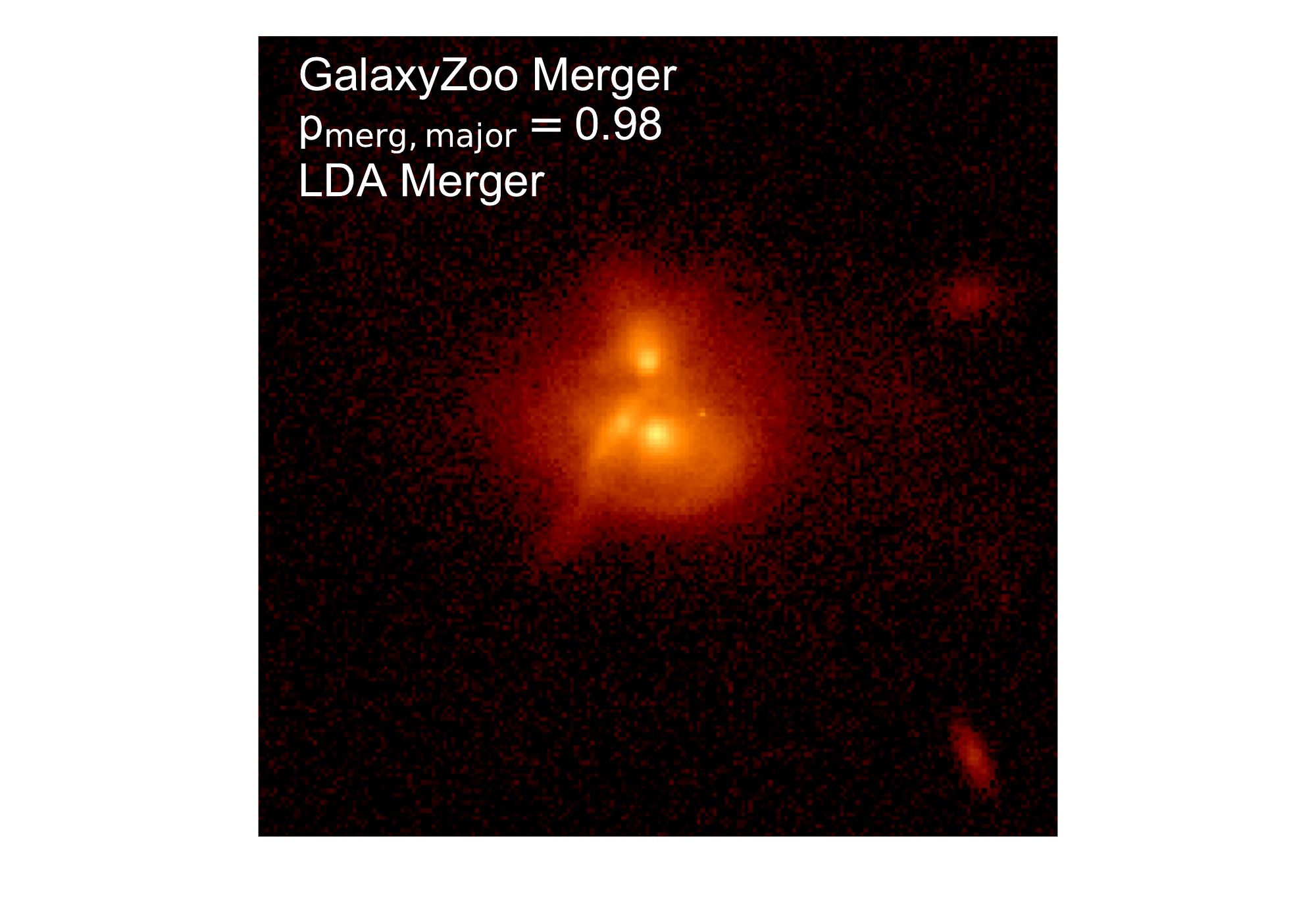}\includegraphics[scale=0.5,trim={3.5cm 0 3.5cm 0},clip]{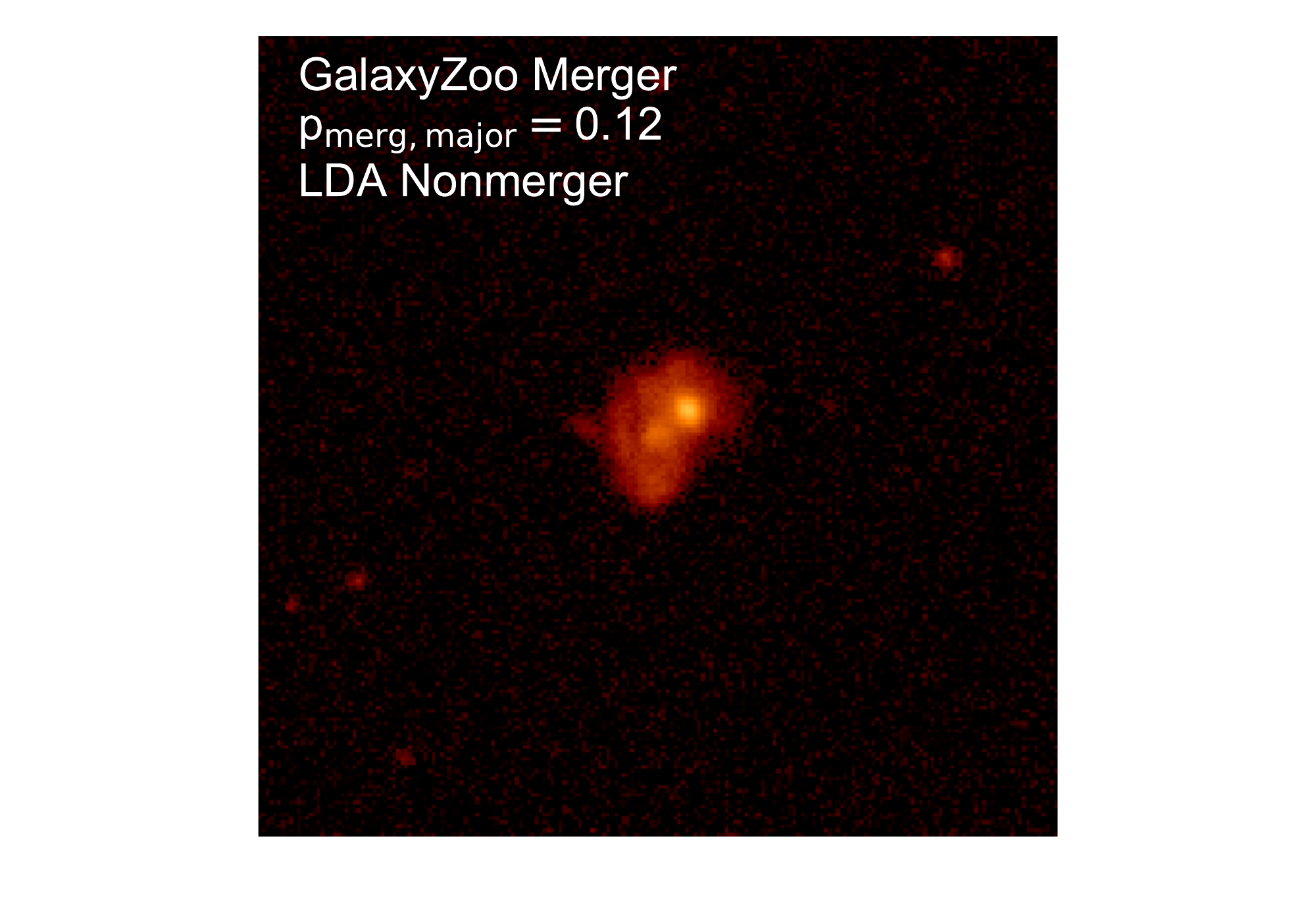}
    
    \includegraphics[scale=0.5,trim={3.5cm 0 3.5cm 0},clip]{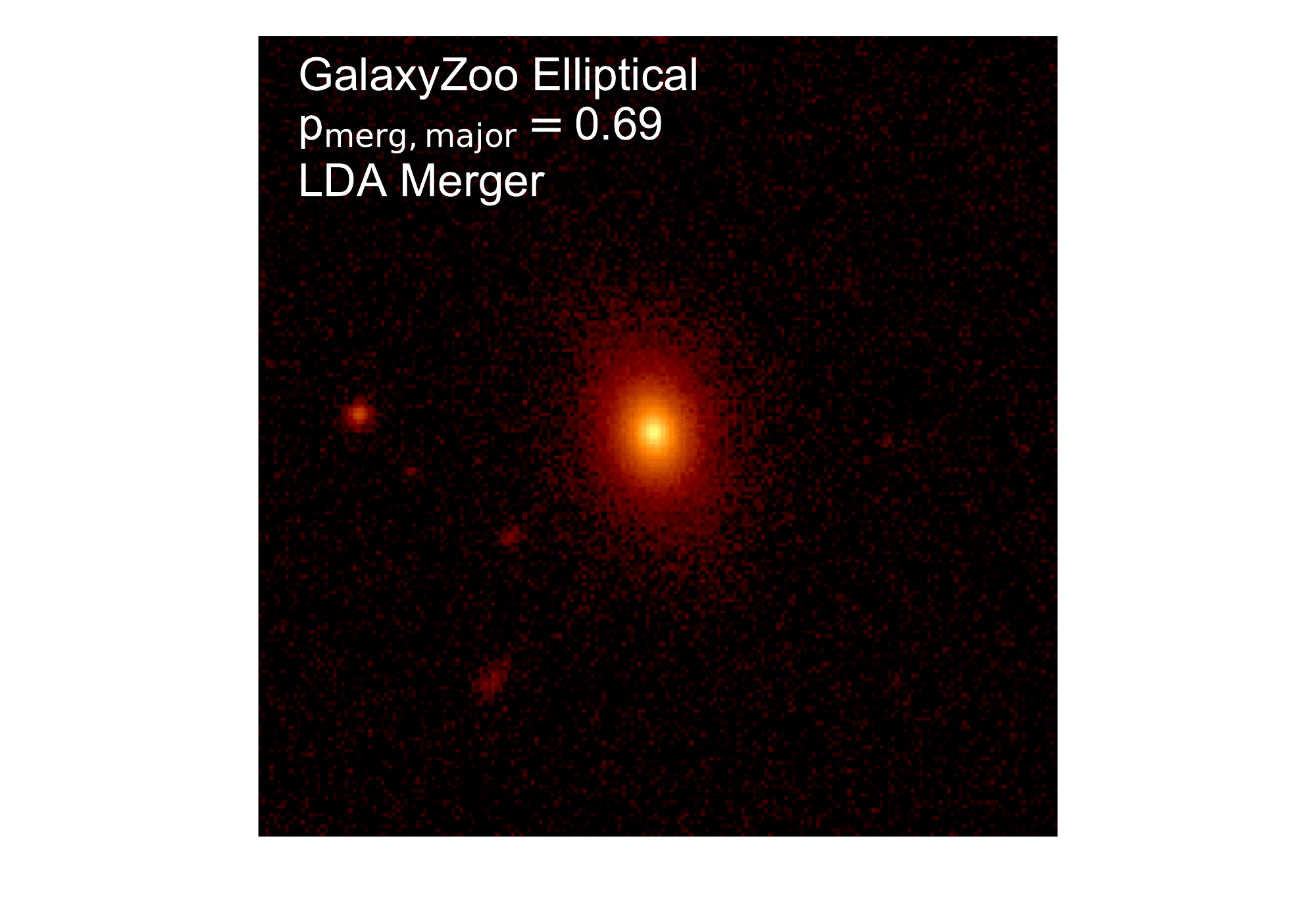}\includegraphics[scale=0.5,trim={3.5cm 0 3.5cm 0},clip]{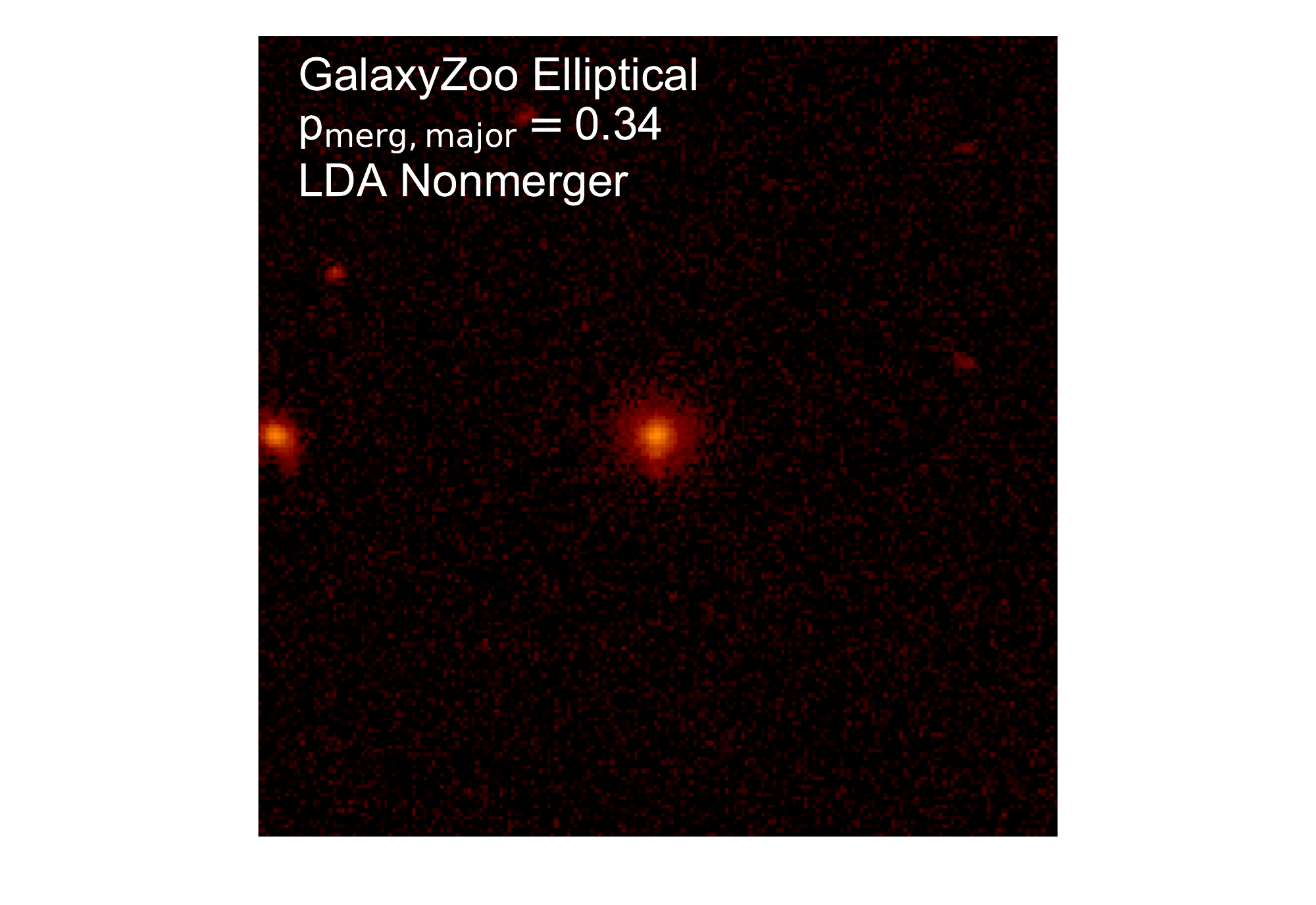}
    
    \includegraphics[scale=0.5,trim={3.5cm 0 3.5cm 0},clip]{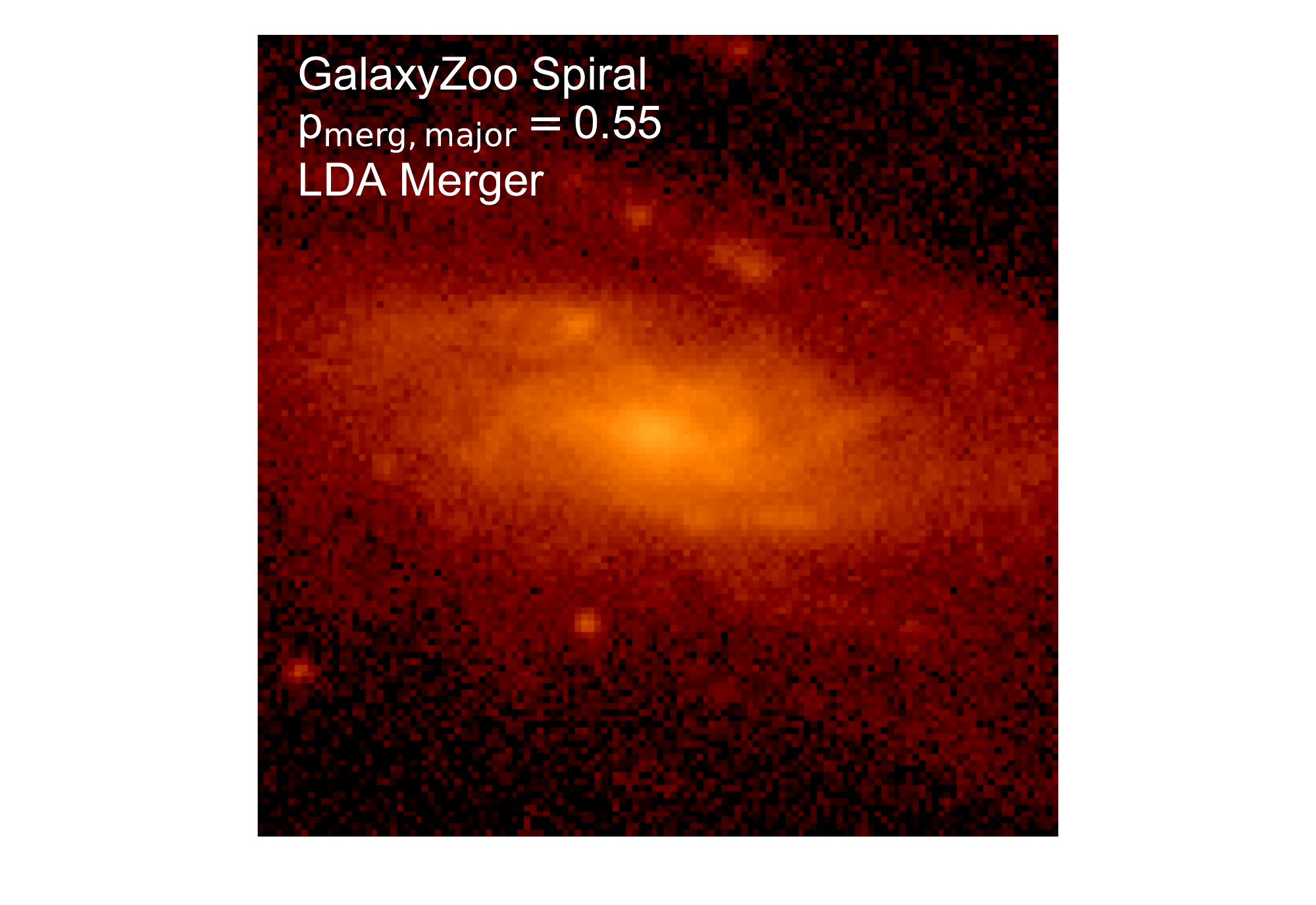}\includegraphics[scale=0.5,trim={3.5cm 0 3.5cm 0},clip]{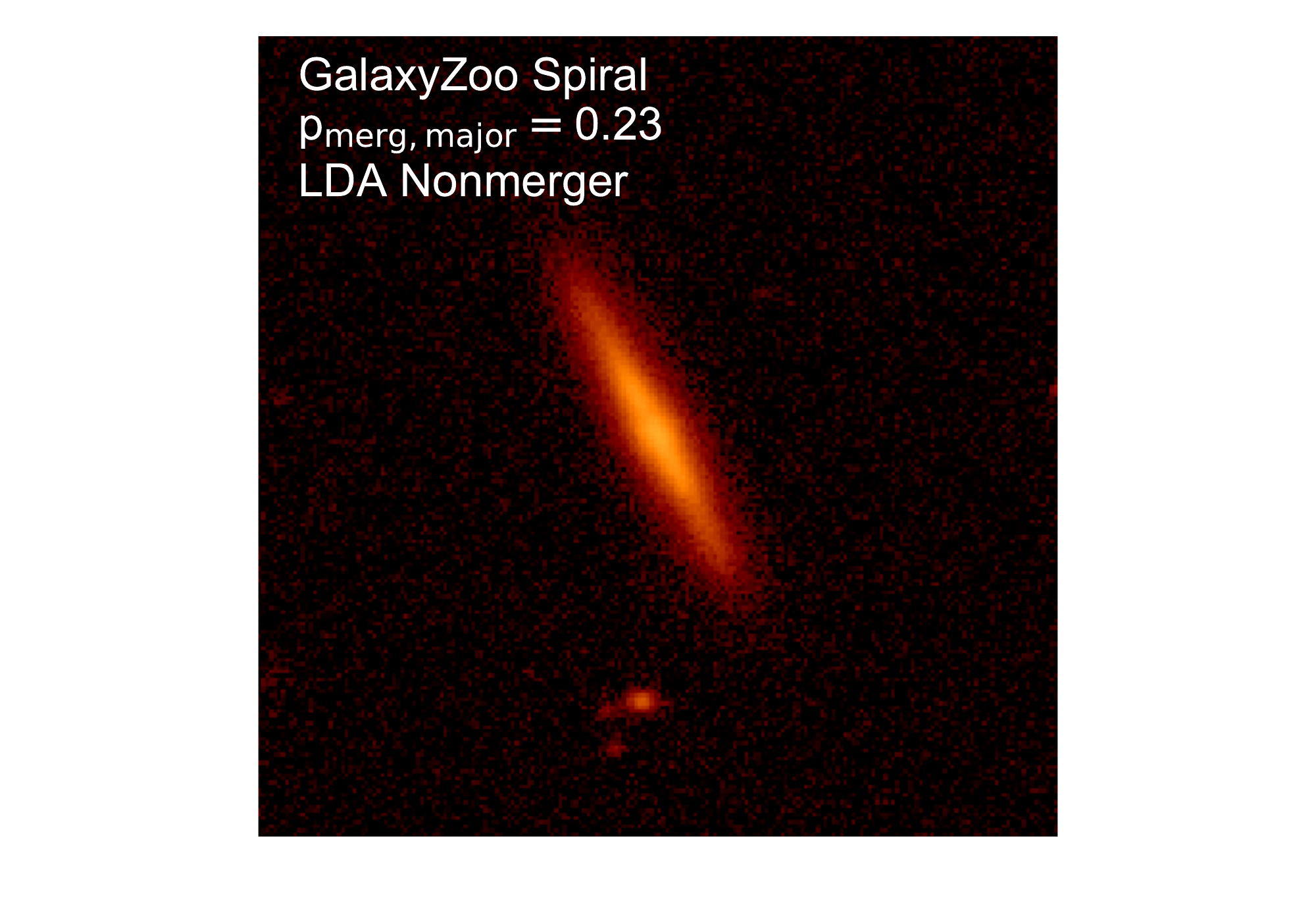}
    \caption{$r-$band images of GalaxyZoo mergers (top), elliptical galaxies (middle), and spiral galaxies (bottom) that are classified as mergers (left column) and nonmergers (right column) by the major merger classification technique. For each case, the probability that the galaxy is a merger as classified by the major merger technique ($p_{\mathrm{merg, major}}$) is given. The major merger technique identifies 86\% of GalaxyZoo mergers as mergers, 73\% of GalaxyZoo ellipticals as mergers, and 14\% of GalaxyZoo spirals as mergers.}
    \label{galzoo}
\end{figure*}

\begin{figure*}
\centering
    \includegraphics[scale=0.5,trim={3.5cm 0 3.5cm 0},clip]{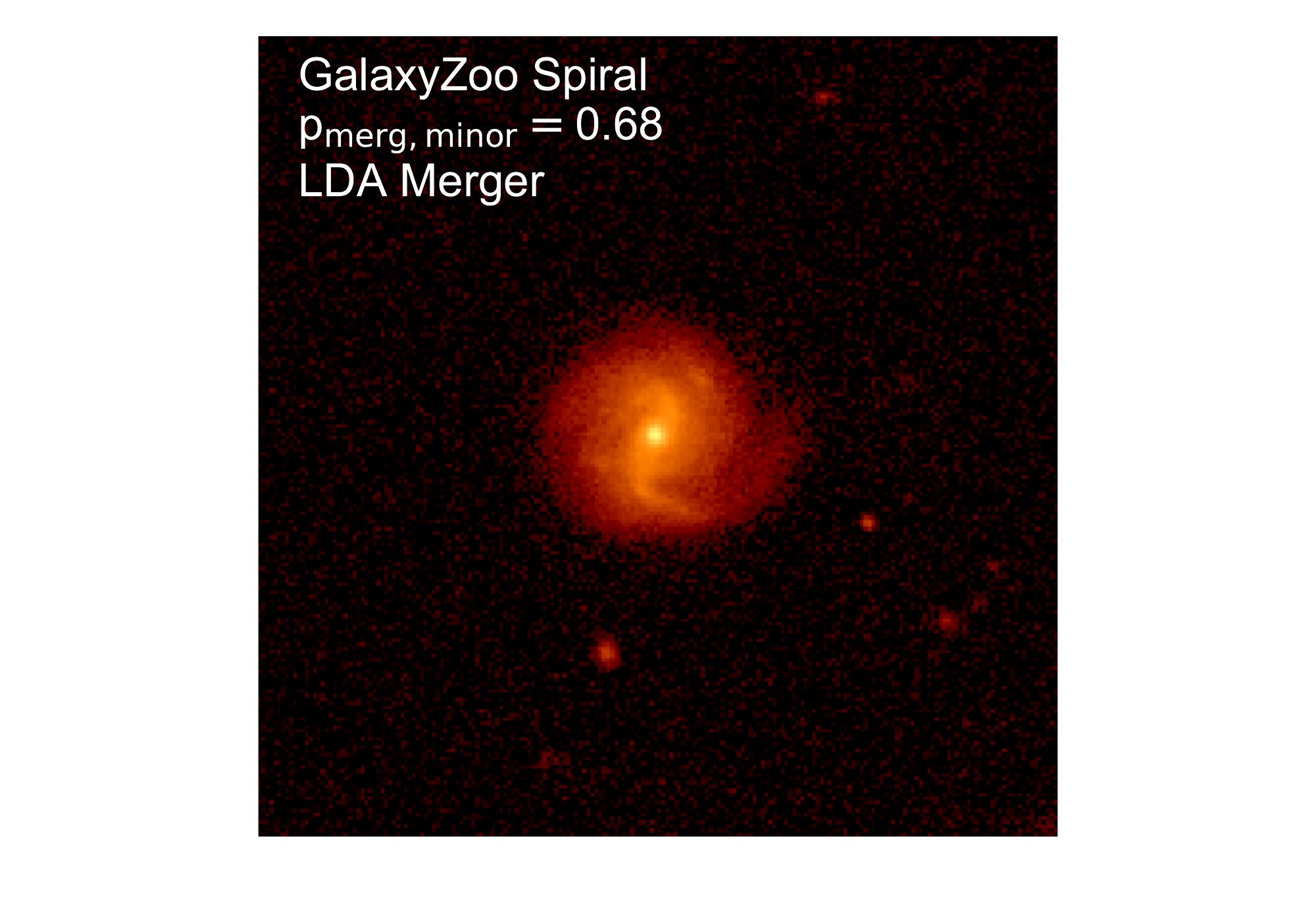}\includegraphics[scale=0.5,trim={3.5cm 0 3.5cm 0},clip]{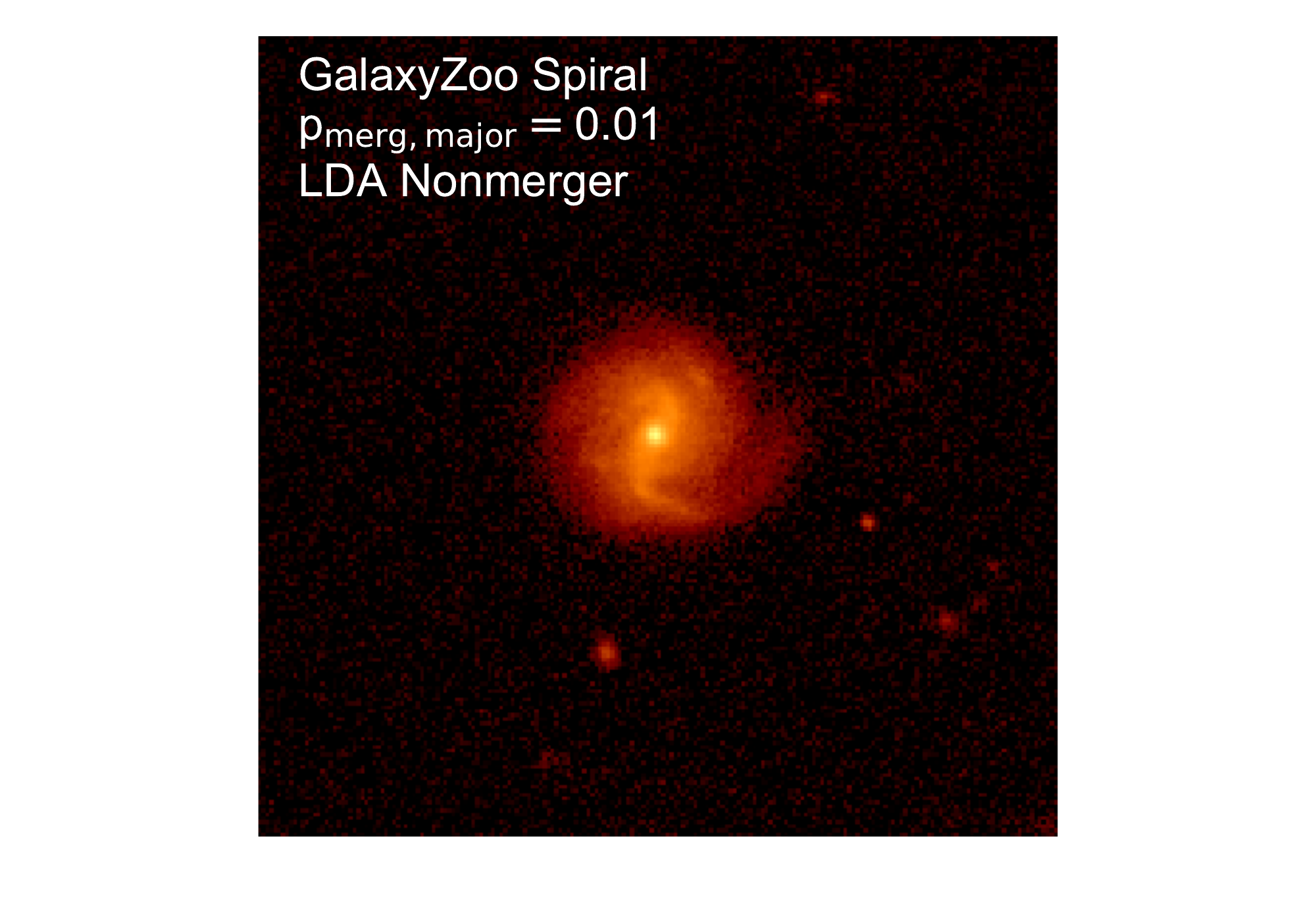}
    \caption{$r-$band images of a GalaxyZoo spiral galaxy that is identified as a merger by the minor merger technique (left) and a nonmerger by the major merger technique (right). The probabilities that the galaxy is a merger are listed.}
    \label{minormajor}
\end{figure*}
To preliminarily test the performance of the LDA technique on real images of galaxies, we apply the major and minor merger classification techniques to a sample of SDSS galaxies that have been identified as mergers, spirals, and ellipticals in GalaxyZoo (\citealt{Lintott2008,Lintott2011}).
We randomly select 50 galaxies from each galaxy morphology classification (merger, spiral, and elliptical) using a `superclean' cutoff value of $p_{\mathrm{merg}}, p_{\mathrm{el}}, p_{\mathrm{CS}} > 0.95$, where $p_{\mathrm{merg}}$ is the probability that the galaxy is a merger, $p_{\mathrm{el}}$ is the probability that the galaxy is classified as elliptical, and $p_{\mathrm{CS}}$ is the probability that the galaxy is classified under the umbrella classification of `combined spirals'. The probabilities are the percentage of GalaxyZoo users that selected a given morphology type. We require that these galaxies exceed the <S/N> cutoff value of 2.5.

We apply the major and minor merger combined LDA techniques to each of the three subsamples of galaxy types and determine the fraction of galaxies that are classified by the LDA technique as merging and nonmerging. We show some example classifications for the major merger classification tool in Figure \ref{galzoo}. For the major merger classification tool, we find that 86\% of the GalaxyZoo mergers are identified as mergers, 73\% of the GalaxyZoo elliptical galaxies are identified as mergers, and 14\% of the Galaxy Zoo spirals are classified as mergers. We show images of these classification categories in Figure \ref{galzoo} for the major merger classifier. For the minor merger classification, 93\% of GalaxyZoo mergers are classified as mergers, 61\% of the GalaxyZoo ellipticals are classified as mergers, and 43\% of the GalaxyZoo spirals are classified as mergers.

The difficulty in using the GalaxyZoo sample as a test sample is that we do not know a priori which galaxies are merging. We are able to use the GalaxyZoo merger sample as `true' mergers since they are obvious visual mergers (classified as such by GalaxyZoo users), so the fraction of true positives and false negatives is reliable. The GalaxyZoo users were conservative, only reluctantly classifying the most obvious mergers as such (\citealt{Darg2010}). However, we are unable to adequately establish the relative fractions of false positives and true negatives since the samples of GalaxyZoo ellipticals and spirals may be contaminated by merging galaxies that lack the obvious visual signs of clear major mergers such as tidal tails. Therefore, our discussion mainly relies upon the fraction of true positives and false negatives from the classification of the GalaxyZoo mergers and only briefly discusses true negatives and false positives from the GalaxyZoo ellipticals and spirals population. We plan to delve into this discussion in more depth in future work that presents the classification of real galaxies (Nevin et al. (2019, in prep)).

The major merger classifier recovers $\sim$86\% of the GalaxyZoo mergers. This fraction agrees with the fraction of true positives and false negatives from the simulation measured in Appendix \ref{accuracy}, where 82\% of true mergers are identified as such. Figure \ref{galzoo} shows a failure mode of the major merger classifier. The classifier fails to identify the GalaxyZoo merger in the top right of the figure as a merger because while it appears to be two separate galaxies by eye, the two galaxies are symmetrically aligned in such a way that $A$ and $A_S$ are low. We plan to investigate the failure modes of the technique in more detail in Nevin et al. (2019, in prep). 

The minor merger classification identifies 93\% of the GalaxyZoo mergers as such which is more than predicted (62\%) when testing on simulated galaxies. It is important to note that the GalaxyZoo mergers are more likely to be major mergers in their early phases with clear visual disturbances. Therefore, the minor merger tool performs better than expected since it is applied to galaxies that are more easily identified than most minor mergers in the simulated galaxy sample.

The fraction of nonmergers (according to GalaxyZoo) that are identified as mergers by the LDA technique is low for the spiral galaxies and high for the elliptical galaxies. For instance, the major merger classifier tool identifies 73\% (14\%) of the GalaxyZoo elliptical (spiral) galaxies as mergers. The spiral galaxy false positive rate is closer to the 3\% false positive rate from the major merger classifier when it is tested on simulated galaxies. However, there are an excess of galaxies identified as mergers in the GalaxyZoo elliptical galaxy population. It is not obvious that these galaxies have undergone mergers recently from visual inspection. While it is possible that some of these galaxies have recently merged, many of them may be false positives. It is important to note that the galaxies that are classified as mergers among the GalaxyZoo elliptical galaxies have $p_{merg}$ values that are barely above the 0.5 threshold we define here for the LDA classification. In future work, we plan to set multiple probability thresholds to eliminate false positives amongst our merger samples. Finally, since the LDA technique was constructed from disk-dominated galaxies, it is most accurate and precise at classifying galaxies that most closely match the specifications of the simulated sample. Therefore, it may not be as accurate for elliptical galaxies. We plan to address this concern in future work (Nevin et al. (2019, in prep)) when we apply the classification to SDSS galaxies.

There is an interesting discrepancy between the false positive rate of the major and minor merger technique for the sample of GalaxyZoo spirals. The major merger technique identifies 14\% of the sample as merging whereas the minor merger technique identifies 43\% of the sample as merging. Of the GalaxyZoo spirals, there are 16 galaxies that are classified as merging by the minor merger technique that are classified as nonmerging by the major merger technique. We visually inspect these galaxies, and find that $\sim$50\% could be classified as spirals with disturbed structure while $\sim$40\% have a secondary point source component that could either be a star forming region or a stellar bulge. More follow up work is required here, but preliminarily, it appears that the minor merger technique is identifying some possible minor mergers that have been missed by GalaxyZoo users. It is possible that it could also be identifying star forming regions because while we prescribe a 10\% flux threshold for the fitting of $n$, it could be fitting bright star forming regions that exceed this threshold. More work is required to distinguish between these two possibilities. It is expected that the major merger technique also misses some of these minor mergers. We show an example of a possible minor merger that was detected by the minor merger technique but not by the major merger technique in Figure \ref{minormajor}. This galaxy has disturbed spiral structure and some possible secondary point sources that could be star forming regions.

\section{Conclusion}
\label{conclude}

We create a suite of merging and nonmerging disk-dominated galaxies with different gas fractions, mass ratios, and bulge-to-total mass ratios using \texttt{GADGET-3} hydrodynamics simulations. We use the dust radiative transfer code \texttt{SUNRISE} to produce resolved dust-attenuated optical spectra from the simulations, from which we extract SDSS $r-$band images at $\sim0.1$ Gyr intervals. These snapshots cover the early, late, and post-coalescence stages of the simulated mergers. We then `SDSS-ize' the simulated images of galaxies, introducing residual noise and convolving to the seeing limit of the SDSS survey. We use these `SDSS-ized' images to measure seven different imaging predictors ($Gini$, $M_{20}$, concentration ($C$), asymmetry ($A$), clumpiness ($S$), S\'ersic index ($n$), and shape asymmetry ($A_S$)), which we combine to create a Linear Discriminant Analysis (LDA) classification scheme. This classification technique is able to accurately identify merging galaxies over a range of mass ratios, gas fractions, viewing angles, and merger stages. We create two overall classifications, one for major mergers and one for minor mergers, that we will apply to classify SDSS galaxies, assigning each galaxy a posterior probability of being a merging galaxy. Based on these results we make the following conclusions:

\begin{itemize}

\item The LDA technique outperforms previous merger identification methods such as $Gini-M_{20}$, $A$, and $A_S$ in terms of accuracy and precision. While the precisions of $Gini-M_{20}$, $A$, and $A_S$ are high with few false positives, the accuracies vary between $40-90$\%, and change with merger mass ratio. The LDA technique improves upon this with accuracies of 85\% (81\%) and precisions of 97\% (94\%) for the combined major (minor) merger simulations. The LDA accuracy and precision varies little with merger initial conditions (< 10\%), indicating that the LDA technique is more stable and accurate than individual predictor merger identification techniques.

\item The LDA technique lengthens the timescale of observability of merging galaxies (> 2 Gyr) and the galaxy mergers are identified at all stages (early, late, and post-coalescence) of a merger. The observability timescales for $Gini-M_{20}$, $A$, and $A_S$ are $0.2-0.8$, < 0.1, and $2.2 - 7.8$ Gyr, respectively. The LDA technique incorporates many imaging predictors and is therefore able to combine the strengths of all these imaging predictors to be sensitive to all stages of the galaxy mergers.

\item The predictor coefficients of LD1 change little with gas fraction and are most affected by the mass ratio of the merging galaxies. For instance, $A$ and $A_S$ are important for major mergers due to their visually disturbed morphology while $C$ and $Gini$ are more important for minor mergers, since they show consistent enhancement in light concentration as the merger progresses. This supports the idea that even minor mergers can build stellar bulges. $A$ is an important coefficient for a range of mass ratios, identifying major and minor mergers alike.

\end{itemize}

We plan to apply this imaging LDA technique to the SDSS galaxies (Nevin et al. (2019, in prep)). Additionally, since the MaNGA survey is an imaging and IFS survey, we will incorporate several kinematic predictors based on the stellar velocity and stellar velocity dispersion maps from the hydrodynamics simulations into this analysis to improve the classification.

\section{Acknowledgements}
We would like to thank the anonymous referee whose comments greatly improved the quality of this manuscript.

R. N. would like to thank Aaron Stemo, Tom Nummy, Adalyn Fyhrie, Dan Gole, Hilary Egan, Mike Stefferson, and Kate Rowlands; this paper would not have been possible without your excellent statistical advice, expertise on machine learning techniques, and help with supercomputing. Additionally, R. N. would like to thank Kevin Bundy, Kyle Westfall, Michael Blanton, and David Law for their help with understanding the MaNGA survey and imaging parameters. Finally, R. N. thanks Vicente Rodriguez-Gomez for his invaluable help with \texttt{statmorph} and measuring imaging predictors of mock images.

R. N. and J. M. C. are supported by NSF AST-1714503.

L. B. acknowledges support by NSF award \#1715413.

This work used the Extreme Science and Engineering Discovery Environment (XSEDE), which is supported by National Science Foundation grant number ACI-1548562. We specifically utilized Comet and Oasis through the XSEDE allocation for `An Imaging and Kinematic Approach for Improved Galaxy Merger Identifications' (TG-AST130041). We would also like to acknowledge the help of Martin Kandes, who assisted with the optimization of the LDA tool.

This work utilized the RMACC Summit supercomputer, which is supported by the National Science Foundation (awards ACI-1532235 and ACI-1532236), the University of Colorado Boulder, and Colorado State University. The Summit supercomputer is a joint effort of the University of Colorado Boulder and Colorado State University.

Funding for the Sloan Digital Sky Survey IV has been provided by the Alfred P. Sloan Foundation, the U.S. Department of Energy Office of Science, and the Participating Institutions. SDSS-IV acknowledges
support and resources from the Center for High-Performance Computing at
the University of Utah. The SDSS web site is www.sdss.org.

SDSS-IV is managed by the Astrophysical Research Consortium for the 
Participating Institutions of the SDSS Collaboration including the 
Brazilian Participation Group, the Carnegie Institution for Science, 
Carnegie Mellon University, the Chilean Participation Group, the French Participation Group, Harvard-Smithsonian Center for Astrophysics, 
Instituto de Astrof\'isica de Canarias, The Johns Hopkins University, 
Kavli Institute for the Physics and Mathematics of the Universe (IPMU) / 
University of Tokyo, Lawrence Berkeley National Laboratory, 
Leibniz Institut f\"ur Astrophysik Potsdam (AIP),  
Max-Planck-Institut f\"ur Astronomie (MPIA Heidelberg), 
Max-Planck-Institut f\"ur Astrophysik (MPA Garching), 
Max-Planck-Institut f\"ur Extraterrestrische Physik (MPE), 
National Astronomical Observatories of China, New Mexico State University, 
New York University, University of Notre Dame, 
Observat\'ario Nacional / MCTI, The Ohio State University, 
Pennsylvania State University, Shanghai Astronomical Observatory, 
United Kingdom Participation Group,
Universidad Nacional Aut\'onoma de M\'exico, University of Arizona, 
University of Colorado Boulder, University of Oxford, University of Portsmouth, 
University of Utah, University of Virginia, University of Washington, University of Wisconsin, 
Vanderbilt University, and Yale University.

\bibliographystyle{apj}
\bibliography{library_overleaf}

\appendix

\section{Initial Conditions}
\label{ICs}

We vary the initial masses, mass ratios, gas fractions, and B/T (bulge-to-total mass) ratios of the merging galaxy models based upon previous work with similar merger simulations (e.g., \citealt{Cox2008,Lotz2008,blecha11,Blecha2018}). Additionally, we select the values for these initial conditions based upon the range of observed values for present day galaxies in SDSS as in \citet{Cox2008}. Our goal is to produce simulated mergers that are typical of merging galaxies in SDSS and also comparable to previous work with the imaging predictors of simulated galaxies (e.g., \citealt{Cox2008,Lotz2008}).

Our simulations span a range in total stellar mass $10.6 < \mathrm{log} \ M_{\star} (M_{\odot}) < 10.8$, which agrees well with the fiducial models used in \citet{Cox2008} that have a range in total mass $9.0 <\mathrm{log} \ M_{\star} (M_{\odot}) < 10.7$. SDSS galaxies span a range in stellar mass (for individual galaxies) of $9 < \mathrm{log} \ M_{\star} (M_{\odot}) < 11$. When we compare this to the total mass of the merger simulations (which combine two galaxies), we find that the simulated galaxies are in the middle of the expected mass range for SDSS galaxies.

We vary the total mass ratio between 1:2 and 1:10 to capture three major merger simulations and two minor merger simulations. We are able to compare to \citet{Lotz2010b} and \citet{Cox2008}, who choose mass ratio ranges of 1:1-1:20 and 1:1-1:22.7, respectively. 

We select gas fractions between $0.1-0.3$. This range is typical of the SDSS galaxies, which have gas fractions between 0 and 0.5 (\citealt{Catinella2010}). The mean gas fraction of the SDSS population is $\sim0.2-0.3$, which is in good agreement with our choice to run most galaxy simulations with a gas fraction of 0.3. Additionally, \citet{Cox2008} and \citet{Lotz2010a} vary the gas fraction between $0.2-0.4$ and $0.2-0.5$, respectively, providing a good amount of overlap for comparison of results.

Most of our simulations do not have stellar bulges, but we do include bulges in the minor merger simulations. \citet{Cox2008} demonstrate that bulges act to stabilize the disk of the galaxy for large mass ratio mergers (they see this effect primarily for 1:5 to 1:20 mass ratio mergers), leading to less disturbed morphology than bulgeless mergers. Since this effect is most prominent for minor mergers, we include stellar bulges in the progenitor galaxies for these simulations. We lack the computation resources to additionally investigate this effect for major mergers.

The B/T ratio depends on the total stellar mass of a galaxy and SDSS galaxies range between $9 < \mathrm{log} \ M_{\star} (M_{\odot}) < 11$. Measured B/T ratios for this stellar mass range for galaxies in SDSS span $0-0.6$ (\citealt{Bluck2014}). 
Therefore, our choice of a B/T ratio of 0.2 for the minor merger simulations is typical of SDSS galaxies. We use a matched sample of isolated galaxies so that the slightly enhanced $C$ and $n$ values for the minor mergers (relative to the major mergers) are accounted for in the LDA technique.

\section{Merging Galaxy Priors}
\label{Apriors}
LDA requires a prior to characterize the dataset if the relative numbers of objects in each class are not representative of the overall population. If the frequency of merging and nonmerging galaxies in our simulated dataset exactly matched the frequency of merging and nonmerging galaxies in nature, our priors would be [0.5, 0.5] and would not need to be specified. However, since we have a lower frequency of nonmerging galaxies in our inputs to LDA relative to the frequency of nonmerging galaxies in reality, we use the fraction of merging galaxies in the universe ($f_{\mathrm{merg}}$) as our prior.

We use a different fraction for major and minor mergers. For major mergers, we use the fraction of merging galaxies, $f_{\mathrm{merg}} = 0.1$ from \citet{Lotz2011}. This is an average, calculated from different $f_{\mathrm{merg}}$ measurements that rely upon $Gini$-$M_{20}$ and $A$ to determine merger fractions for galaxies in the local universe ( \citealt{Lotz2008,Conselice2009,Lopez-Sanjuan2009,Shi2009}). We choose not to use pair fractions to determine $f_{\mathrm{merg}}$ as they tend to underestimate the fraction of merging galaxies since pair studies are only sensitive to the early stages of a merger. 

It is important to note that \citet{Lotz2011} use individual predictors (such as $Gini-M_{20}$ or $A$ alone) to identify mergers, and find short timescales of observability ($\sim0.2-0.6$ Gyr). As discussed in Section \ref{timescale}, we find timescales of observability $>$ 2 Gyr from the LDA technique and therefore $f_{\mathrm{merg}} = 0.1$ is a conservative estimate. In reality, observed merger fractions may be underestimated in the local universe (particularly for minor mergers) since the observability timescales of past imaging methods are short. Using merging galaxies in the Millennium simulation, \citet{Bertone2009} find that the estimate of $f_{\mathrm{merg}}$ for minor and major mergers increases by a factor of $2-10$ if the observability timescale is increased from 0.4 Gyr to 1 Gyr.

While the fraction of minor mergers is less certain, studies have indicated that it is $3-5$ times greater than the major merger rate, so we use $f_{\mathrm{merg}} = 0.3$ for the minor merger simulations (e.g., \citealt{Bertone2009,Lotz2011}).

For comparison purposes, we also estimate $f_{\mathrm{merg}}$ from the Illustris simulation. Using estimations of the timescale of the merger, we convert from merger rate (measured directly from Illustris to be $\sim0.1$ Gyr$^{-1}$ (\citealt{Rodriguez-Gomez2015})) to the merger fraction of galaxies in the local universe. If we multiply this rate by the $\sim2$ Gyr timescale estimate from our work, we find $f_{\mathrm{merg}}$ = 0.2, which is in good agreement with the 0.1 value for $f_{\mathrm{merg}}$ from observations of merging galaxies in the literature. If we use the $0.2-0.6$ Gyr timescale from \citet{Lotz2008}, we find a much lower merger fraction of $f_{\mathrm{merg}}=0.02-0.06$.

\begin{figure*}
\centering
\includegraphics[scale=0.7]{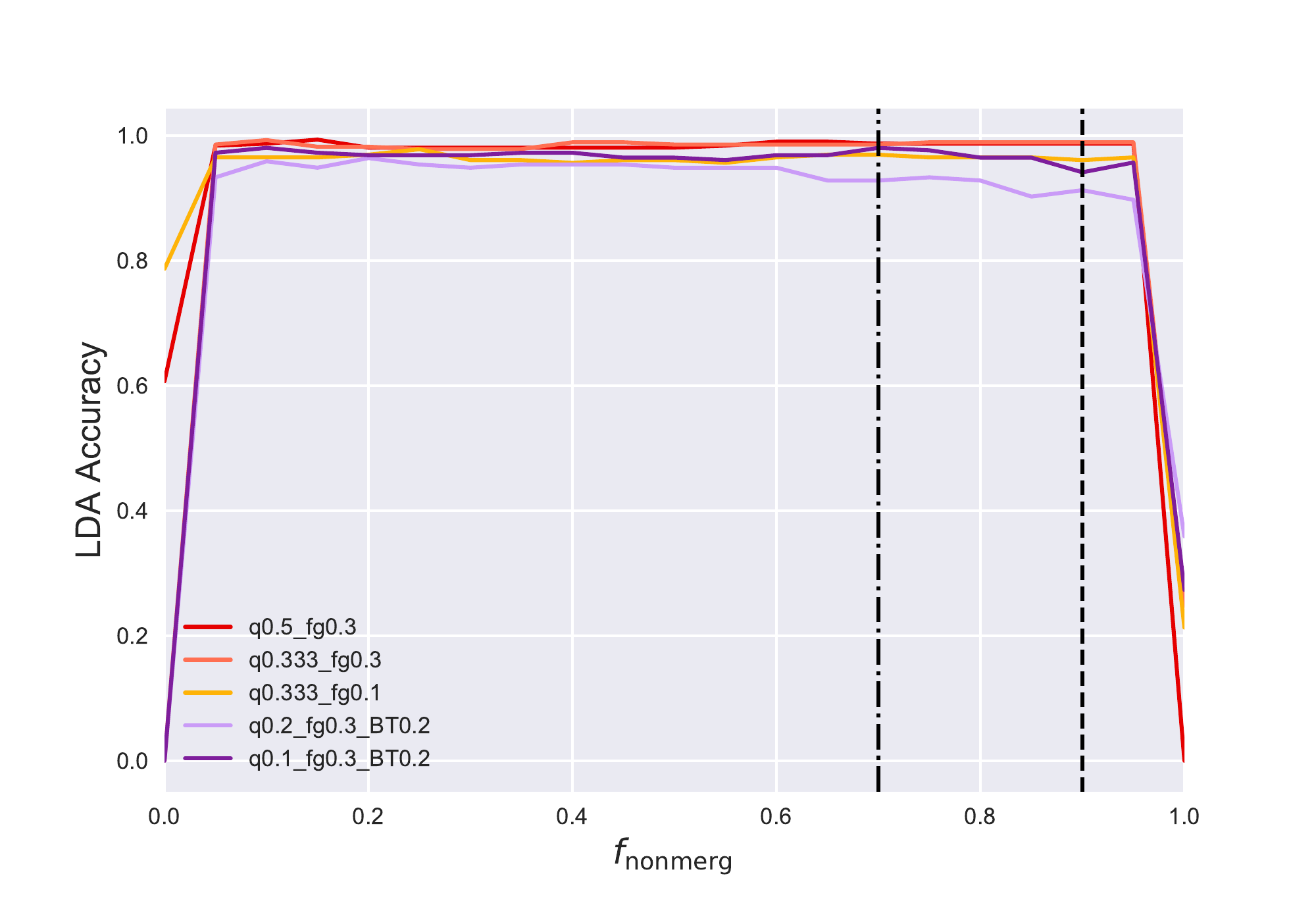}
\caption{Accuracy of the LDA with varying choices for the priors, or the fraction of galaxies that are merging and nonmerging. The vertical dashed lines represent our choice of priors for minor ($f_{\mathrm{nonmerg}}$ = 0.7) and major mergers ($f_{\mathrm{nonmerg}}$ = 0.9).}
\label{fig:priors}
\end{figure*}

Since the estimates of $f_{\mathrm{merg}}$ are so uncertain, we compare the results of using different values for $f_{\mathrm{nonmerg}}$ on the outcome of the LDA in Figure \ref{fig:priors}. For each simulation, we have more snapshots of merging galaxies than nonmerging, which is not reflective of reality. We use the LDA accuracy to measure the sensitivity of the technique to the input priors. We find that the LDA is relatively insensitive to prior selection within a range of priors on the fraction of merging and nonmerging galaxies. This range exists from $0.1 < f_{\mathrm{nonmerg}} < 0.9$. As we increase $f_{\mathrm{nonmerg}}$ above 0.9, we start to see the accuracy decline from $\sim80-90$\% correct identifications to $60-70$\%. While this is a significant decline, the decline is somewhat asymptotic. Therefore, at our chosen prior for major mergers (0.9), the accuracy has declined to around 90\% for the three major merger simulations pictured, which is still very high. Additionally, while the minor merger simulations are less accurate as $f_{\mathrm{nonmerg}}$ increases, the prior for minor mergers is 0.7, so they do not fall off as fast in accuracy at this point.

\section{Testing the Assumptions of LDA with Multivariate Analysis}
\label{mva}

We carry out a simple multiple linear regression analysis to test the assumptions of LDA and examine the input predictors. Many of these techniques as well as an introduction to LDA are covered in \citet{James2013}. The key assumptions of LDA include multivariate normality, homoscedasticity (that the covariance among groups is equal), and an absence of multicollinearity. However, it should also be noted that LDA is relatively robust to slight violations of these assumptions and can still be reliable even when certain assumptions are violated (\citealt{Duda2001,TaoLi2006}). We conduct a preliminary multivariate analysis of the input predictors to screen for multicollinearity and violations of normality and homoscedasticity. We present our results for the major merger and minor merger combined simulations and show plots just for the major merger combined simulation.

\begin{figure*}
\centering
\includegraphics[scale=0.65]{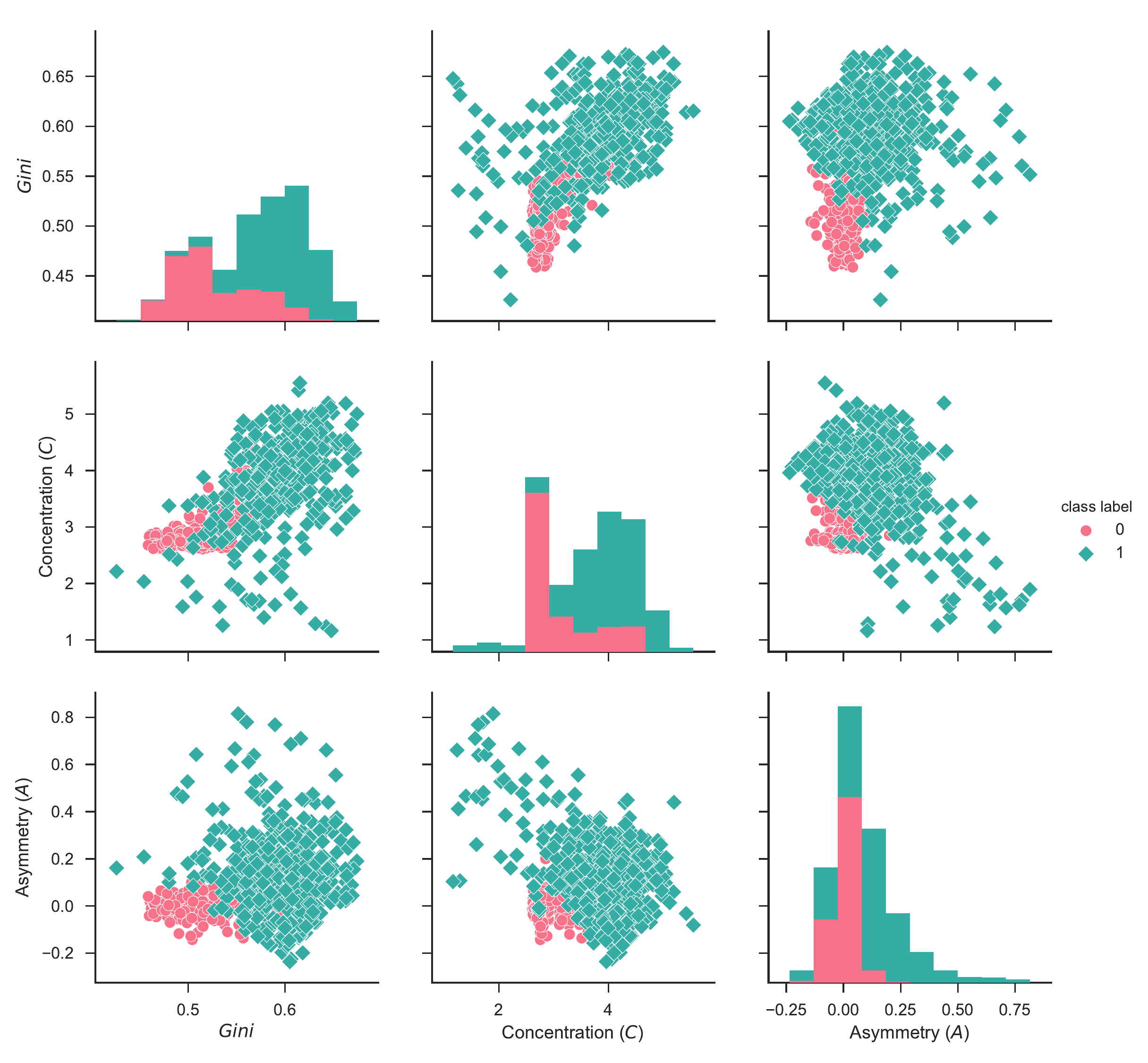}
\caption{Scatter plots and marginalized histograms of the distribution for three input predictors ($Gini$, $C$, and $A$) for the major merger combined simulation. The nonmerging class is pink (0) and the merging class is green (1). The predictors violate normality and homoscedasticity assumptions.}
\label{correlated}
\end{figure*}

We first address the multivariate normality assumption by examining the individual histograms of the input predictors for both of the combined simulations (the major merger combined simulation is Figure \ref{correlated}). Visually, the predictors do not seem to be drawn from a normal distribution. We conduct Shapiro-Wilke and Kolmogorov Smirnov tests for normality and in both cases we are able to reject the null hypothesis that the data are drawn from a Gaussian multivariate normal distribution for the majority of predictors. 

We also address the homoscedasticity assumption in Figure \ref{correlated}. By examining the distributions of the values for each class for a cross-section of input predictors, we are able to determine that the covariances for each class are not equal. We conduct a Levene test and confirm that we can reject the null hypothesis and that the covariance matrices are not equal. This is unsurprising given that multivariate normality is also violated.

\begin{figure*}
\includegraphics[scale=0.8]{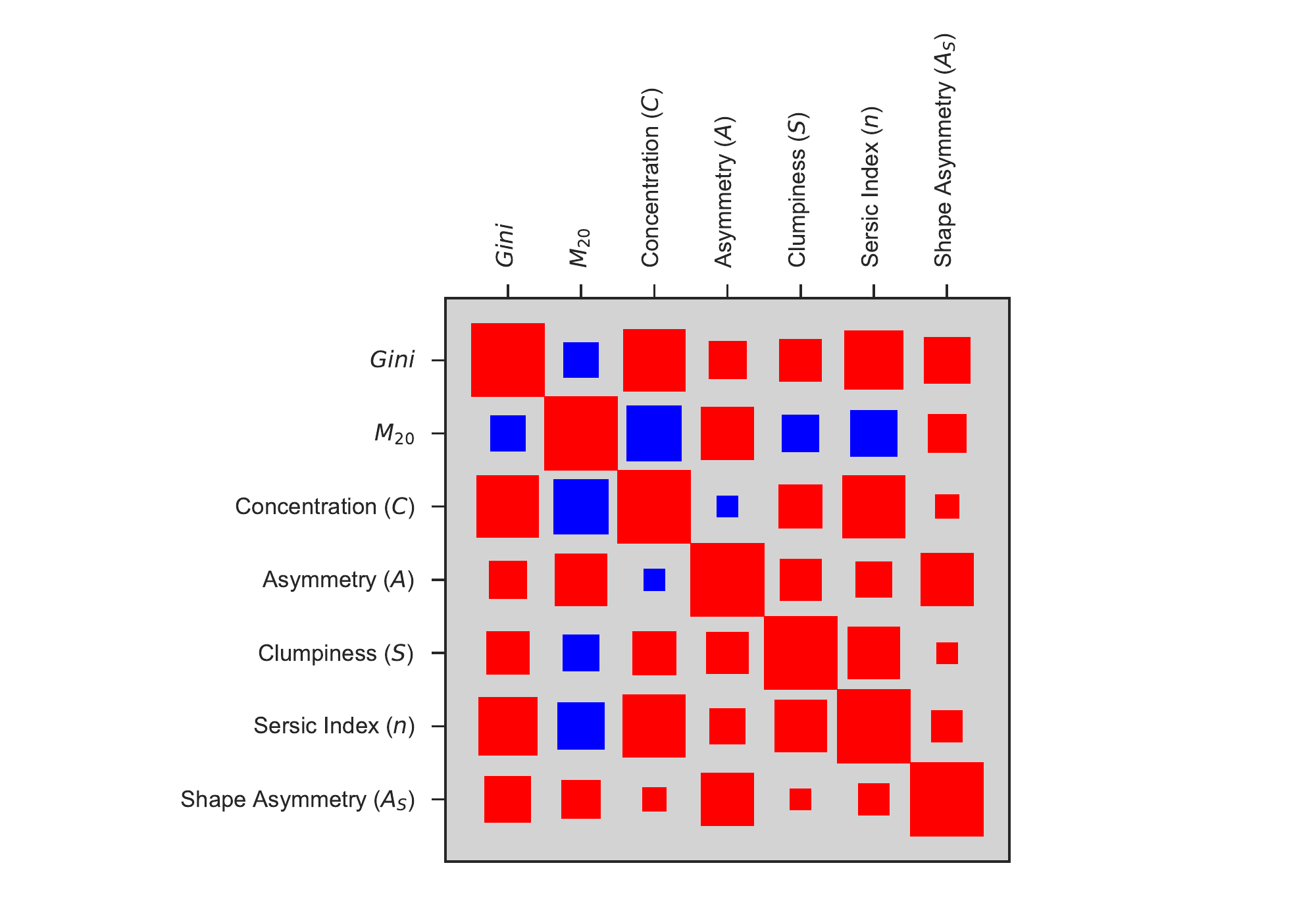}
\caption{Hinton visual correlation matrix for the major merger combined simulation. A red box represents a positive correlation and a blue box is a negative correlation. The size of the box is the strength of Pearson's r correlation coefficient. For scale, the diagonal boxes have a Pearson's r values of one.}
\label{hinton}
\end{figure*}

We next examine the relationships between predictors to determine if the predictors demonstrate multicollinearity. A violation of multicollinearity could lead to a decrease in the predictive ability of LDA. We screen for multicollinearity visually in Figure \ref{hinton} using a Hinton visual diagram where the size and color of the boxes indicate the strength and sign (positive is red and negative is blue) of the correlation. We find that many predictors have a large positive correlation. We further examine the strength of the correlation and find that $n$ and $C$ have the most significant correlation for the major mergers with a Pearson's r value of 0.72. For the minor mergers, $M_{20}$ and $A$ have the largest correlation with an r value of 0.66. All Pearson's r values are below 0.99, the threshold value for multicollinearity, so we can rule out multicollinearity in this dataset. However, we must still deal with these correlations using interaction terms, which remove cross-correlation effects from the coefficients of LD1 so that we can individually assess trends with the seven main coefficients (\citealt{James2013}).

To screen for outliers we use box and whisker plots. Outliers can affect the LDA classification, dominating the analysis. We find that a few inputs are greater than 1.5 times the interquartile range (as indicated by the extent of the whiskers). However, overall, there are very few outliers. We also calculate the z scores for each predictor and find that none are outside 3$\sigma$ from the sample mean.

To verify that each predictor is necessary in the LDA, we conduct Ordinary Least Squares (OLS) fitting. We first linearly regress each predictor against the class label (a binary variable for merger/nonmerger classification). While a logit regression would be a better tool with more than two classes, a linear regression is appropriate here since the response is binary (\citealt{James2013}). Additionally, for a binary response variable, LDA is quite similar to multiple linear regression. We find that almost all predictors have an p-value for the t-test below 0.05, indicating that there is a significant relationship between the predictor and the class, or in other words, that the predictor is required for classification. The only predictors that fail this test are the $Gini$, $M_{20}$, and $S$ predictor, which fail for the minor merger simulations, and the $n$ predictor, which fails for the major merger simulation. We also find in our LDA modeling that the $Gini$, $M_{20}$, and $S$ predictors are fairly unimportant for the minor merger simulation and that $n$ is unimportant for the major merger simulation. We also run the OLS fitting for all simulations and find that there are no predictors that are unimportant across the board. Therefore, we include all of the predictors in the LDA classification. We ultimately discover that all predictors are important according to the forward stepwise selection for certain simulations, so we cannot eliminate them prior to the LDA.

Overall, the data violate multivariate normality and homoscedasticity while passing the tests for multicollinearity and extreme outliers. For classification purposes, LDA is very robust to varying distributions of the data and can still achieve good performance even when the covariance matrices are not equal and normality assumptions are violated (\citealt{Duda2001, TaoLi2006}). 

One alternative approach to LDA is to utilize Quadratic Discriminant Analysis, which does not rely on the equality of covariance matrices. We test the accuracy, precision, recall, and $F_1$ score of both an LDA and a QDA method (see Appendix \ref{accuracy} for the accuracy of LDA). We find that the LDA classifier performs very well and while the QDA classifier increases accuracy, recall, and $F_1$ score by $\sim5$\%, the LDA classifier is still above $\sim85$\% for accuracy on all simulations. A downside to using QDA is that it is inherently nonlinear and does not allow us to directly interpret each predictor individually. We choose to use LDA since it allows for a more practical interpretation of the imaging predictors and since it does an adequate job of separating the merging and nonmerging classes across all simulations.

Additional ways to prepare the input data for better classification include increasing the number of observations (number of galaxy snapshots) and standardizing the data. We already have at least 20-30 snapshots per class and are therefore more robust to violations of normality and homogeneity of covariance. Also, when we combine all the individual runs of LDA for the final combined major and minor merger runs, we increase the sample size to at least 100 observations per class. This final analysis is robust to violations of normality and homoscedasticity. 

We also find that the predictors require standardization prior to the input to LDA; they have very different means and standard deviations, which could affect the output of LDA. For instance, if one variable has a large mean, it could dominate the first discriminant axis (LD1). We standardize the input predictors to all have a mean of zero and a standard deviation of one prior to our LDA classification.

\section{Forward Stepwise Variable Selection and k-fold Cross-Validation}
\label{Akfold}

\begin{figure*}
\centering

\includegraphics[scale=0.3]{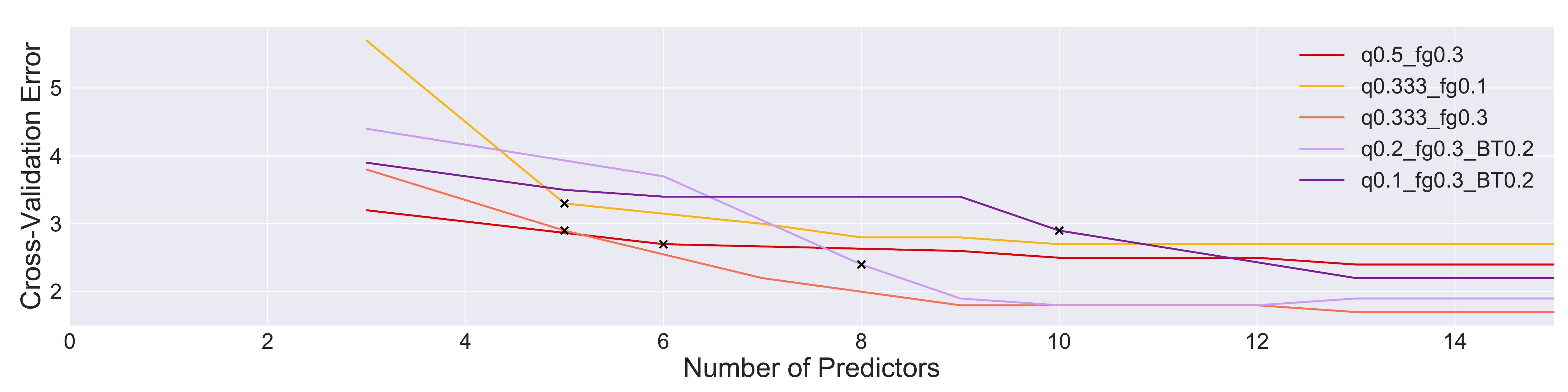}
\caption{Forward stepwise selection of the number of predictors for each run of LDA. We mark the minimum number of `required' predictors for each run with black xs. This point is within one standard error of the minima of the cross-validation error curve for each run. We run LDA for each simulation using the predictors selected from this method.}
\label{cv}
 \end{figure*}

A limitation of LDA is that there are no standard errors on the LDA predictor coefficients. We use the stratified $k-$fold cross-validation method with ten folds to return an estimate of the underlying distribution of possible values for the LDA coefficients given the data. $k-$fold cross-validation functions by dividing the sample into $k$ equal sized samples, where the first $k$-1 samples will be used as the training set and the $k$th sample will be used as the test set. We repeat the LDA $k$ times and estimate the mean and standard deviation of the LDA coefficients from the data. Stratified cross-validation specifically requires that the test sample includes a number of snapshots from each class that are representative of the overall sample. 

This method is effective for minimizing bias and variance given the correct choice of $k$ (\citealt{James2013}). For instance, \citet{Efron1983} prove that $k-$fold cross-validation is `almost unbiased' if $k$ is large and the sampling is random (this approaches leave one out cross-validation (LOOCV), when $k=n$ where $n$ is the sample size). However, LOOCV has a high variance since it involves finding the variance of $n$ fitted models which are trained on nearly identical data. The mean of highly correlated quantities has a higher variance, so we choose an intermediate value of $k$ so that we avoid high bias and high variance. \citet{Kohavi1995} find that stratified $k-$fold cross-validation with ten folds is the most effective at model selection even if computation power allows for more folds. We also find that $k$=10 is a good choice to ensure that the number of misclassifications (cross-validation error) is minimized and that the mean and standard error of the LD1 coefficients is stable. 

We use forward stepwise selection with $k-$fold cross-validation to determine which predictors are necessary to build LD1 for each simulation. The purpose of this process is to avoid introducing excess predictors that are unnecessary to the separation of the merging and nonmerging galaxies along LD1. We choose forward stepwise selection since it is less computationally expensive than best subset selection (\citealt{James2013}). 

Forward stepwise selection begins with a model without predictors. It then determines which predictor to add by comparing the cross-validation error for each predictor. The cross-validation error is also the number of misclassifications corresponding to each model. For instance, the first step of forward stepwise selection is comparing the cross-validation error from a model with only one predictor (i.e., $Gini$) to all other possible models with only one term (i.e., a model with $M_{20}$, a model with $M_{20}*Gini$, etc.). The one-term model with the lowest number of misclassifications is selected. Next, the forward stepwise election iteratively attempts to add all remaining predictors to to the model. It chooses to add a term only if the cross-validation error of the overall model is less than that of the previous step. Again, it adds the term that minimizes the cross-validation error as compared to all other possible terms. 

The forward stepwise selection proceeds until no more terms are required to decrease the cross-validation error. We refer to the predictors in the final model as the `required' predictors. We show the process in Figure \ref{cv}, where we determine the number of predictors necessary for the LD1 for each simulation by minimizing the cross-validation error with forward stepwise selection. We additionally use the one-standard-error rule from \citet{James2013} to select the best overall model. This allows us to select the simplest model for which the estimated cross-validation error is within one standard error of the lowest point on the curve in Figure \ref{cv}. The standard error of the cross-validation error is the standard deviation of the number of misclassifications for all ten $k-$folds.

\section{LDA Performance: Accuracy and Precision of the Classifier}
\label{accuracy}

\begin{figure*}
\hspace{-0.5cm}
\includegraphics[scale=0.45]{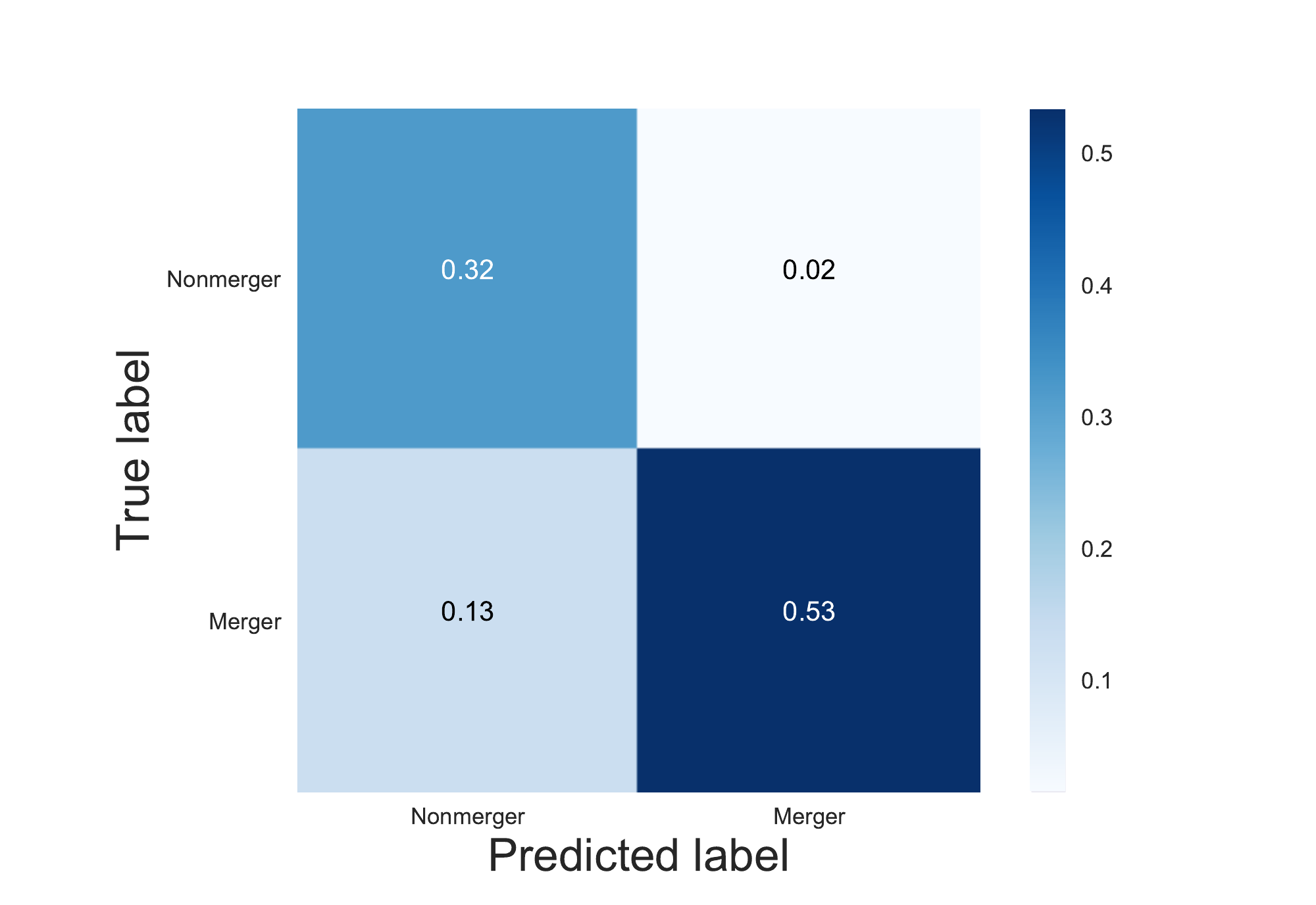}
\hspace{-1.5cm}
\includegraphics[scale=0.45]{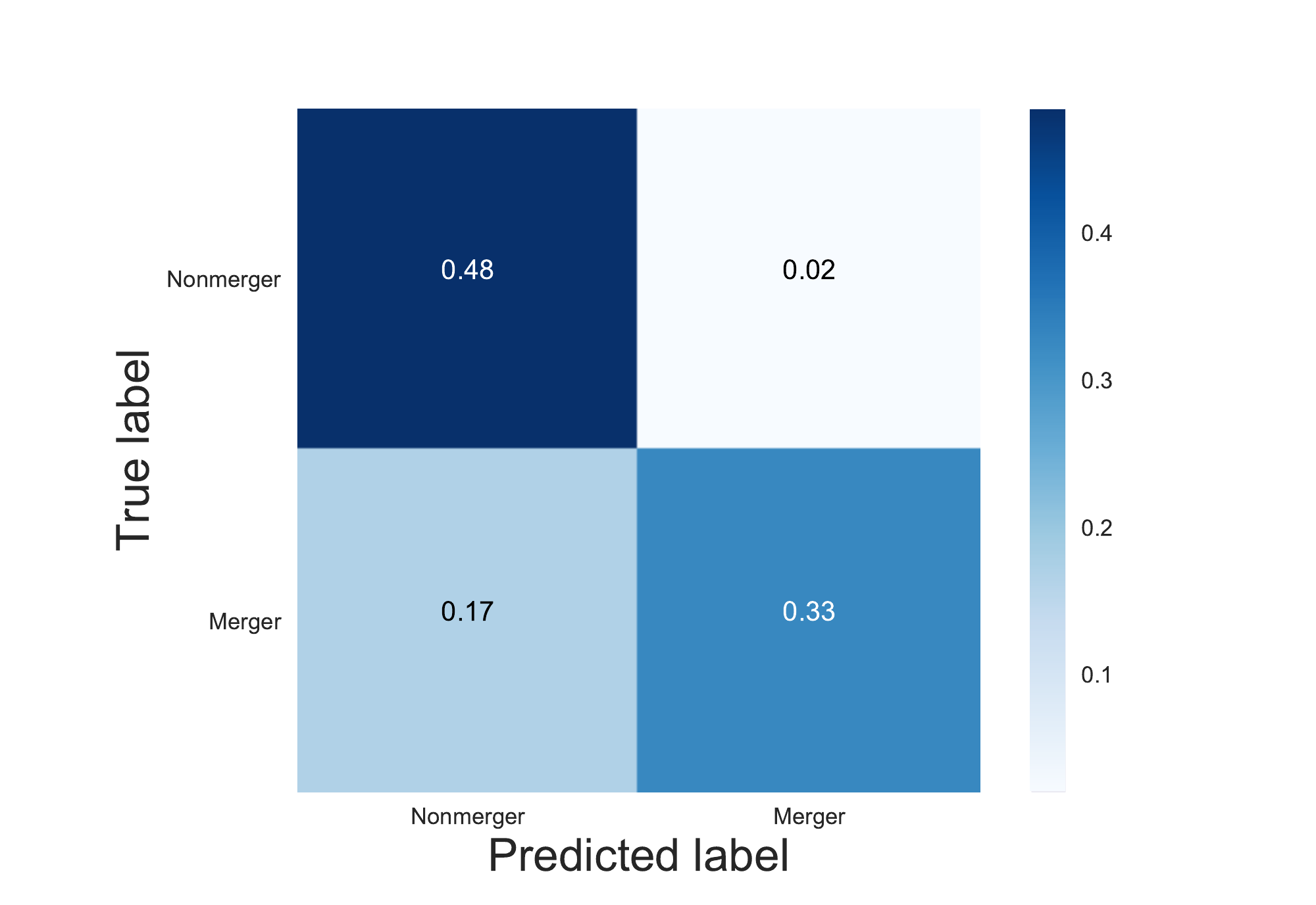}
\hspace{-0.5cm}
\caption{Confusion matrices for the major (left) and minor (right) combined simulations. The y axis represents the true categories of the test set of galaxies from the simulation. The x axis is the predicted category from the LD1 classifier for the test set of galaxies.}
\label{confusion}
\end{figure*}

We investigate the performance of the classifier using the confusion matrix (Figure \ref{confusion}). The confusion matrix is constructed using the ten randomized test and training sets of galaxies, which were created in the $k$-fold method described in Appendix \ref{Akfold}. It is the mean confusion matrix from the ten $k$-fold runs. 

The confusion matrix shows the number of nonmerging galaxies from the test set that were correctly classified as nonmerging (upper left) and the number of merging galaxies that were correctly identified as merging (lower right), which are referred to as `True Negatives' (TN) and `True Positives' (TP), respectively. The top right corner of the confusion matrix is the galaxies that were classified as merging although they are in fact nonmerging. These are the `False Positives' or FP. The bottom left square is the galaxies that are merging but were not correctly identified as merging, or the `False Negatives' or FN. In Figure \ref{confusion} we show the normalized percentage of galaxies that fall into each category for the final combined major and minor merger runs of LDA.

From the confusion matrix, we quantify accuracy, precision, recall, and $F_1$ score of the LDA classification in order to assess the overall performance of the LDA method for each simulation. 

The accuracy is the number of correct classifications divided by the total number of classifications:

$$A = \frac{TN + TP}{TN + TP + FN + FP}$$

We assess the precision of the LDA, or the fraction of correct positive predictions:

$$P = \frac{TP}{TP+FP}$$

Recall is the fraction of true positives that are classified as such:
$$R = \frac{TP}{TP+FN} $$

The $F_1$ score is the harmonic mean of recall and precision:
$$F_1 = \frac{2TP}{2TP + FN + FP}$$

We collect the accuracy, precision, recall, and $F_1$ score values for all simulations in Table \ref{tab:pa}. We find that the LDA classification performs well, with all performance metrics around or above $\sim0.7-0.9$. Accuracy ranges from $0.85-0.91$ while precision is between $0.89-0.98$ for all runs. This confirms our discussion from Section \ref{mva}; the LDA method is accurate, and therefore we are not concerned that our violations of normality and homoscedasticity are detrimental to the classification. 

The LDA has a very high precision value. This indicates that it does an excellent job of identifying all merging galaxies as merging. This is critical to the next phase of analysis, which will include classifying SDSS galaxies; we do not wish to misidentify mergers and are more tolerant with false positives than false negatives, given that this initial classification is created from simulated galaxies.

\begin{table*}[h!]
  \begin{center}
    \caption{LDA performance. We list the accuracy, precision, recall, and $F_1$ score as defined in Appendix \ref{accuracy} for all runs of the LDA classification.}
    \label{tab:pa}
    \begin{tabular}{c|c|c|c|c}

      Simulation & Accuracy & Precision & Recall & $F_1$ Score  \\
      \hline
All Major& 0.85& 0.97 & 0.80& 0.88\\
All Minor & 0.81 & 0.94 & 0.66 & 0.78\\

	q0.5\_fg0.3 &0.91& 0.98 & 0.82 & 0.90 \\
    q0.333\_fg0.3 & 0.90 & 0.97 & 0.87 & 0.92\\
      q0.333\_fg0.1 & 0.86 & 0.96 & 0.83 & 0.89\\
      
      q0.2\_fg0.3\_BT0.2 & 0.88& 0.96& 0.78& 0.86\\
 	q0.1\_fg0.3\_BT0.2 & 0.89&0.89 &0.79 &0.84 \\

    \end{tabular}
  \end{center}
\end{table*}

\end{document}